\documentclass{article}
\usepackage{amssymb}
\usepackage{color,graphicx}
\graphicspath{{Eps/}}

\def\axexy{
\bpic(65,65)
  \put(0,30){\vector(1,0){65}}
  \put(60,25){\makebox(5,5){$x$}}
  \put(30,0){\vector(0,1){65}}
  \put(30,60){\makebox(5,5){$y$}}
\epic
}

\def\objet{
\bpic(40,40)
  \put(0,15){\line(0,1){10}}
  \put(0,25){\line(1,2){5}}
  \put(5,35){\line(1,1){5}}
  \put(10,40){\line(1,0){15}}
  \put(25,40){\line(1,-1){10}}
  \put(35,30){\line(0,-1){15}}
  \put(35,15){\line(-1,-1){10}}
  \put(25,5){\line(-1,0){10}}
  \put(15,5){\line(-1,1){5}}
  \put(10,10){\line(-2,1){10}}
\epic
}

\def\proj{
\bpic(27,27)
  \put(0,0){\line(1,2){2}}
  \put(2,4){\line(0,1){6}}
  \put(2,10){\line(1,1){4}}
  \put(6,14){\line(0,1){5}}
  \put(6,19){\line(1,1){4}}
  \put(10,23){\line(1,0){6}}
  \put(16,23){\line(1,1){4}}
  \put(20,27){\line(1,0){7}}
  \put(0,0){\line(1,1){27}}
\epic
}

\def\bdoc{\begin{document}}             \def\edoc{\end{document}}
\def\babs{\begin{abstract}}             \def\eabs{\end{abstract}}
\def\bcc{\begin{center}}                \def\ecc{\end{center}}
\def\ben{\begin{enumerate}}             \def\een{\end{enumerate}}
\def\bit{\begin{itemize}}               \def\eit{\end{itemize}}
\def\beq{\begin{equation}}              \def\eeq{\end{equation}}
\def\btabu{\begin{tabular}}             \def\etabu{\end{tabular}}
\def\barr{\begin{array}}                \def\earr{\end{array}}
\def\beqn{\begin{eqnarray}}             \def\eeqn{\end{eqnarray}}
\def\beqnx{\begin{eqnarray*}}           \def\eeqnx{\end{eqnarray*}}
\def\bfig{\begin{figure}}               \def\efig{\end{figure}}
\def\bpic{\begin{picture}}              \def\epic{\end{picture}}
\def\bm#1{\mbox{\boldmath $#1$}}
\def\wt#1{\widetilde{#1}}
\def\wh#1{\widehat{#1}}

\def\ab{{\bm a}}\def\Ab{{\bm A}}\def\Ac{{\cal A}}
\def\bb{{\bm b}}\def\Bb{{\bm B}}\def\Bc{{\cal B}}
\def\cb{{\bm c}}\def\Cb{{\bm C}}\def\Cc{{\cal C}}
\def\db{{\bm d}}\def\Db{{\bm D}}\def\Dc{{\cal D}}
\def\eb{{\bm e}}\def\Eb{{\bm E}}\def\Ec{{\cal E}}
\def\fb{{\bm f}}\def\Fb{{\bm F}}\def\Fc{{\cal F}}
\def\gb{{\bm g}}\def\Gb{{\bm G}}\def\Gc{{\cal G}}
\def\hb{{\bm h}}\def\Hb{{\bm H}}\def\Hc{{\cal H}}
\def\ib{{\bm i}}\def\Ib{{\bm I}}\def\Ic{{\cal I}}
\def\jb{{\bm j}}\def\Jb{{\bm J}}\def\Jc{{\cal J}}
\def\kb{{\bm k}}\def\Kb{{\bm K}}\def\Kc{{\cal K}}
\def\lb{{\bm l}}\def\Lb{{\bm L}}\def\Lc{{\cal L}}
\def\mb{{\bm m}}\def\Mb{{\bm M}}\def\Mc{{\cal M}}
\def\nb{{\bm n}}\def\Nb{{\bm N}}\def\Nc{{\cal N}}
\def\ob{{\bm o}}\def\Ob{{\bm O}}\def\Oc{{\cal O}}
\def\pb{{\bm p}}\def\Pb{{\bm P}}\def\Pc{{\cal P}}
\def\qb{{\bm q}}\def\Qb{{\bm Q}}\def\Qc{{\cal Q}}
\def\rb{{\bm r}}\def\Rb{{\bm R}}\def\Rc{{\cal R}}
\def\sb{{\bm s}}\def\Sb{{\bm S}}\def\Sc{{\cal S}}
\def\tb{{\bm t}}\def\Tb{{\bm T}}\def\Tc{{\cal T}}
\def\ub{{\bm u}}\def\Ub{{\bm U}}\def\Uc{{\cal U}}
\def\vb{{\bm v}}\def\Vb{{\bm V}}\def\Vc{{\cal V}}
\def\wb{{\bm w}}\def\Wb{{\bm W}}\def\Wc{{\cal W}}
\def\xb{{\bm x}}\def\Xb{{\bm X}}\def\Xc{{\cal X}}
\def\yb{{\bm y}}\def\Yb{{\bm Y}}\def\Yc{{\cal Y}}
\def\zb{{\bm z}}\def\Zb{{\bm Z}}\def\Zc{{\cal Z}}
\def\abh{\wh{\ab}} \def\Abh{\wh{\Ab}}
\def\bbh{\wh{\bb}} \def\Bbh{\wh{\Bb}}
\def\cbh{\wh{\cb}} \def\Cbh{\wh{\Cb}}
\def\dbh{\wh{\db}} \def\Dbh{\wh{\Db}}
\def\ebh{\wh{\eb}} \def\Ebh{\wh{\Eb}}
\def\fbh{\wh{\fb}} \def\Fbh{\wh{\Fb}}
\def\gbh{\wh{\gb}} \def\Gbh{\wh{\Gb}}
\def\hbh{\wh{\hb}} \def\Hbh{\wh{\Hb}}
\def\ibh{\wh{\ib}} \def\Ibh{\wh{\Ib}}
\def\jbh{\wh{\jb}} \def\Jbh{\wh{\Jb}}
\def\kbh{\wh{\kb}} \def\Kbh{\wh{\Kb}}
\def\lbh{\wh{\lb}} \def\Lbh{\wh{\Lb}}
\def\mbh{\wh{\mb}} \def\Mbh{\wh{\Mb}}
\def\nbh{\wh{\nb}} \def\Nbh{\wh{\Nb}}
\def\obh{\wh{\ob}} \def\Obh{\wh{\Ob}}
\def\pbh{\wh{\pb}} \def\Pbh{\wh{\Pb}}
\def\qbh{\wh{\qb}} \def\Qbh{\wh{\Qb}}
\def\rbh{\wh{\rb}} \def\Rbh{\wh{\Rb}}
\def\sbh{\wh{\sb}} \def\Sbh{\wh{\Sb}}
\def\tbh{\wh{\tb}} \def\Tbh{\wh{\Tb}}
\def\ubh{\wh{\ub}} \def\Ubh{\wh{\Ub}}
\def\vbh{\wh{\vb}} \def\Vbh{\wh{\Vb}}
\def\wbh{\wh{\wb}} \def\Wbh{\wh{\Wb}}
\def\xbh{\wh{\xb}} \def\Xbh{\wh{\Xb}}
\def\ybh{\wh{\yb}} \def\Ybh{\wh{\Yb}}
\def\zbh{\wh{\zb}} \def\Zbh{\wh{\Zb}}
\def\abt{\wt{\ab}} \def\Abt{\wt{\Ab}}
\def\bbt{\wt{\bb}} \def\Bbt{\wt{\Bb}}
\def\cbt{\wt{\cb}} \def\Cbt{\wt{\Cb}}
\def\dbt{\wt{\db}} \def\Dbt{\wt{\Db}}
\def\ebt{\wt{\eb}} \def\Ebt{\wt{\Eb}}
\def\fbt{\wt{\fb}} \def\Fbt{\wt{\Fb}}
\def\gbt{\wt{\gb}} \def\Gbt{\wt{\Gb}}
\def\hbt{\wt{\hb}} \def\Hbt{\wt{\Hb}}
\def\ibt{\wt{\ib}} \def\Ibt{\wt{\Ib}}
\def\jbt{\wt{\jb}} \def\Jbt{\wt{\Jb}}
\def\kbt{\wt{\kb}} \def\Kbt{\wt{\Kb}}
\def\lbt{\wt{\lb}} \def\Lbt{\wt{\Lb}}
\def\mbt{\wt{\mb}} \def\Mbt{\wt{\Mb}}
\def\nbt{\wt{\nb}} \def\Nbt{\wt{\Nb}}
\def\obt{\wt{\ob}} \def\Obt{\wt{\Ob}}
\def\pbt{\wt{\pb}} \def\Pbt{\wt{\Pb}}
\def\qbt{\wt{\qb}} \def\Qbt{\wt{\Qb}}
\def\rbt{\wt{\rb}} \def\Rbt{\wt{\Rb}}
\def\sbt{\wt{\sb}} \def\Sbt{\wt{\Sb}}
\def\tbt{\wt{\tb}} \def\Tbt{\wt{\Tb}}
\def\ubt{\wt{\ub}} \def\Ubt{\wt{\Ub}}
\def\vbt{\wt{\vb}} \def\Vbt{\wt{\Vb}}
\def\wbt{\wt{\wb}} \def\Wbt{\wt{\Wb}}
\def\xbt{\wt{\xb}} \def\Xbt{\wt{\Xb}}
\def\ybt{\wt{\yb}} \def\Ybt{\wt{\Yb}}
\def\zbt{\wt{\zb}} \def\Zbt{\wt{\Zb}}
\def\alphab{\bm{\alpha}}
\def\betab{\bm{\beta}}
\def\gammab{\bm{\gamma}}            \def\Gammab{\bm{\Gamma}}
\def\deltab{\bm{\delta}}            \def\Deltab{\bm{\Delta}}
\def\epsilonb{\bm{\epsilon}}
\def\varepsilonb{\bm{\varepsilon}}
\def\etab{\bm{\eta}}
\def\thetab{\bm{\theta}}            \def\Thetab{\bm{\Theta}}
\def\varthetab{\bm{\vartheta}}
\def\iotab{\bm{\iota}}
\def\kappab{\bm{\kappa}}
\def\varkappab{\bm{\vrkappa}}
\def\lambdab{\bm{\lambda}}          \def\Lambdab{\bm{\Lambda}}
\def\mub{\bm{\mu}}
\def\nub{\bm{\nu}}
\def\xib{\bm{\xi}}                  \def\Xib{\bm{\Xi}}
\def\pib{\bm{\pi}}                  \def\Pib{\bm{\Pi}}
\def\varpib{\bm{\varpi}}
\def\taub{\bm{\tau}}
\def\phib{\bm{\phi}}
\def\varpib{\bm{\varpi}}
\def\rhob{\bm{\rho}}
\def\varrhob{\bm{\varrho}}
\def\sigmab{\bm{\sigma}}            \def\Sigmab{\bm{\Sigma}}
\def\varsigmab{\bm{\varsigma}}
\def\phib{\bm{\phi}}                \def\Phib{\bm{\Phi}}
\def\varphib{\bm{\varphi}}
\def\chib{\bm{\chi}}
\def\psib{\bm{\psi}}                \def\Psib{\bm{\Psi}}
\def\omegab{\bm{\omega}}            \def\Omegab{\bm{\Omega}}
\def\taub{\bm{\tau}}
\def\uepsilonb{\bm{\uepsilon}}      \def\Uepsilonb{\bm{\Uepsilon}}
\def\zetab{\bm{\zeta}}
\def\gammabh{\wh{\gammab}}            \def\Gammabh{\wh{\Gammab}}
\def\deltabh{\wh{\deltab}}            \def\Deltabh{\wh{\Deltab}}
\def\epsilonbh{\wh{\epsilonb}}
\def\varepsilonbh{\wh{\varepsilonb}}
\def\etabh{\wh{\etab}}
\def\thetabh{\wh{\thetab}}            \def\Thetabh{\wh{\Thetab}}
\def\varthetabh{\wh{\varthetab}}
\def\iotabh{\wh{\iotab}}
\def\kappabh{\wh{\kappab}}
\def\varkappabh{\wh{\vrkappab}}
\def\lambdabh{\wh{\lambdab}}          \def\Lambdabh{\wh{\Lambdab}}
\def\mubh{\wh{\mub}}
\def\nubh{\wh{\nub}}
\def\xibh{\wh{\xib}}                  \def\Xibh{\wh{\Xib}}
\def\pibh{\wh{\pib}}                  \def\Pibh{\wh{\Pib}}
\def\varpibh{\wh{\varpib}}
\def\taubh{\wh{\taub}}
\def\phibh{\wh{\phib}}
\def\varpibh{\wh{\varpib}}
\def\rhobh{\wh{\rhob}}
\def\varrhobh{\wh{\varrhob}}
\def\sigmabh{\wh{\sigmab}}
\def\varsigmabh{\wh{\varsigmab}}
\def\phibh{\wh{\phib}}
\def\varphibh{\wh{\varphib}}
\def\chibh{\wh{\chib}}
\def\psibh{\wh{\psib}}                \def\Psibh{\wh{\Psib}}
\def\omegabh{\wh{\omegab}}            \def\Omegabh{\wh{\Omegab}}
\def\taubh{\wh{\taub}}
\def\uepsilonbh{\wh{\uepsilonb}}      \def\Uepsilonbh{\wh{\Uepsilonb}}
\def\zetabh{\wh{\zetab}}
\def\gammabt{\wt{\gammab}}            \def\Gammabt{\wt{\Gammab}}
\def\deltabt{\wt{\deltab}}            \def\Deltabt{\wt{\Deltab}}
\def\epsilonbt{\wt{\epsilonb}}
\def\varepsilonbt{\wt{\varepsilonb}}
\def\etabt{\wt{\etab}}
\def\thetabt{\wt{\thetab}}            \def\Thetabt{\wt{\Thetab}}
\def\varthetabt{\wt{\varthetab}}
\def\iotabt{\wt{\iotab}}
\def\kappabt{\wt{\kappab}}
\def\varkappabt{\wt{\vrkappab}}
\def\lambdabt{\wt{\lambdab}}          \def\Lambdabt{\wt{\Lambdab}}
\def\mubt{\wt{\mub}}
\def\nubt{\wt{\nub}}
\def\xibt{\wt{\xib}}                  \def\Xibt{\wt{\Xib}}
\def\pibt{\wt{\pib}}                  \def\Pibt{\wt{\Pib}}
\def\varpibt{\wt{\varpib}}
\def\taubt{\wt{\taub}}
\def\phibt{\wt{\phib}}
\def\varpibt{\wt{\varpib}}
\def\rhobt{\wt{\rhob}}
\def\varrhobt{\wt{\varrhob}}
\def\sigmabt{\wt{\sigmab}}
\def\varsigmabt{\wt{\varsigmab}}
\def\phibt{\wt{\phib}}
\def\varphibt{\wt{\varphib}}
\def\chibt{\wt{\chib}}
\def\psibt{\wt{\psib}}                \def\Psibt{\wt{\Psib}}
\def\omegabt{\wt{\omegab}}            \def\Omegabt{\wt{\Omegab}}
\def\taubt{\wt{\taub}}
\def\uepsilonbt{\wt{\uepsilonb}}      \def\Uepsilonbt{\wt{\Uepsilonb}}
\def\zetabt{\wt{\zetab}}

\def\d#1{\,\mbox{d} #1}
\def\pth#1{\left(#1\right)}
\def\acc#1{\left\{#1\right\}}
\def\cro#1{\left[#1\right]}
\def\rem#1{}

\def\zerob{{\bm 0}}
\def\oneb{{\bm 1}}

\def\intg{\int\kern-1.1em\int}
\def\expf#1{\exp\left[ {#1} \right]}

\def\bs{\bar{s}}
\def\us{\underline{s}}
\def\fmin{f_{\mbox{min}}}
\def\fmax{f_{\mbox{max}}}
\def\expf#1{\mbox{exp}\left\{#1\right\}}
\def\argmin#1#2{\mbox{arg}\min_{#1}\left\{#2\right\}}
\def\argmax#1#2{\mbox{arg}\max_{#1}\left\{#2\right\}}

\def\lra{\longrightarrow}
\def\fh{\widehat{f}}
\def\fbh{\widehat{\fb}}
\def\Rbh{\widehat{\Rb}}
\def\Sigmabh{\widehat{\Sigmab}}

\def\rbh{\widehat{\rb}}
\def\qbh{\widehat{\qb}}
\def\xbh{\widehat{\xb}}
\def\tbh{\widehat{\tb}}
\def\thetabhk{\widehat{\thetab}^{(k)}}
\def\thetabhkp{\widehat{\thetab}^{(k+1)}}

\def\disp{\displaystyle}
\def\vsm{\vspace*{-12pt}}
\def\hsm{\hspace*{-3em}}

\def\pmatrix#1{\left[\begin{matrix}#1\end{matrix}\right]}
\def\pmat#1{\left(\barr{c} #1 \earr\right)}
\def\pmata#1#2{\left(\barr{c} #1 \\ #2 \earr\right)}
\def\pmatb#1#2#3#4{\left(\barr{cc} #1 & #2 \\ #3 & #4 \earr\right)}

\def\th{\widehat{t}}
\def\xh{\widehat{x}}
\def\lambdah{\widehat{\lambda}}
\def\muh{\widehat{\mu}}
\def\sigmah{\widehat{\sigma}}
\def\rhoh{\widehat{\rho}}
\def\betah{\widehat{\beta}}
\def\alphah{\widehat{\alpha}}
\def\tbh{\widehat{\tb}}
\def\mubh{\widehat{\mub}}
\def\sigmabh{\widehat{\sigmab}}
\def\rhobh{\widehat{\rhob}}
\def\oneb{\mbox{\bf 1}}

\def\ie{i.e.,}
\def\iid{i.i.d.}
\def\apost{\emph{a posteriori }}
\def\aprio{\emph{a priori }}

\def\sumi{\sum_{i=1}^{M} }
\def\sumj{\sum_{j=1}^{N} }
\def\lambdah{\widehat{\lambda}}
\def\lambdabh{\widehat{\lambdab}}

\def\det#1{\mbox{det}\left(#1\right)}
\def\defined{\,\shortstack{$\triangle$\\ =}\,}
\def\sign{\hbox{sign}}

\def\ss{source separation }
\def\pdf{probability density function }
\def\mm{mixing matrix }
\def\sm{separating matrix }

\def\sbh{\widehat{\sb}}
\def\abh{\widehat{\ab}}
\def\Sbh{\widehat{\Sb}}
\def\Abh{\widehat{\Ab}}
\def\Bbh{\widehat{\Bb}}

\def\Lambdab{\bm{\Lambda}}
\def\xbh{\widehat{\xb}}

\def\pxta{p\left( \Xb | \Ab \right)}
\def\pxas{p\left( \Xb | \Ab, \Sb \right)}
\def\pasx{p\left( \Ab,\Sb | \Xb \right)}
\def\pax{p\left( \Ab | \Xb \right)}
\def\psx{p\left( \sb | \Xb \right)}
\def\pbx{p\left( \Bb | \Xb \right)}
\def\pxb{p\left( \Xb | \Bb \right)}

\def\ria{r_i\left( [\Ab^{-1} \xb]_i \right)}
\def\rib{r_i\left( [\Bb \xb]_i \right)}
\def\riat{r_i\left( [\Ab^{-1} \xb(t)]_i \right)}
\def\ribt{r_i\left( [\Bb \xb(t)]_i \right)}

\def\pia{p_i\left( [\Ab^{-1} \xb]_i \right)}
\def\pib{p_i\left( [\Bb \xb]_i \right)}
\def\piat{p_i\left( [\Ab^{-1} \xb(t)]_i \right)}
\def\pibt{p_i\left( [\Bb \xb(t)]_i \right)}

\def\zmat{\left[0\right]}
\def\lra{\longrightarrow}
\def\dpdx#1#2{\frac{\partial #1}{\partial #2}}
\def\diag#1{\mbox{diag}\left\{#1\right\}}

\def\KL#1#2{\mbox{KL}\left(#1~:~#2\right)}

\def\esp#1{\mbox{E}\left\{ #1 \right\}}
\def\espx#1#2{\mbox{E}_{#1}\left\{ #2 \right\}}
\def\Prob#1{\mbox{P}\left(#1\right)}
\def\var#1{\mbox{Var}\left(#1\right)}
\def\cov#1{\mbox{Cov}\left(#1\right)}
\def\corr#1{\mbox{Corr}\left(#1\right)}
\def\trace#1{\mbox{Tr}\left(#1\right)}
\def\rang#1{\mbox{rang}\left(#1\right)}

\def\sigmae{\sigma_{\epsilon}}
\def\Sigmae{\Sigmab_{\epsilon}}

\def\fraco#1#2{\left({#1}\right)\,/\,\left({#2}\right)}

\def\mjk{m_{jk}}
\def\pjk{p_{jk}}
\def\pjz{p_{j0}}
\def\pjkl{p_{jk}}
\def\sigmajk{\sigma_{jk}}
\def\Rbe{\Rb_{\epsilon}}

\def\zerob{{\bm 0}}
\def\oneb{{\bm 1}}
\def\intg{\int\kern-1.1em\int}
\def\expf#1{\exp\left[ {#1} \right]}

\def\gbu{\underline{\gb}}
\def\fbu{\underline{\fb}}
\def\zbu{\underline{\zb}}
\def\epsilonbu{\underline{\epsilonb}}
\def\thetabu{\underline{\thetab}}

\def\sigmae{{\sigma_{\epsilon}}}
\def\Sigmae{{\Sigmab_{\epsilonb}}}
\def\Sigmaf{{\Sigmab_{\fb}}}
\def\Sigmag{{\Sigmab_{\gb}}}

\def\fbuh{\widehat{\fbu}}
\def\zbuh{\widehat{\fbu}}
\def\thetabuh{\widehat{\thetabu}}

\def\fbh{\widehat{\fb}}
\def\Pbh{\widehat{\Pb}}
\def\Sigmabh{\widehat{\Sigmab}}
\def\Abh{\widehat{\Ab}}
\def\thetabh{\widehat{\thetab}}

\def\Rjk{{R_j}_k}
\def\fbik{{\fb_i}_k}
\def\fbjk{{\fb_j}_k}

\def\mik{{m_i}_k}
\def\sigmaik{{\sigma_i^2}_k}
\def\Sigmaik{{\Sigmab_i}_k}
\def\alphaik{{\alpha_i}_k}
\def\betaik{{\beta_i}_k}

\def\muik{{\mu_i}_k}
\def\vik{{v_i}_k}

\def\mjk{{m_j}_k}
\def\sigmajk{{\sigma_j^2}_k}
\def\Sigmajk{{\Sigmab_j}_k}
\def\alphajk{{\alpha_j}_k}
\def\betajk{{\beta_j}_k}

\def\mujk{{\mu_j}_k}
\def\vjk{{v_j}_k}
\def\njk{{n_j}_k}

\def\alphae{\alpha^{\epsilon}}
\def\betae{\beta^{\epsilon}}
\def\alphaez{\alpha^{\epsilon}_{0}}
\def\betaez{\beta^{\epsilon}_{0}}
\def\alphaei{\alpha^{\epsilon}_{i}}
\def\betaei{\beta^{\epsilon}_{i}}
\def\alphaeiz{\alpha^{\epsilon}_{i0}}
\def\betaeiz{\beta^{\epsilon}_{i0}}

\def\mbz{{{\mb}_z}}
\def\Sigmabz{{{\Sigmab}_z}}
\def\diag#1{\mbox{diag}\left[#1\right]}

\def\blue#1{{\color{blue}#1}}
\def\red#1{{\color{red}#1}}
\def\green#1{{\color{green}#1}}
\def\magenta#1{{\color{magenta}#1}}
\def\cyan#1{{\color{cyan}#1}}
\def\REM#1{}

\def\intd{\int\kern-.8em\int}

\def\cresta{
\thicklines\setlength{\unitlength}{.5mm}
\begin{picture}(120,120)
 \multiput(0,27)(0,20){5}{\line(1,0){80}}
 \multiput(0,27)(20,0){5}{\line(0,1){80}}
 \put(61,33){\red{$f_N$}}
 \put(0,92){\red{$f_1$}}
 \put(40,72){\red{$f_j$}}
 \put(88,50){\blue{$g_{i}$}}
 \put(38,120){\blue{$H_{ij}$}}
 \put(-5,115){\line(3,-2){90}}
 \put(42,85){\blue{\line(3,-2){20}}}
\end{picture}}

\def\tomoXaz{
\thicklines\setlength{\unitlength}{.96mm}
\begin{picture}(70,68)
  \put(15,35){\makebox(20,15){$\red{f(x,y)}$}}
  \put(-1,0){\axexy}
  \put(10,10){\objet}
  \put(5,5){\vector(1,1){55}}
  \put(55,55){\makebox(5,10){$r$}}
  \put(35,28){\makebox(5,10){$\phi$}}
  \put(60,8){\makebox(10,15){$\bullet$D}}
  \put(60,0){\makebox(15,10){$\blue{g(r,\phi)}$}}
  \put(10,55){\makebox(10,10){S$\bullet$}}
  \put(62,15){\line(-1,1){45}}
  \put(40,-5){\proj}
  \blue{\multiput(10,14)(5,0){9}{\line(0,1){40}}}
  \blue{\multiput(8,14)(0,5){9}{\line(1,0){40}}}
\end{picture}
}

\title{Inverse problems in imaging systems and the general Bayesian inversion frawework}

\author{Ali Mohammad-Djafari \\ 
Laboratoire des signaux et syst\`emes \\
Unit\'e mixte de recherche n° 8506 (CNRS, , Sup\'elec, Univ. Paris Sud)\\
Sup\'elec, plateau de Moulon, 3 rue Joliot-Curie,  \\   
91192   GIF-SUR-YVETTE   Cedex, France}

\begin{document}
\maketitle
\begin{abstract}
In this paper, first a great number of inverse problems which arise in instrumentation, in computer imaging systems and in computer vision are presented. Then a common general forward modeling for them is given and the corresponding inversion problem is presented. Then, after showing the inadequacy of the classical analytical and least square methods for these ill posed inverse problems, a Bayesian estimation framework is presented which can handle, in a coherent way, all these problems. 
One of the main steps, in Bayesian inversion framework is the prior modeling of the unknowns. For this reason, a great number of such models and in particular the compound hidden Markov models are presented. 
Then, the main computational tools of the Bayesian estimation are briefly presented. 
Finally, some particular cases are studied in detail and new results are presented. 
\end{abstract}

\section{Introduction}
Inverse problems arise in many applications in science and engineering. 
The main reason is that, very often we want to measure the distribution of 
an un-observable quantity $f(\rb)$ from the observation of another quantity $g(\sb)$ which is related to it and accessible to the measurement. 
The mathematical relation which gives $g(\sb)$ when $f(\rb)$ is known is called \emph{forward problem}:
\beq\label{eq1}
g(\sb) = [\Hc f(\rb)](\sb) + \epsilon(\sb)
\eeq
where $\Hc$ is the \emph{forward model}. In this relation, $\rb$ and $\sb$ may represent either time $t$, position on a line $x$, position on a surface $\rb=(x,y)$, position in space $\rb=(x,y,z)$ or any combinations of them. 

This forward model is often non linear, but it can be linearized. So, in this paper, we only consider the linear model, which in its general form, can be written as
\beq\label{eq2}
g(\sb) = \intg h(\rb,\sb) f(\rb) \d{\rb} + \epsilon(\sb)
\eeq
where $h(\rb,\sb)$ represents the \emph{measuring system response} and $\epsilon(\sb)$ all the errors (modeling, linearization and the other unmodelled errors often called noise). 
In this paper, we assume that the forward model is known perfectly, or at least, known excepted a few number of parameters. The \emph{inverse problem} is then the task of going back from the observed quantity $g(\sb)$ to $f(\rb)$. The main difficulty is that, very often these problems are \emph{ill-posed}, in opposition to the forward problems which are \emph{well-posed} as defined by Hadamard \cite{Hadamard01}. A problem is mathematically well-posed if the problem has a solution (existance), if the solution exists (uniqueness), and if the solution is stable (stability). A problem is then called ill-posed if any of these conditions are not satisfied \cite{Demoment85}. 

In this paper, we will only consider the algebraic methods of inversion where, in a first step the forward problem is discretized, \ie, the integral equation is approximated by a sum and the input $f$, the output $g$ and the errors $\epsilon$ are assumed to be well represented by the finite dimentional vectors $\fb$, $\gb$ and $\epsilonb$ such that:
\beq\label{eq2b}
g_i = \sum_{j=1}^n H_{ij} f_j + \epsilon_i, i=1,\cdots,n \lra \gb=\Hb\fb+\epsilonb
\eeq
where $g_i=g(\sb_i)$, $\epsilon_i=\epsilon(\sb_i)$,  $f_j=f(\rb_j)$ and  $H_{ij}=h(\rb_j,\sb_i)$ or in a more general case
\beqn
g_i &=& <\phi_i(\sb),g(\sb)>=\intg \phi_i(\sb)\; g(\sb) \d{\sb} \nonumber \\ 
\epsilon_i &=& <\phi_i(\sb),\epsilon(\sb)>=\intg \phi_i(\sb)\; \epsilon(\sb) \d{\sb} \\ 
f_j &=& <\psi_j(\sb),f(\rb)>=\intg \psi_j(\rb)\; f(\rb) \nonumber \d{\rb}  
\eeqn
where $\phi_i(\sb)$ and $\psi_j(\rb)$ are appropriate basis function in their corresponding spaces which means that, we assume

\beqn \label{eq2c}
g(\sb) &\simeq& \sum_{i=1}^m g_i \; \phi_i(\sb) \nonumber \\ 
\epsilon(\sb) &\simeq& \sum_{i=1}^m \epsilon_i \; \phi_i(\sb) \nonumber \\ 
f(\rb) &\simeq& \sum_{j=1}^n f_j \; \psi_i(\rb) \\ 
H_{ij} &\simeq& <\phi_i(\sb),\psi_j(\sb)>=\intg\intg \psi_j(\rb) \; \phi_i(\sb) \d{\rb} \d{\sb}
\nonumber 
\eeqn
But, before going further in details of the inversion methods, we are going to present a few examples. 

\subsection{1D signals}
Any instrument such as a thermometer which tries to measure a non directly measurable quantity $f(t)$ (here the time variation of the temperature) transforms it to the time variation of a measurable quantity $g(t)$ (here the length of the liquid in the thermometer). A perfect instrument has be at least lineair. Then the relation between the \emph{output} $g(t)$ and the \emph{input} $f(t)$ is:

\beq\label{eq3}
g(t) = \int h(t,t') f(t') \d{t} + \epsilon(t)
\eeq
where $h(t,t')$ the \emph{instrument's response}. If this response is invariant in time, then we have a \emph{convolution} forward model:

\beq\label{eq4}
g(t) = \int h(t-t') f(t') \d{t} + \epsilon(t)
\eeq
and the corresponding inverse problem is called \emph{deconvolution}. 

\bfig[htb] 
\btabu{@{}c@{}c@{}c@{}}
\includegraphics[width=50mm,height=30mm]{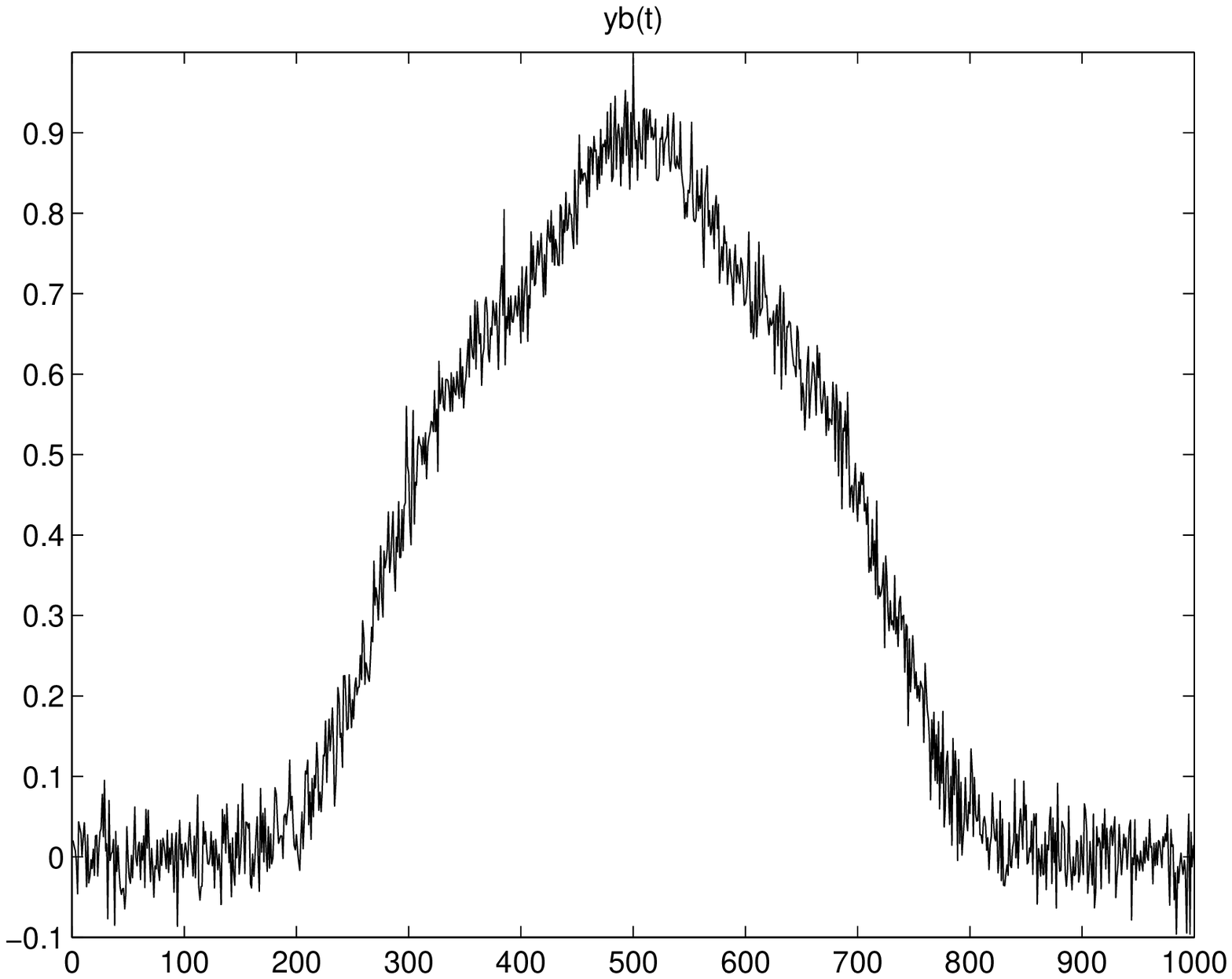}
&
{\Large \btabu[b]{@{}c@{}} {$?$} \\ $\Longrightarrow$ \\ ~\\ ~\\ \etabu  }
&
\includegraphics[width=50mm,height=30mm]{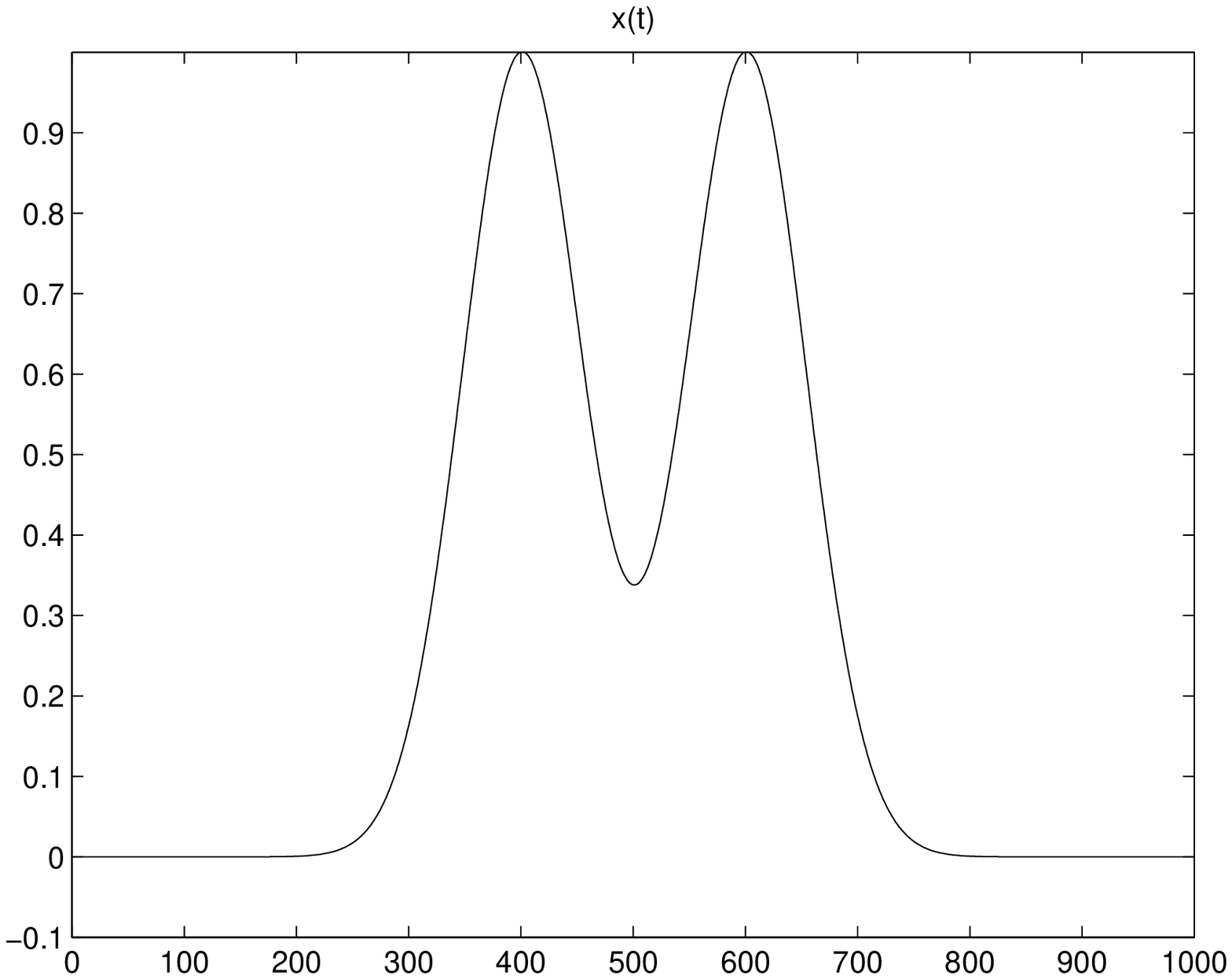}
\etabu
\caption{Deconvolution of 1D signals.}
\label{Fig1}
\efig

The convolution equation (\ref{eq4}) can also be written 
\beq\label{eq4b}
g(t) = \int h(\tau) f(t-\tau) \d{\tau} + \epsilon(t)
\eeq
which is obtained by change of variable $t-t'=\tau$. 
Assuming the sampling interval of $f$, $h$ and $g$ to be equal to 1, the discretized version of the deconvolution equation can then be written: 
\beq
g(i)=\sum_k h(k) f(i-k) + \epsilon(i), \quad i=1,\cdots,T
\eeq
which can be written in the general vector-matrix form:
\beq 
\gb=\Hb\fb+\epsilonb
\eeq
where 
$\gb$ and $\fb$ contains samples of the ouput $g(t)$ and the intput $f(t)$ and the matrix $\Hb$, in this case, is a Toeplitz matrix with a generic ligne composed of the samples of the impulse response $h(t)$. The Toeplitz property is thus identified to the time invariance property of the system response (convolution forward problem).

\subsection{Image restoration}
In this paper, we consider more the case of bivariate signals or images. 
As an example, when the unknown and measured quantities are images, we have
\beq\label{eq5}
g(\rb) = \int h(\rb,\rb') f(\rb') \d{\rb'} + \epsilon(\rb)
\eeq
and if the system response is spatially invariant, we have
\beq\label{eq6}
g(\rb) = \int h(\rb-\rb') f(\rb') + \epsilon(\rb).
\eeq
The case of denoising is the particular case where the point spread function (psf) $h(\rb)$ 
is $h(\rb)=\delta(\rb)$:
\beq\label{eq7}
g(\rb) = f(\rb) + \epsilon(\rb)
\eeq

\bfig[htb] 
\btabu{@{}c@{}c@{}c@{}}
\btabu{c}
\includegraphics[width=50mm,height=50mm]{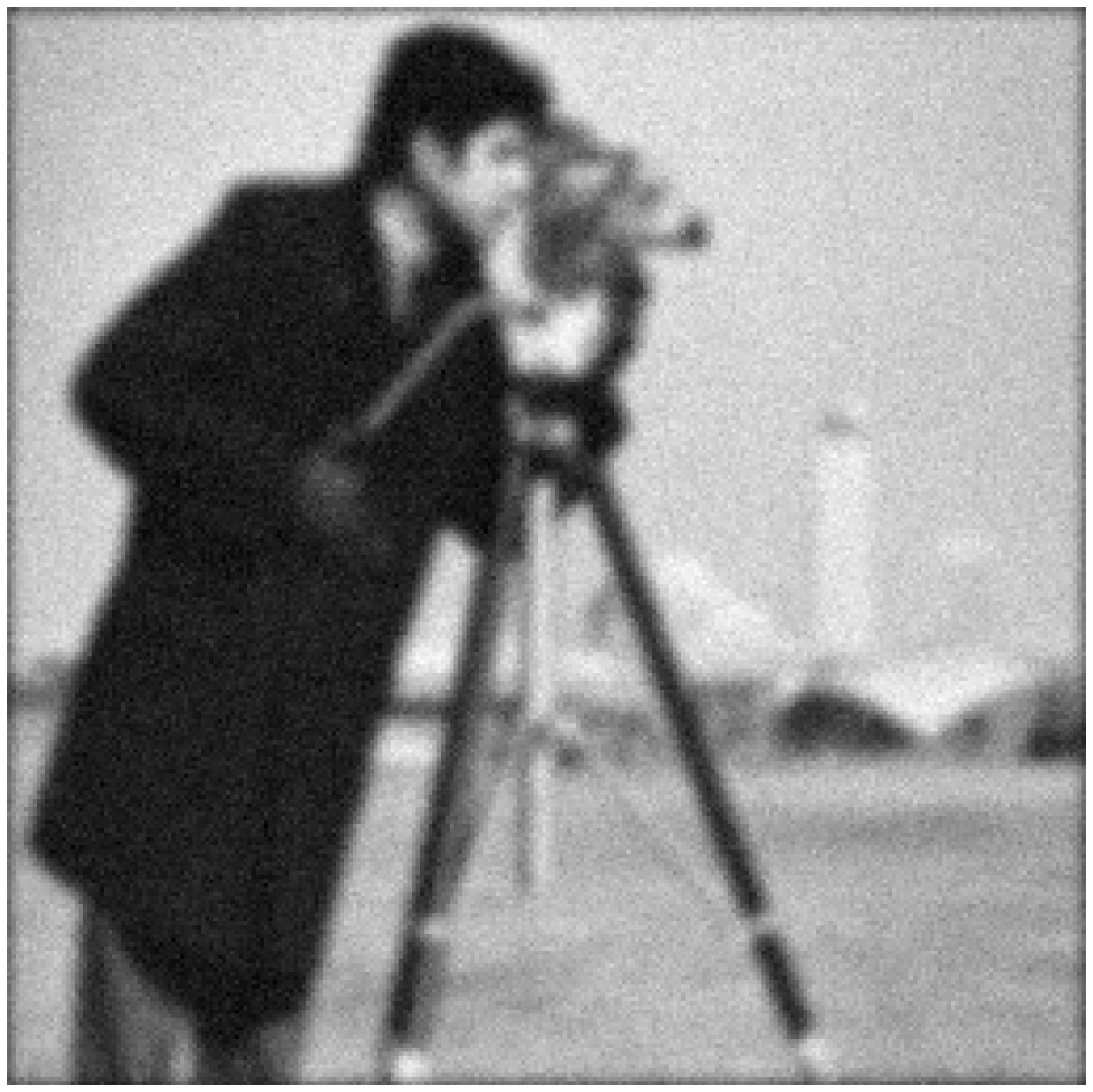}
\etabu
&
{\Large \btabu[b]{@{}c@{}} {$?$} \\ $\Longrightarrow$ \etabu  }
&
\btabu{c}
\includegraphics[width=50mm,height=50mm]{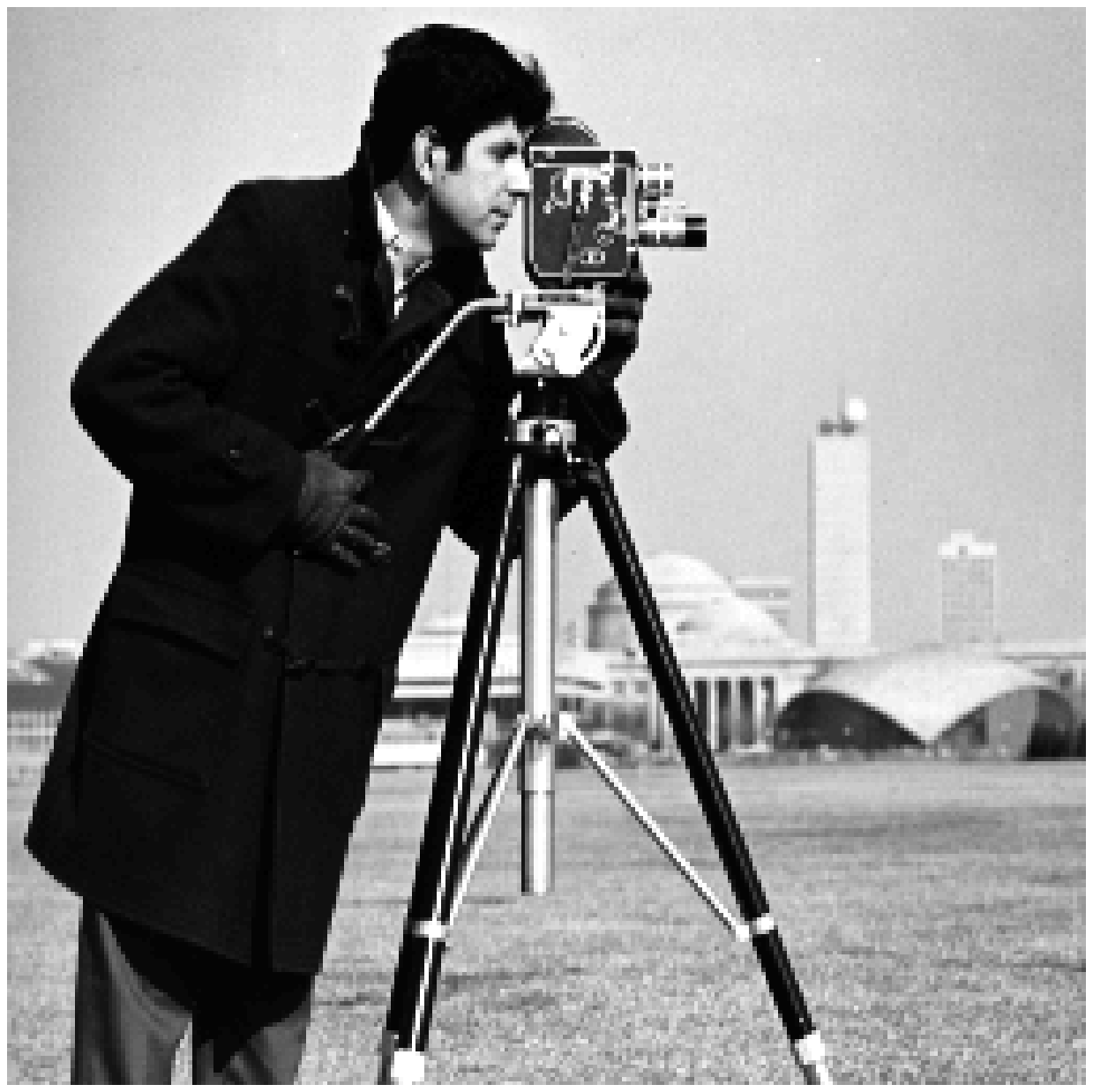}
\etabu
\etabu
\caption{Image restoration as an inverse problem.}
\label{Fig2}
\efig

The discretized version of the 2D deconvolution equation can also be written as  $\gb=\Hb\fb+\epsilonb$ where 
$\gb$ and $\fb$ contains, respectively, the rasterized samples of the ouput $g(\rb)$ and the intput $f(\rb')$, and the matrix $\Hb$ in this case, is a huge dimensional Toeplitz-Bloc-Toeplitz (TBT) matrix with a generic bloc-ligne composed of the samples of the point spread function (PSF) $h(\rb)$. The TBC property is thus identified to the space invariance property of the system response (2D convolution forward problem). For more details on the structure of this matrix refer to the book \cite{Andrews77} and the papers \cite{Hunt71,Hunt72a,Hunt72b}.

\newpage
\subsection{Image reconstruction in computed tomography}
In previous examples, $g(\sb)$ and $f(\rb)$ where defined in the same space. 
The case of image reconstruction in X ray computed tomography (CT) is interesting, because 
the observed data $g(\sb)$ and the unknown image $f(\rb)$ are defined in different 
spaces. The usual forward model in CT is shown in Figure~(\ref{Fig3}). 

In 2D case, the relation between the image to be reconstructed $f(x,y)$ and the projection data $g(r,\phi)=g_{\phi}(r)$ is given by the Radon transform:
\beqn\label{eq10}
g(r,\phi) 
&=& \int_{\Lc_{r,\phi}} f(x,y) \; \d{l} + \epsilon(r,\phi) \nonumber \\
&=& \int\int f(x,y) \; \delta(r-x\cos\phi-y\sin\phi) \d{x}{y}+ \epsilon(r,\phi) 
\eeqn
The discretized version of this forward equation can also be written as  $\gb=\Hb\fb+\epsilonb$ where 
$\gb=[\gb_1,\cdots,\gb_K]$ contains samples of projection datas $g(r,\phi_k)$ for different angles $\phi_k, k=1,\cdots,K$, $\fb=\{f(\rb), \rb\in\Rc\}$ contains the image pixels put in a vector and the elements $H_{ij}$ of the matrix $\Hb$, in this case, represents the length of the $i$-th ray in the $j$-th pixel. This matrix is a very sparse matrix with great number of zero valued elements \cite{Djafari97n,Djafari99a}. 

\bfig[htb]
\bcc
\btabu{@{}cc@{}}
3D & 2D \\ 
\includegraphics[width=60mm,height=60mm]{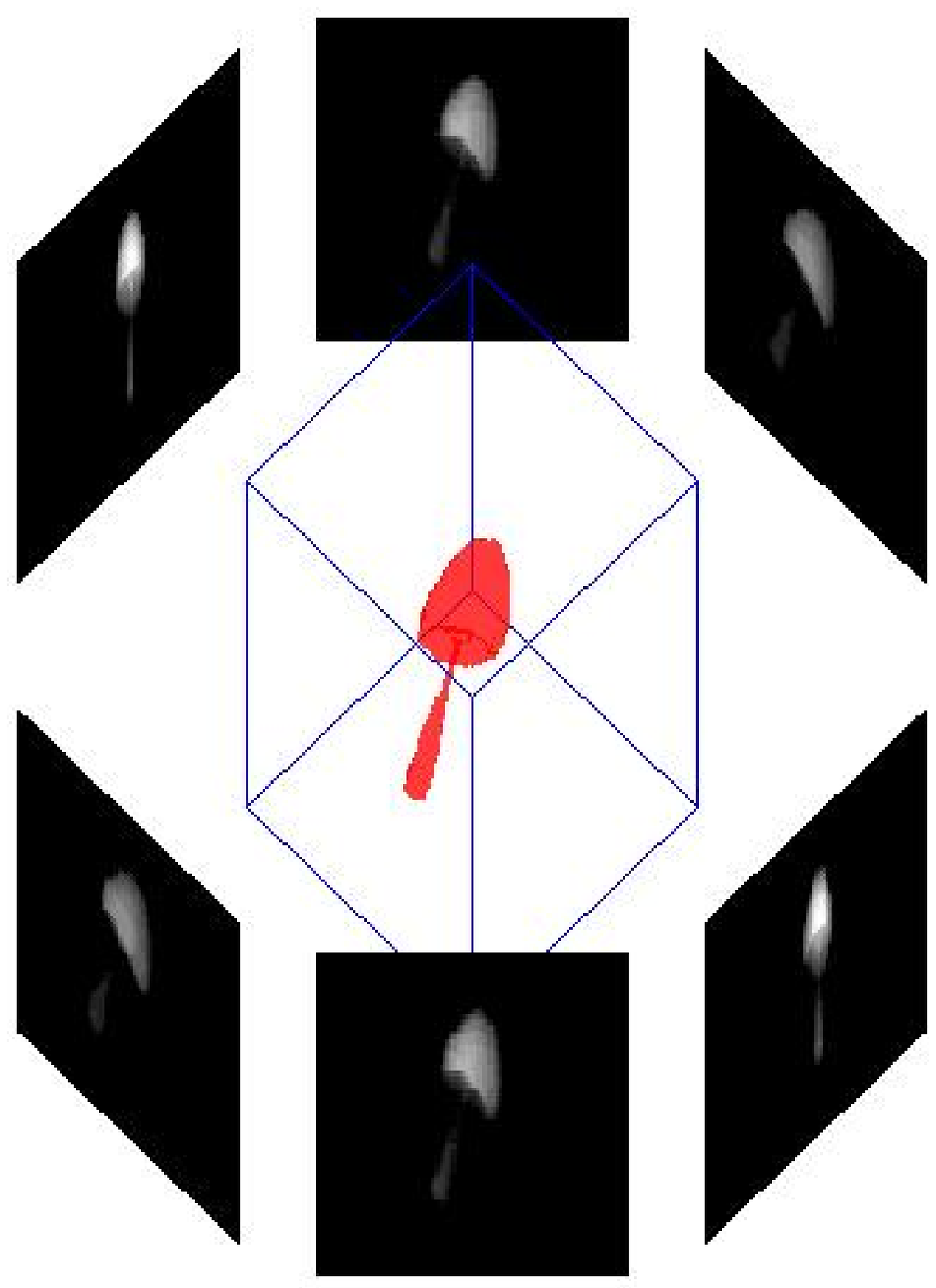} & 
\includegraphics[width=60mm,height=60mm]{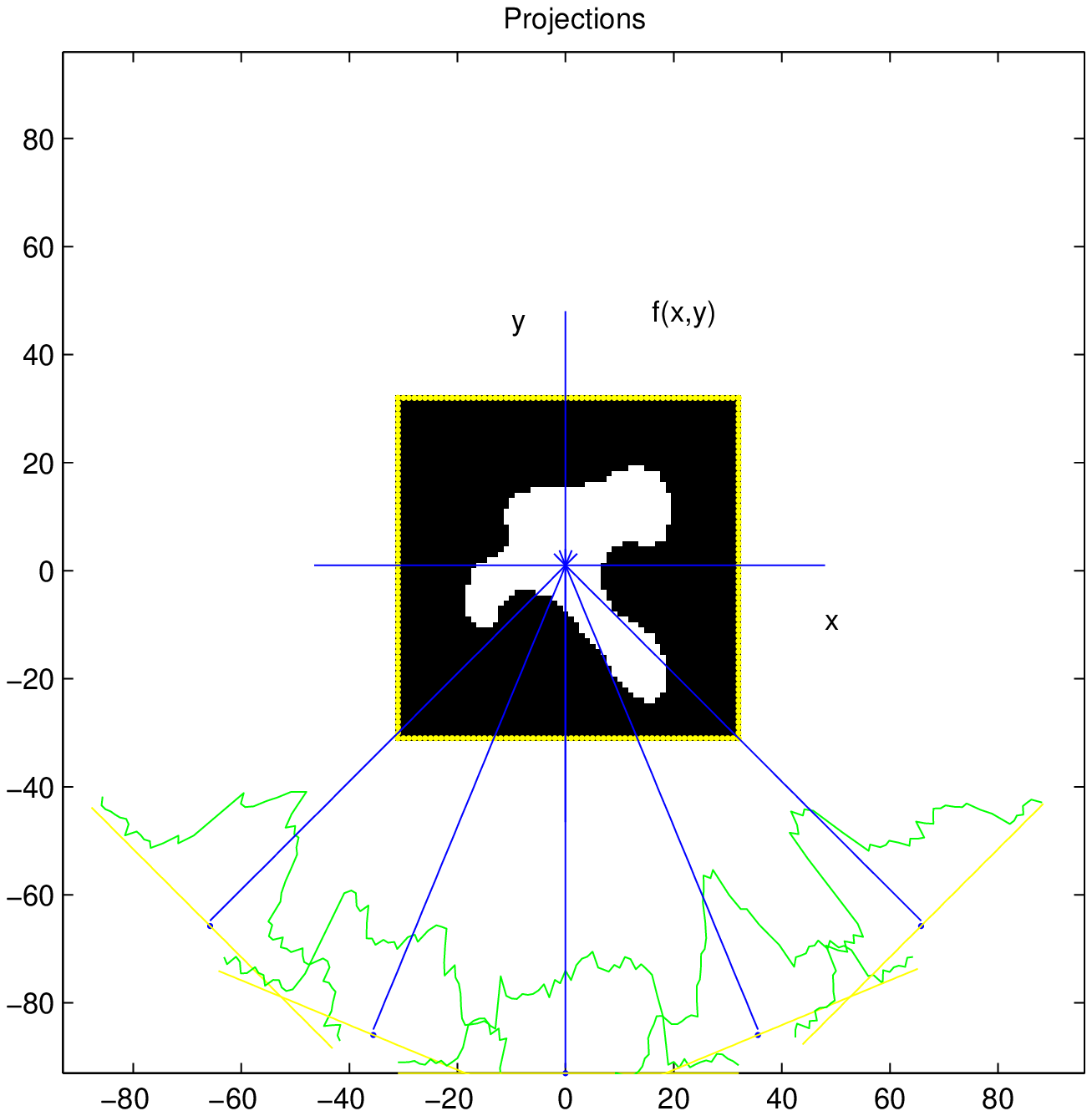} 
\etabu
\\
$\displaystyle{\blue{g_{\phi}(r_1,r_2)}=\int_{\Lc_{r_1,r_2,\phi}} \red{f(x,y,z)} \d{l}}$ \quad 
$\displaystyle{\blue{g_{\phi}(r)}=\int_{\Lc_{r,\phi}} \red{f(x,y)} \d{l}}$ 
\\ 
Forward probelm:  
$\red{f(x,y)}$ or $\red{f(x,y,z)}\lra$ 
$\blue{g_{\phi}(r)}$ or $\blue{g_{\phi}(r_1,r_2)}$
\\
Inverse problem:  
$\blue{g_{\phi}(r)}$ or $\blue{g_{\phi}(r_1,r_2)}\lra$
$\red{f(x,y)}$ or $\red{f(x,y,z)}$  
\ecc
\caption{2D and 3D X ray computed tomography}
\label{Fig3}
\efig

\bfig[h] 
\[
\barr{@{}c@{}c@{}}
\btabu{c}
\tomoXaz
\etabu
& 
\btabu{c} ~\\ 
\cresta
\etabu
\\
\disp{\blue{g(r,\phi)}=\int_{L_{r,\phi}}  \red{f(x,y)} \d{l}}
& 
\disp{\blue{g_i}=\sum_{j=1}^N H_{i,j} \; \red{f_j}} 
\earr
\]
\caption{Discretized 2D X ray computed tomography}
\label{Fig4}
\efig

\bfig[b] 
\btabu{@{}c@{}c@{}c@{}}
\btabu{c}
\includegraphics[width=50mm,height=50mm]{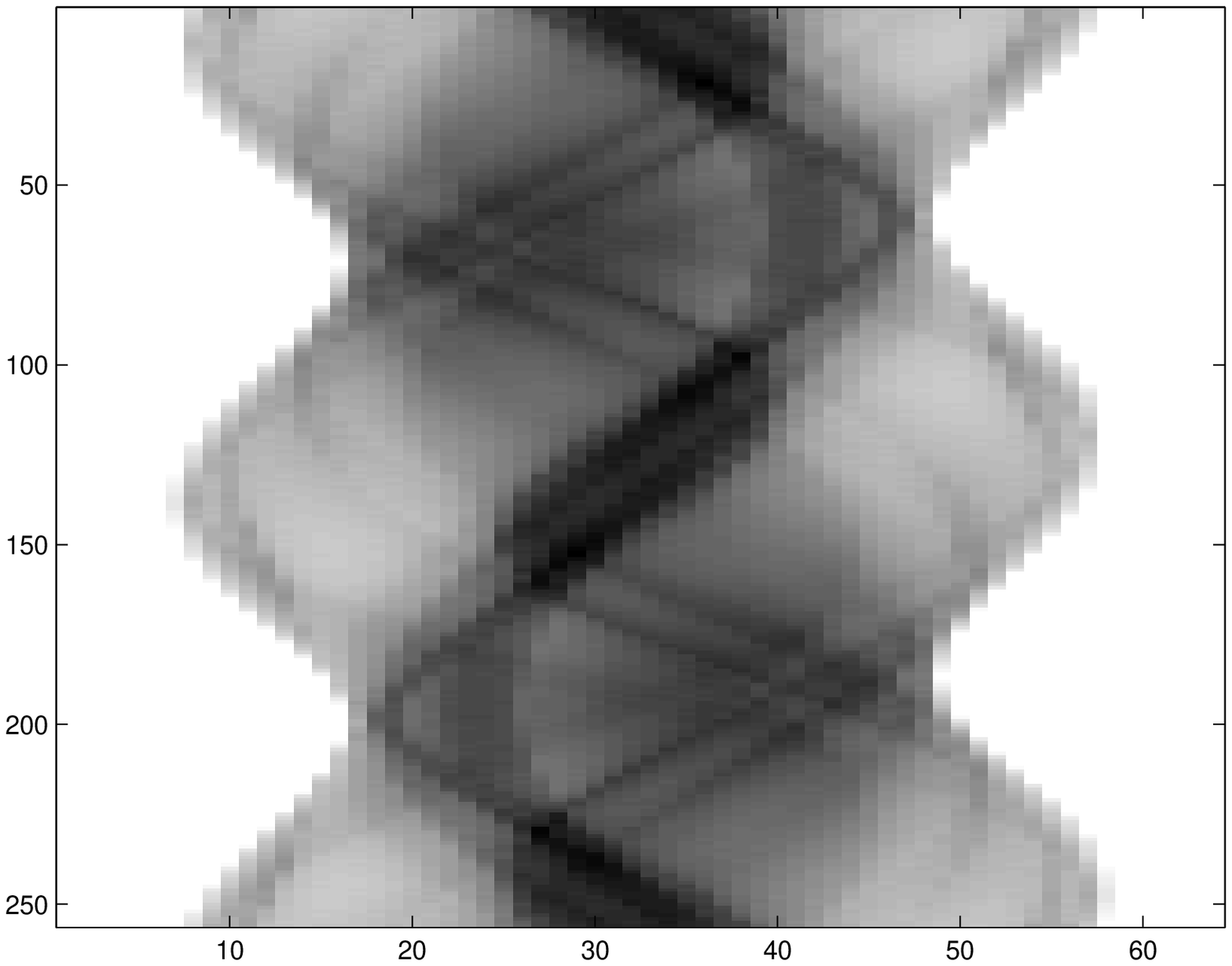}
\etabu
&
{\Large \btabu[b]{@{}c@{}} {$?$} \\ $\Longrightarrow$ \etabu  }
&
\btabu{c}
\includegraphics[width=50mm,height=50mm]{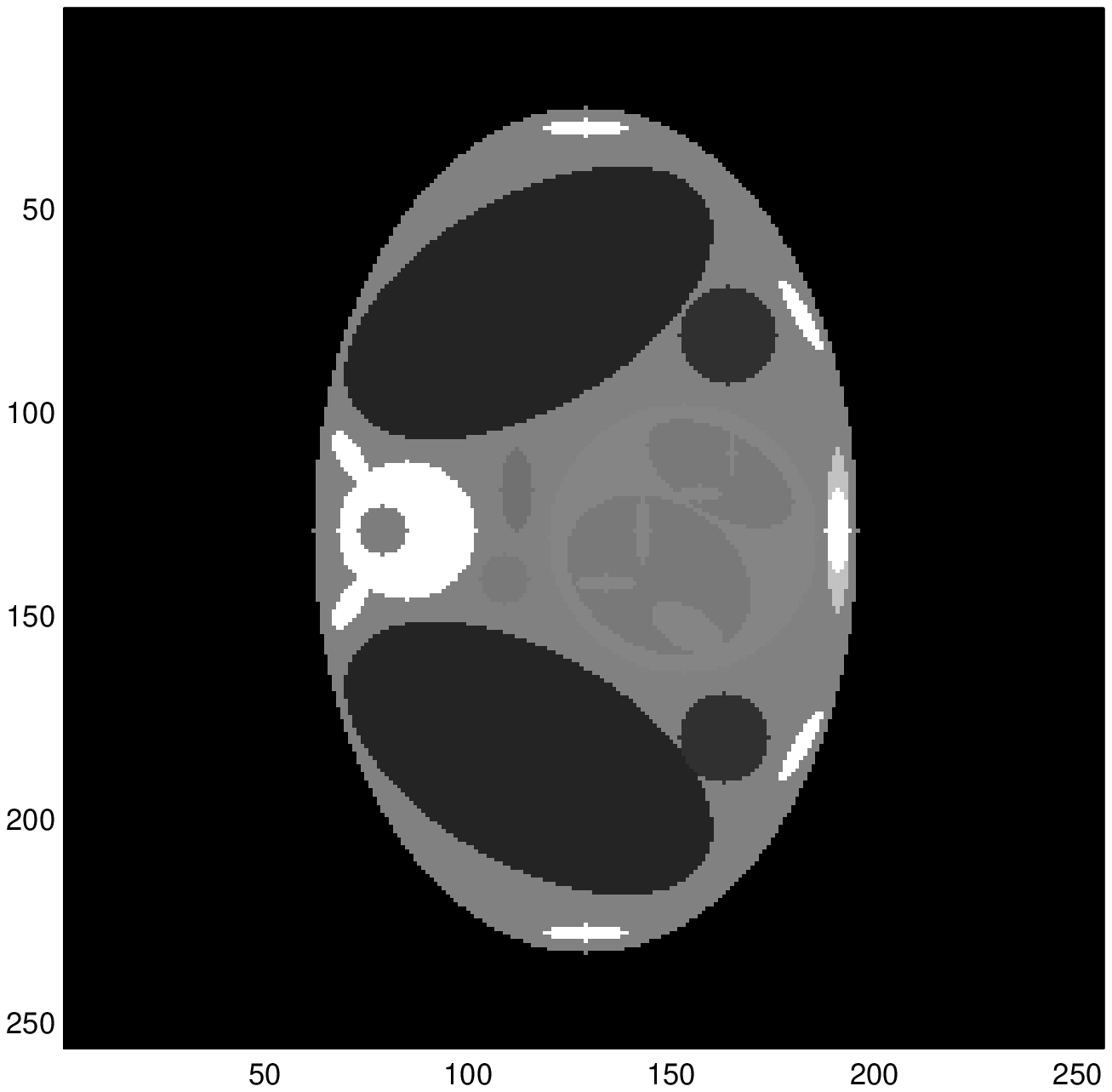}
\etabu
\etabu
\caption{Inverse problem of image reconstruction in X ray computed tomography.}
\label{Fig5}
\efig

\clearpage\newpage
\subsection{Time varying imaging systems}
When the observed and unknown quantities depend on space $\rb$ and time $t$, we have

\beq\label{eq20}
g(\rb,t) = \int h(\rb-\rb', t-t') f(\rb',t') \d{\rb'}\d{t'}+ \epsilon(\rb,t)
\eeq
If the point spread function of the imaging system does not 
depend on time, then we have
\beq\label{eq18}
g(\rb,t) = \int h(\rb-\rb') f(\rb',t) \d{\rb'} + \epsilon(\rb,t)
\eeq
In this case, $t$ can also be considered as an index:
\beq\label{eq19}
g_t(\rb) = \int h(\rb-\rb') f_t(\rb') \d{\rb'} + \epsilon_t(\rb)
\eeq
One example of such problem is the video image restoration shown in Figure~(\ref{Fig6}). 

\bfig[htb] 
\btabu{@{}c@{}c@{}c@{}}
\btabu{l}
\includegraphics[width=25mm,height=25mm]{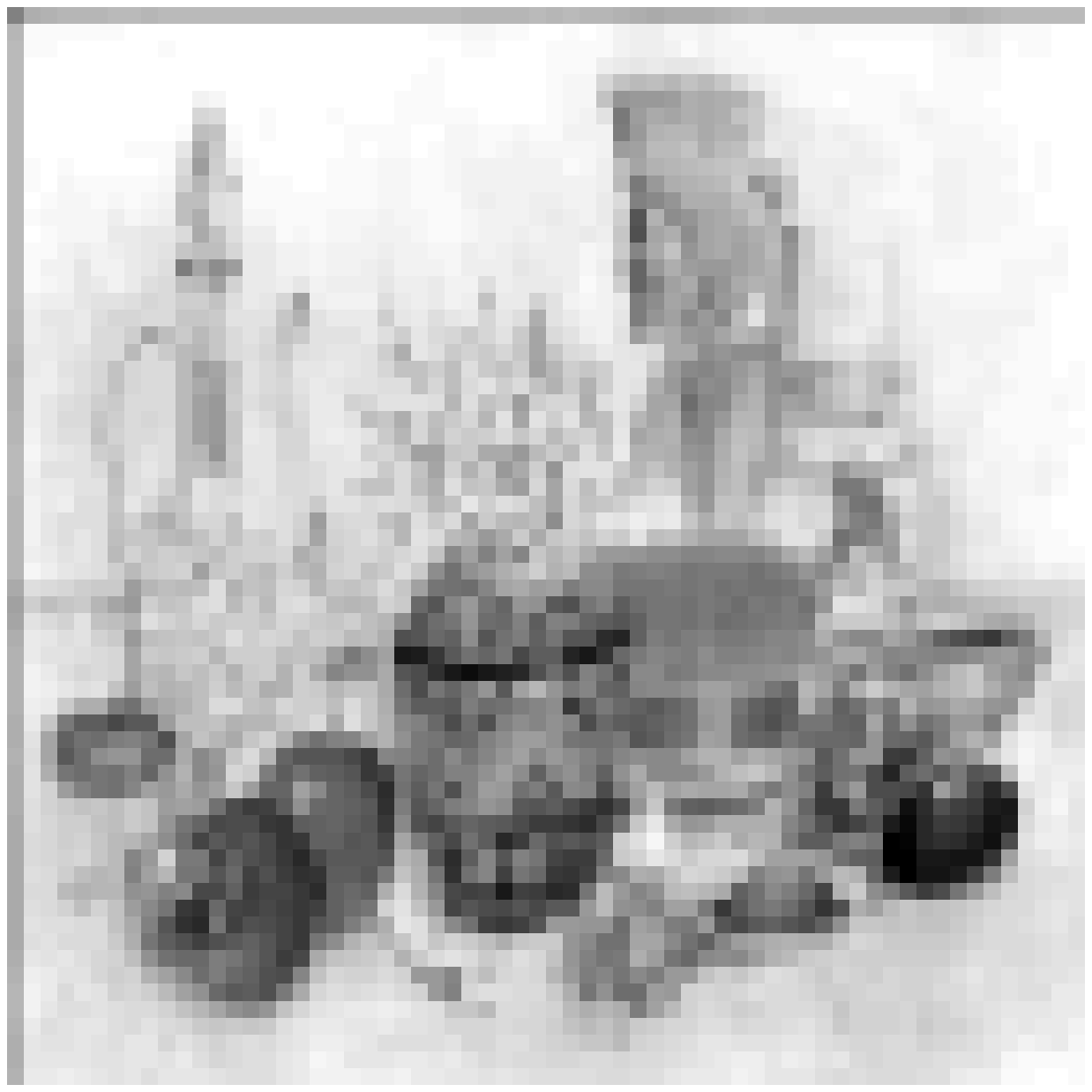} \\[-22mm]\hspace*{5mm} 
\includegraphics[width=25mm,height=25mm]{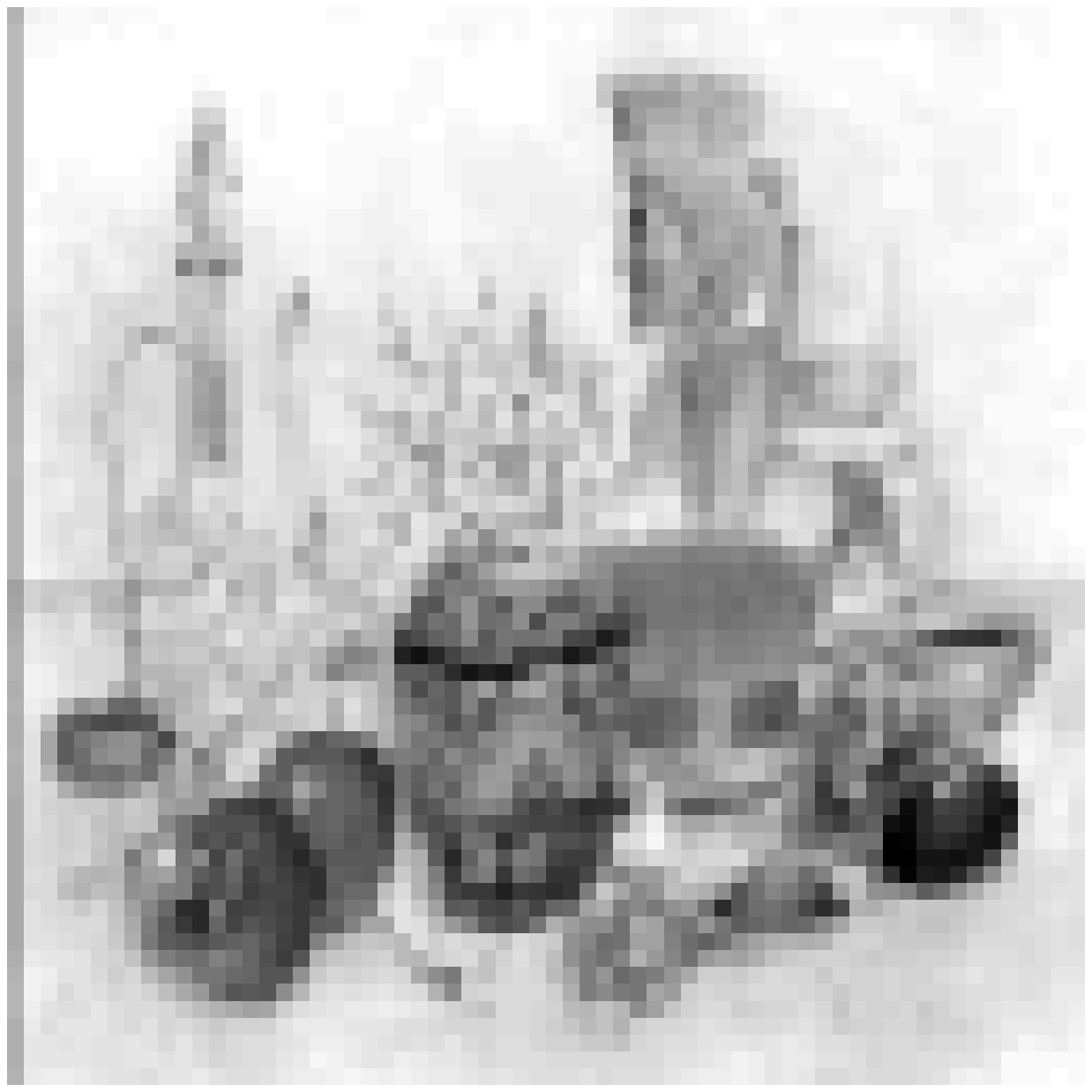} \\[-22mm]\hspace*{10mm}  
\includegraphics[width=25mm,height=25mm]{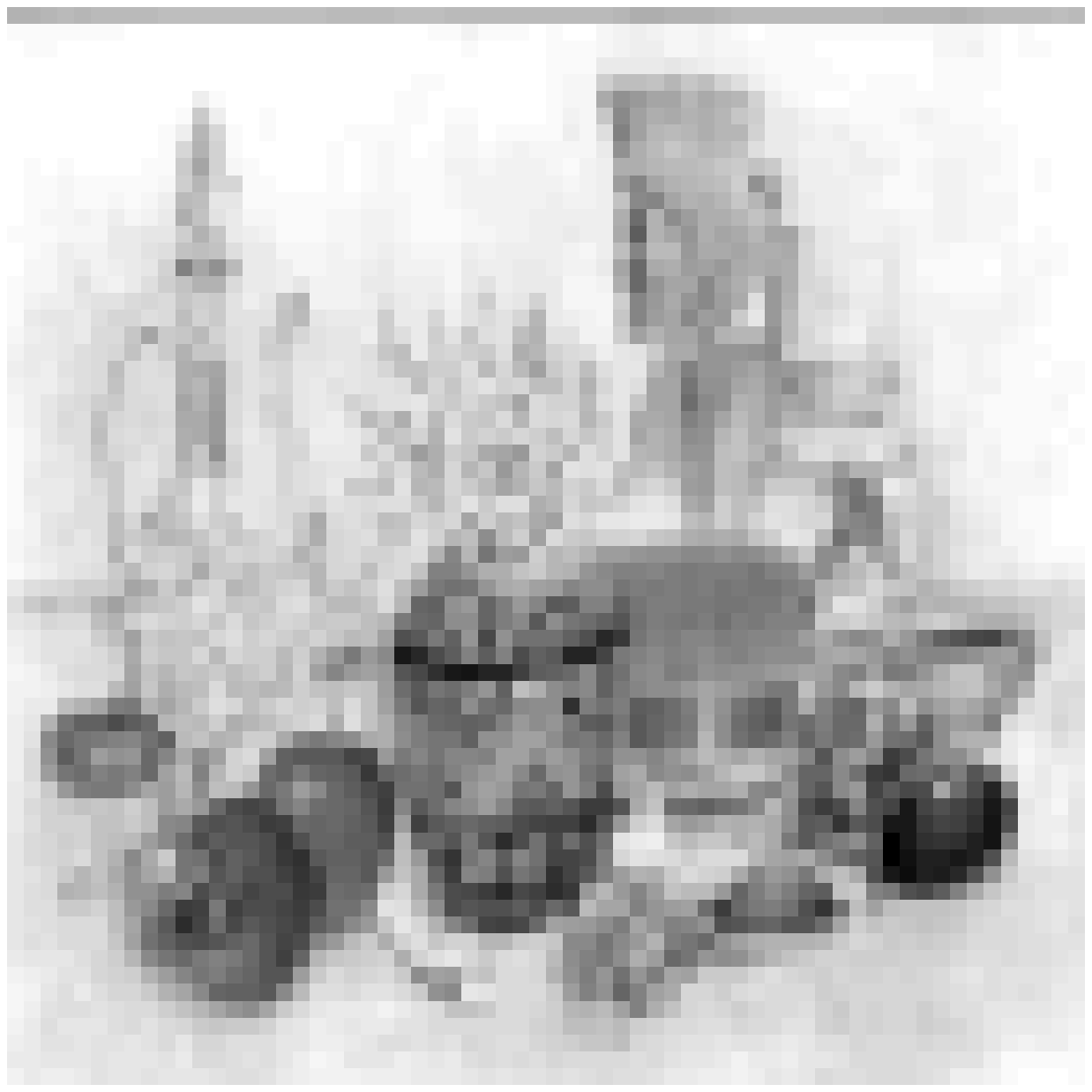} \\[-22mm]\hspace*{15mm}  
\includegraphics[width=25mm,height=25mm]{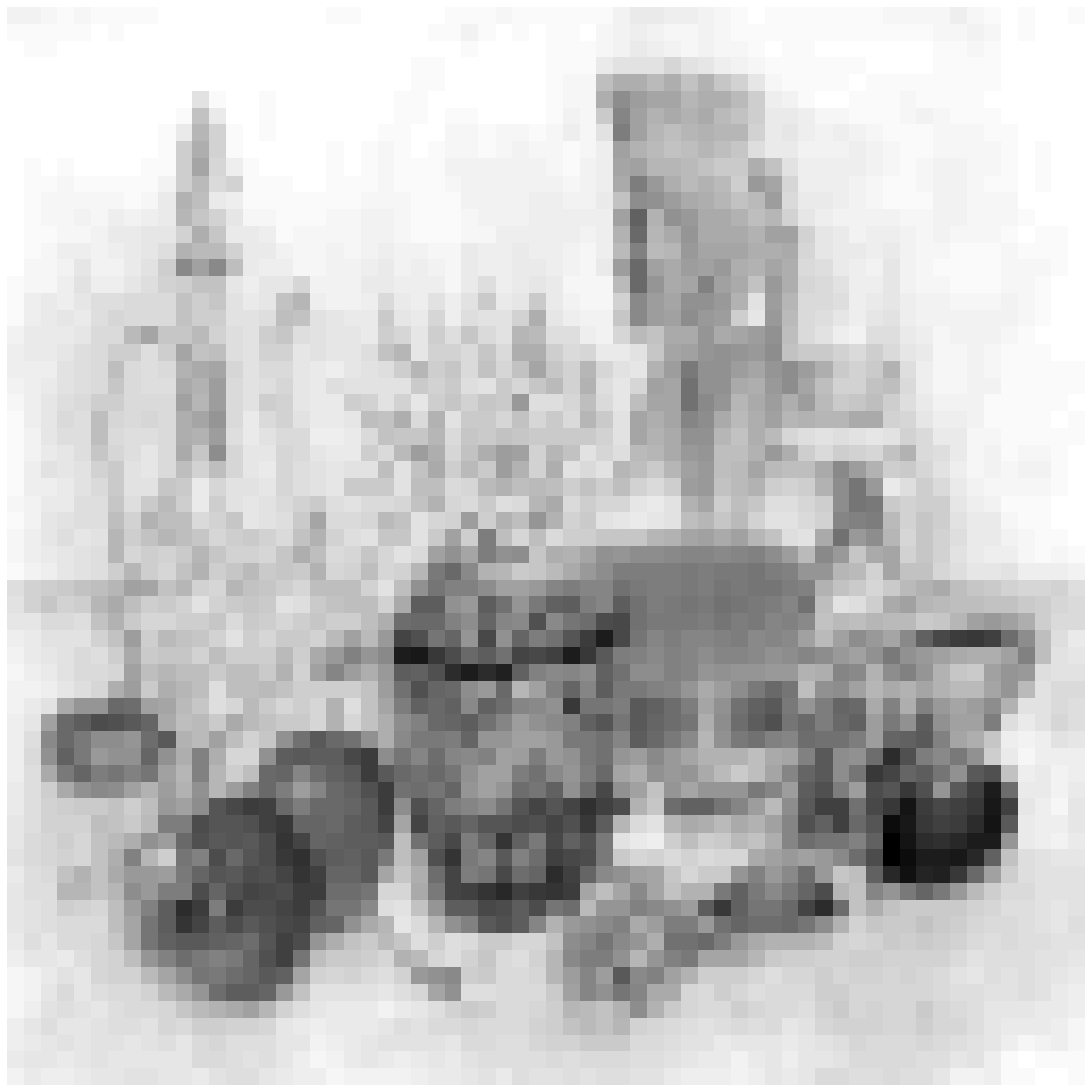} \\[-22mm]\hspace*{20mm}  
\includegraphics[width=25mm,height=25mm]{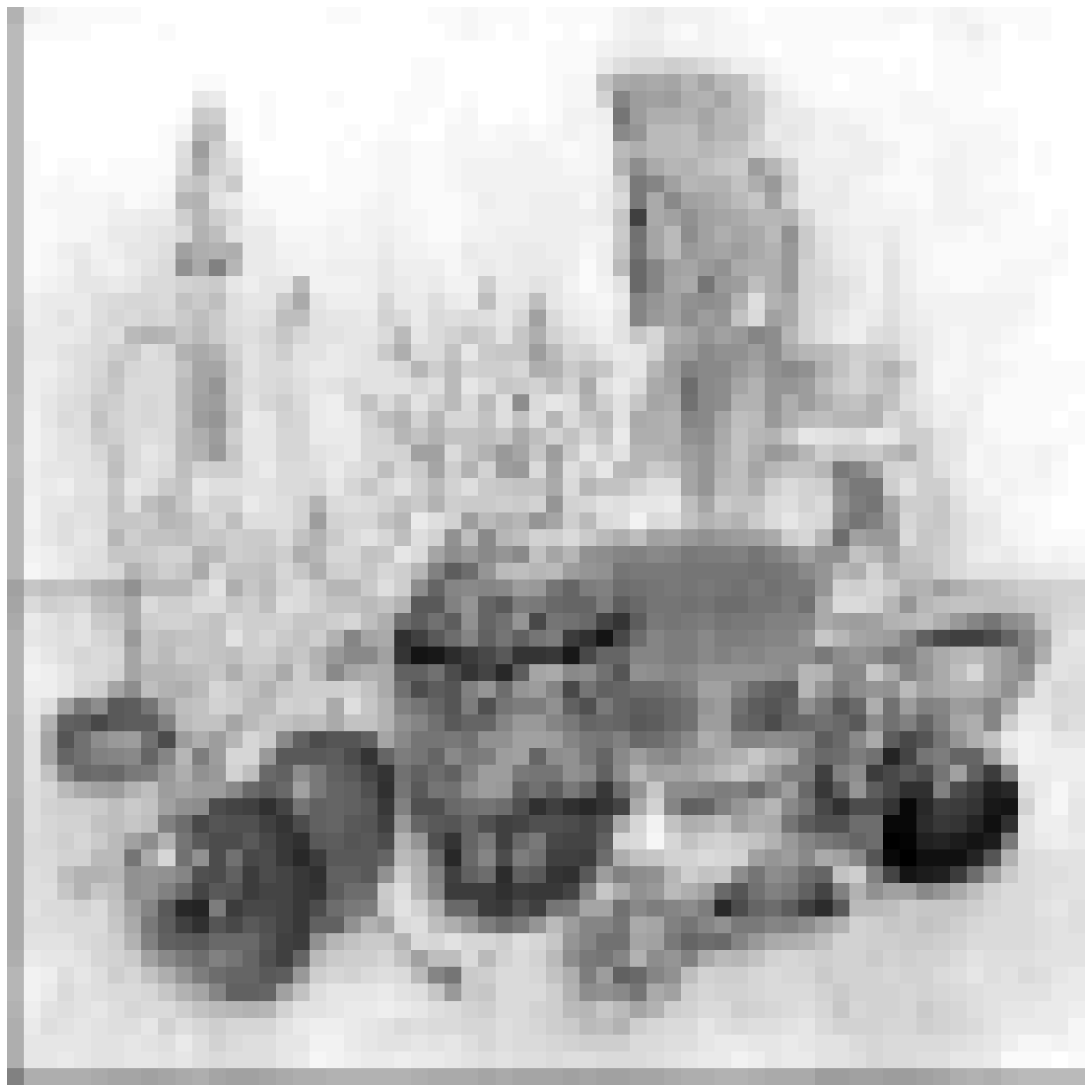} \\[-22mm]\hspace*{25mm}  
\includegraphics[width=25mm,height=25mm]{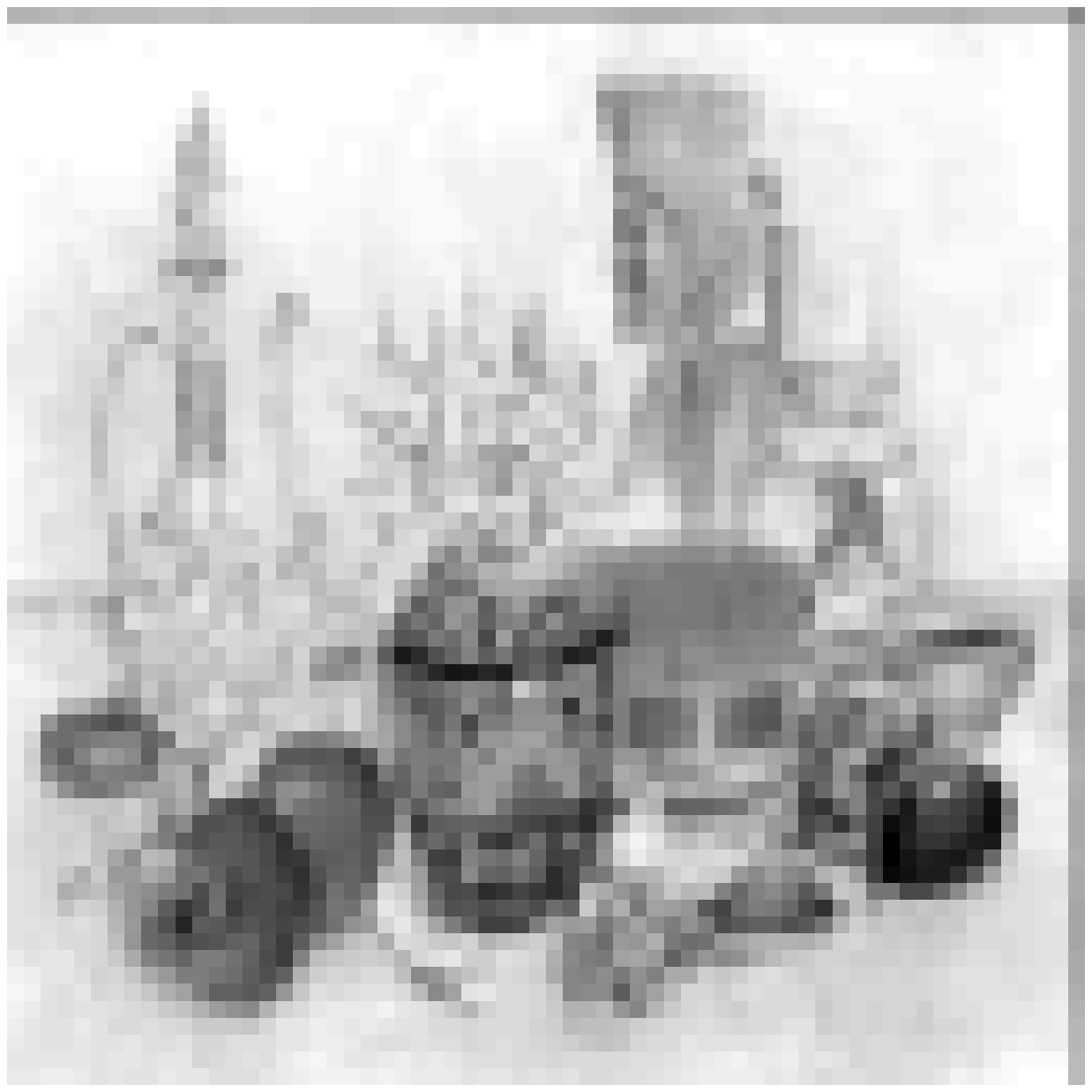}  
\etabu
&
{\Large \btabu{@{}c@{}} {$?$} \\ $\Longrightarrow$ \etabu  }
&
\btabu{l} 
\includegraphics[width=25mm,height=25mm]{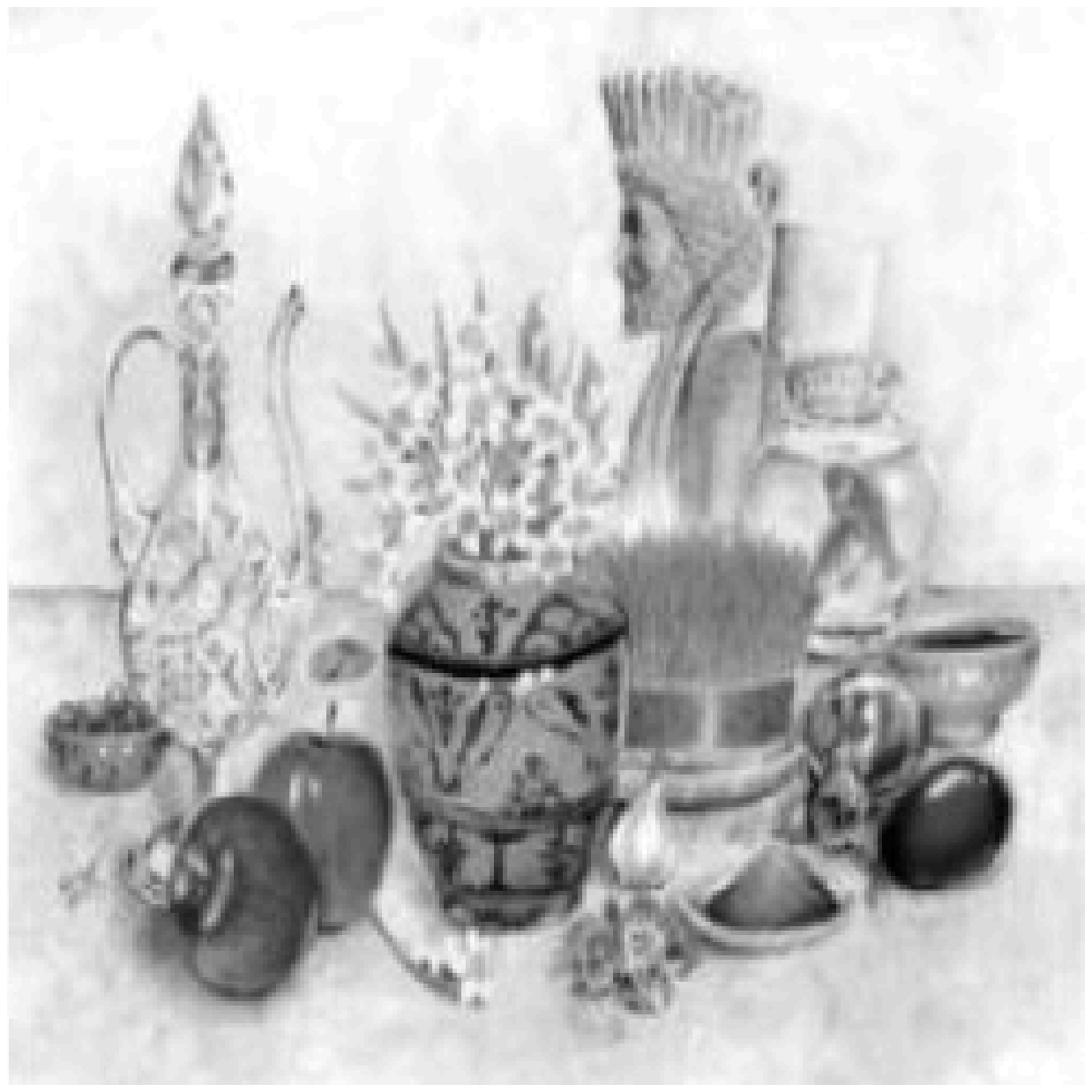} \\[-22mm]\hspace*{5mm} 
\includegraphics[width=25mm,height=25mm]{HR0} \\[-22mm]\hspace*{10mm}  
\includegraphics[width=25mm,height=25mm]{HR0} \\[-22mm]\hspace*{15mm}  
\includegraphics[width=25mm,height=25mm]{HR0} \\[-22mm]\hspace*{20mm}  
\includegraphics[width=25mm,height=25mm]{HR0} \\[-22mm]\hspace*{25mm}  
\includegraphics[width=25mm,height=25mm]{HR0}    
\etabu
\etabu
\caption{Inverse problem of video image restoration}
\label{Fig6}
\efig

The discretized version of this inverse problem can be written as  
\beq
\gb_t=\Hb\fb_t+\epsilonb_t
\eeq
where 
$\gb_t$ and $\fb_t$ contains samples of the ouput $g_t(\rb)$ and the intput $f_t(\rb')$ and the matrix $\Hb$, in this case, is again a Toeplitz-Bloc-Toeplitz (TBT) matrix with a generic bloc-ligne composed of the samples of the point spread function (PSF) $h(\rb)$.

\subsection{Multi Inputs Multi Outputs inverse problems}
Multi Inputs Multi Outputs (MIMO) imaging systems can be modeled as:
\beq\label{eq22}
g_i(\sb) = \sum_{j=1}^N \int h_{ij}(\sb,\rb) f_j(\rb)  \d{\rb} + \epsilon_i(\rb), \quad i=1,\cdots,N
\eeq
\subsubsection{MIMO sources localisation and estimation}
One such example is the case where $n$ radio sources $\{f_j(t), j=1,\cdots,n\}$ emitting in the same time are received by $m$ receivers $\{g_i(t), i=1,\cdots,m\}$, each one receiving a linear combination of delayed and degraded versions of original waves: 
\beq\label{eq23}
g_i(t) = \sum_{j=1}^N \int h_{ij}(t-t') f_j(t'-\tau_{ij})  \d{t'} + \epsilon_i(t), \quad i=1,\cdots,N
\eeq
where $h_{ij}(t)$ is the impulse response of the channel between the $i$-th receiver and the $j$-th source. 
The discretized version of this inverse problem can be written as  
\beq
\gb_i=\Hb_{i,j}\fb_j+\epsilonb_i
\eeq
where 
$\gb_i$ and $\fb_j$ contains samples of the ouput $g_i(t)$ and the intput $f_t(t)$ and the matrices $\Hb_{i,j}$ are Toeplitz matrices described by the impulse responses $h_{i,j}(t)$.

\subsubsection{MIMO deconvolution}
A MIMO image restoration problem is~:
\beq\label{eq24}
g_i(\rb) = \sum_j \int h_{ij}(\rb-\rb') f_j(\rb')  \d{\rb'} + \epsilon_i(\rb)
\eeq
and one such example is the case of color image restoration where each color component can be considered as an input.  

\bfig[htb] 
\btabu{@{}c@{}c@{}c@{}}
\btabu{c}
\includegraphics[width=50mm,height=50mm]{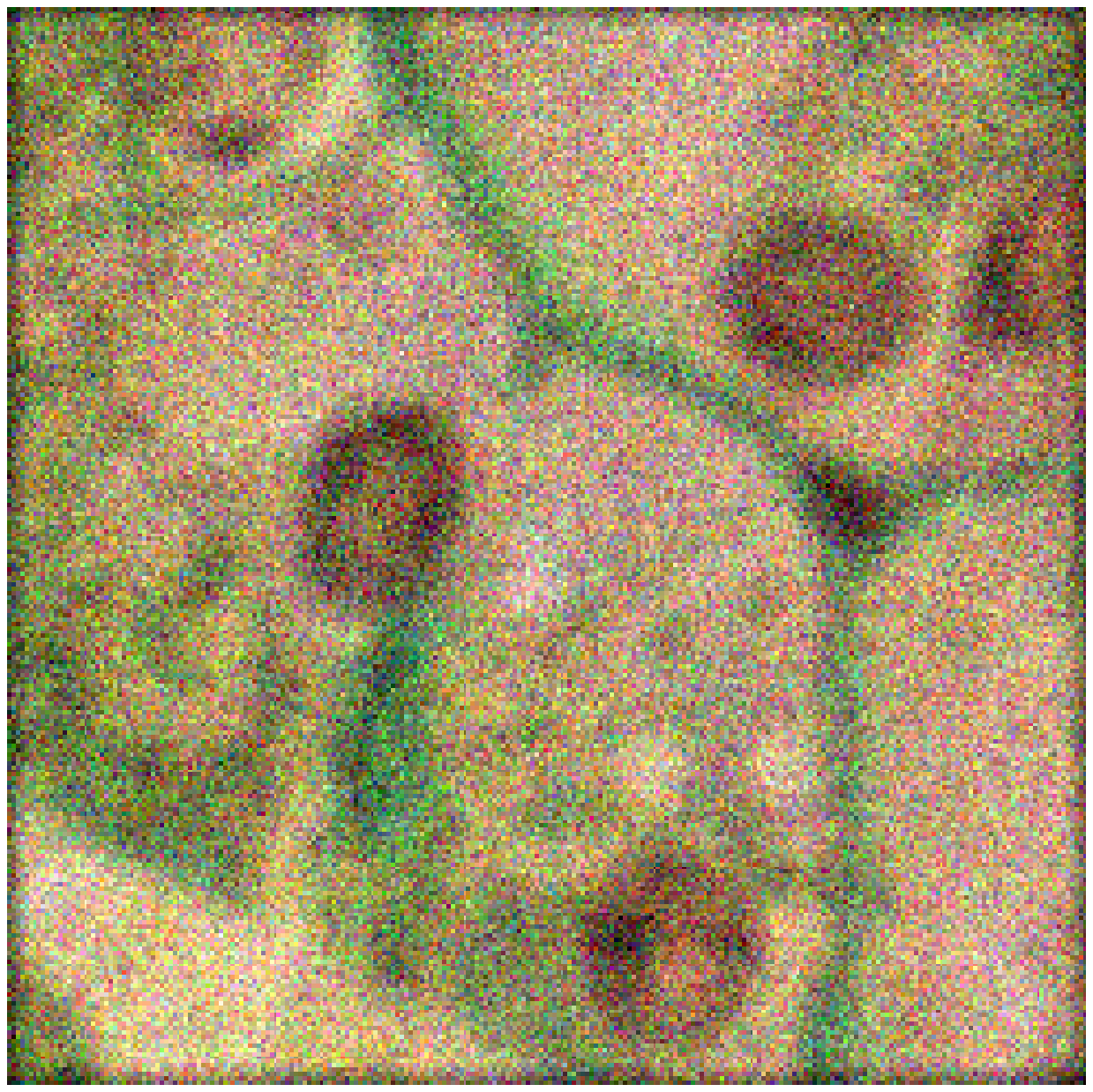}
\etabu
&
{\Large \btabu[b]{@{}c@{}} {$?$} \\ $\Longrightarrow$ \etabu  }
&
\btabu{c}
\includegraphics[width=50mm,height=50mm]{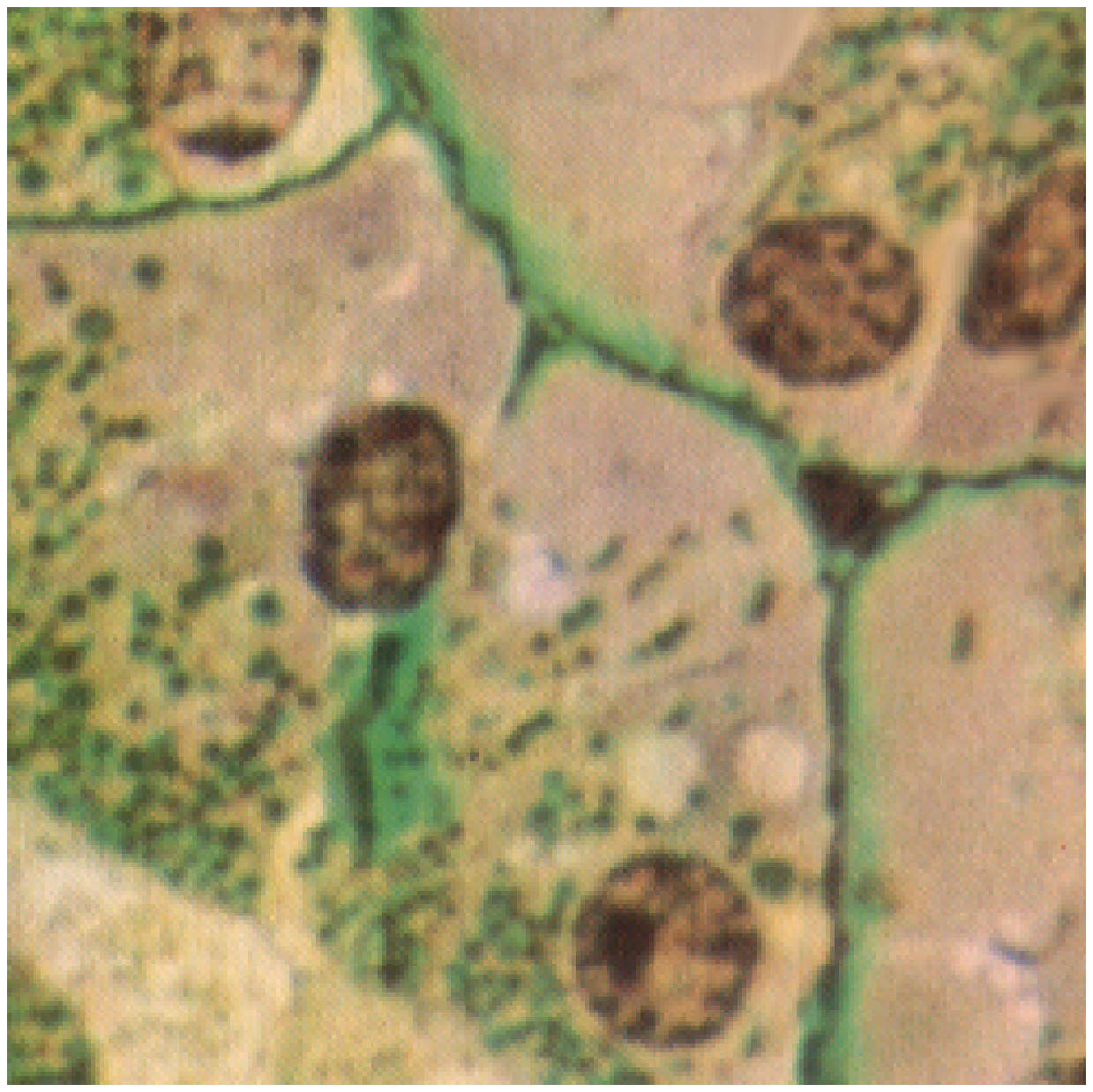}
\etabu
\etabu
\caption{Color image restoration as an example of MIMO inverse problem.}
\label{Fig7}
\efig

\subsection{Source Separation}
A particular case of a MIMO inverse problem is the blind source separation (BSS):
\beq\label{eq25}
g_i(\rb) = \sum_j \int A_{ij} h_{ij}(\rb-\rb') f_j(\rb') \d{\rb'} + \epsilon_i(\rb)
\eeq
and a more particular one is the case of \emph{instantaneous mixing}: 
\beq\label{eq26}
g_i(\rb) = \sum_j A_{ij} f_j(\rb) + \epsilon_i(\rb)
\eeq
The particularity of these problems is that the the mixing matrix $\Ab=\{A_{ij}\}$ is also unknown. 

\bfig[htb] 
\btabu{@{}c@{}c@{}c@{}c@{}c@{}}
\btabu{c}
\includegraphics[width=50mm,height=25mm]{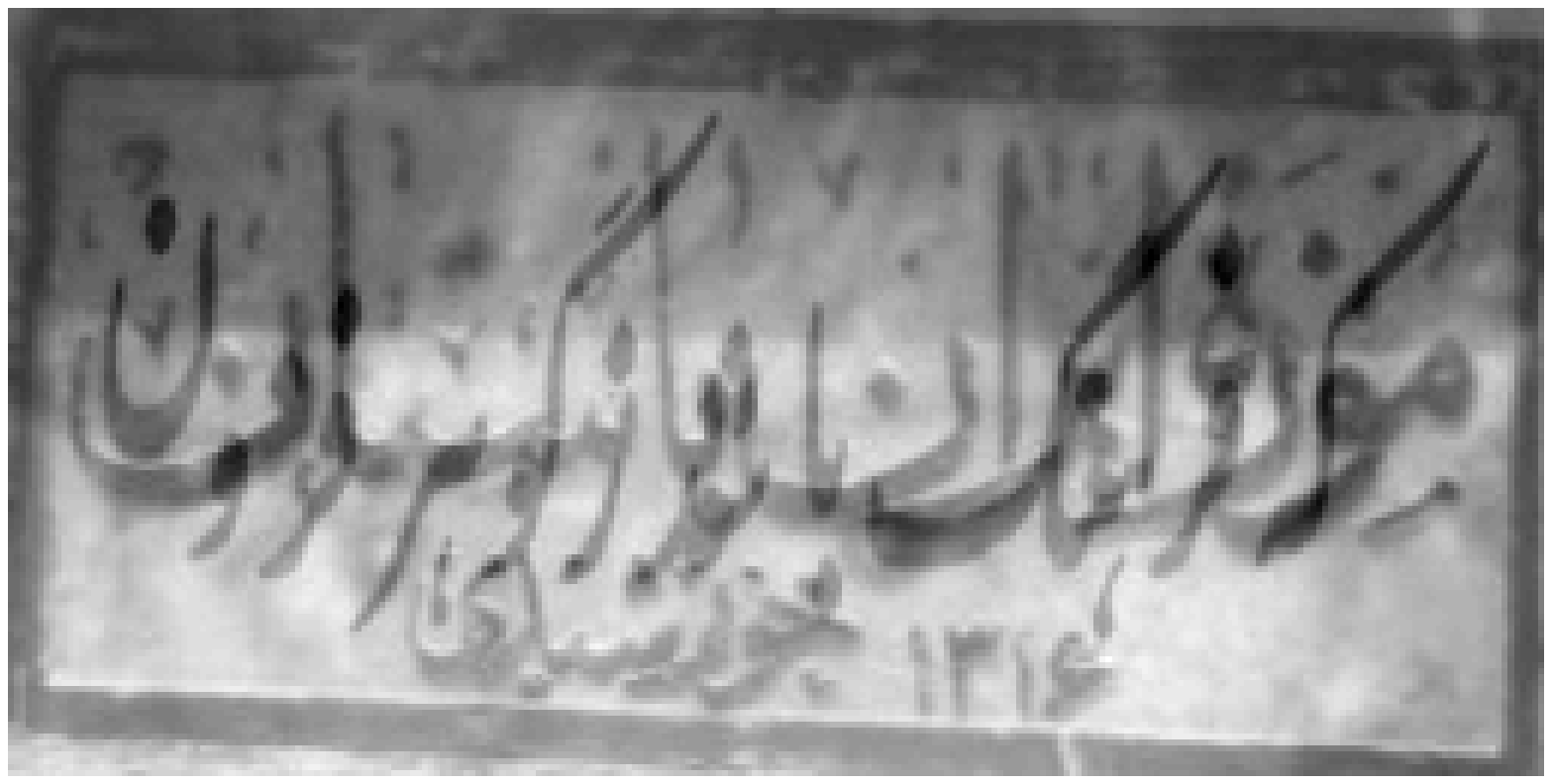} \\
\includegraphics[width=50mm,height=25mm]{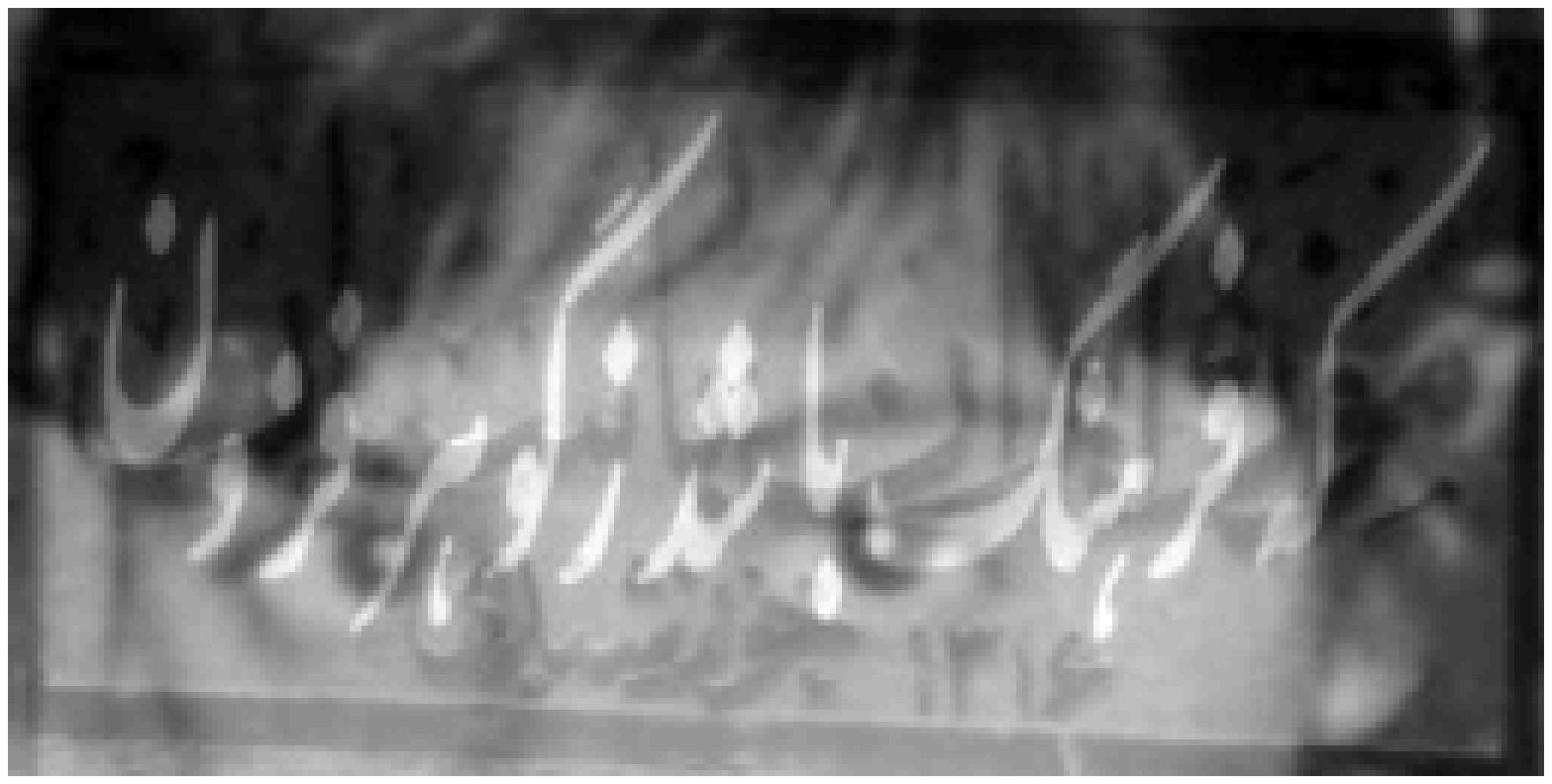} \\
\includegraphics[width=50mm,height=25mm]{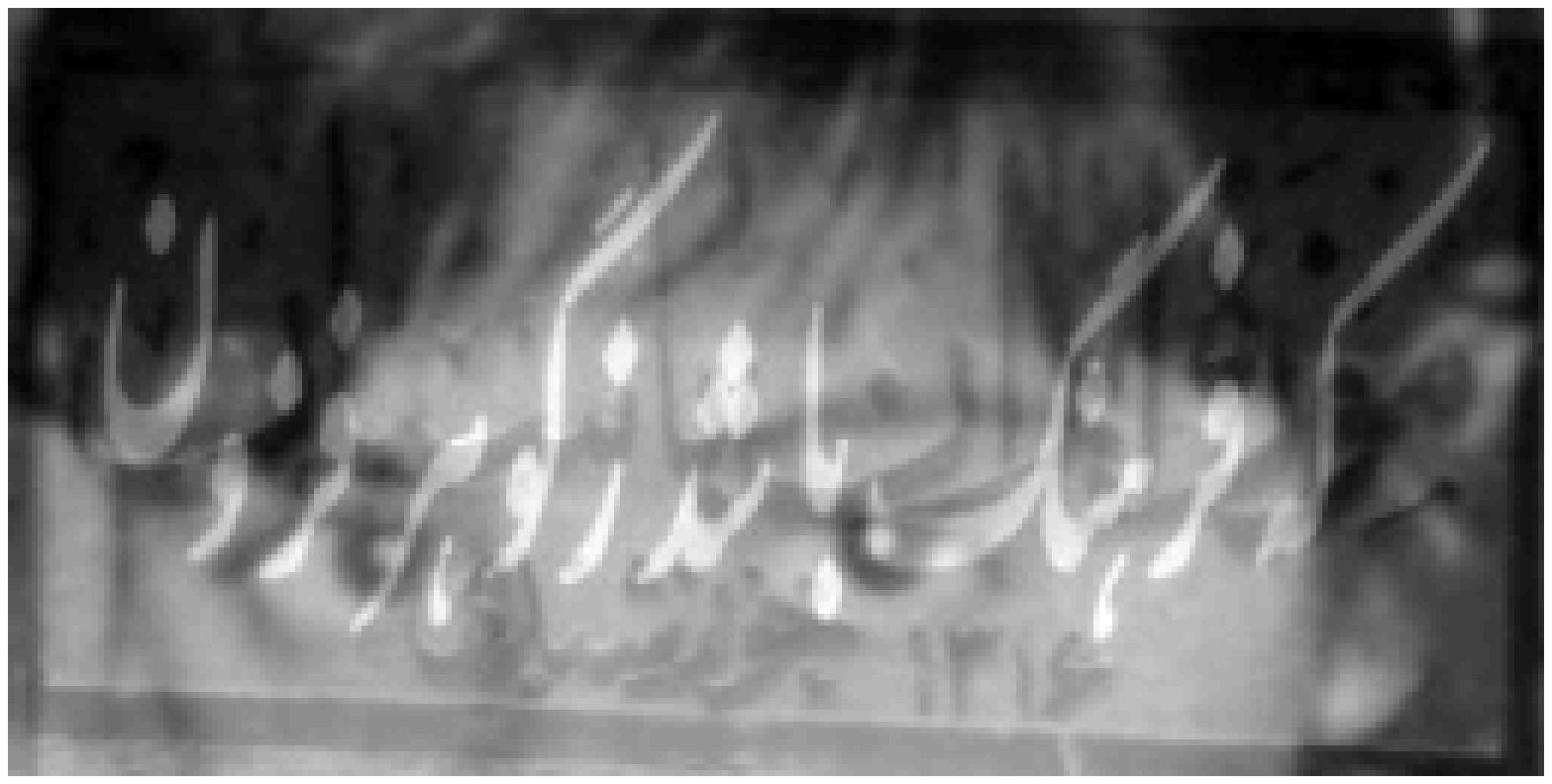}
\etabu
&
{\Large \btabu[b]{@{}c@{}} ~\\ ~\\ ~\\ ~\\ {$?$} \\ $\Longrightarrow$ \etabu  }
&
\btabu{c}
\includegraphics[width=50mm,height=25mm]{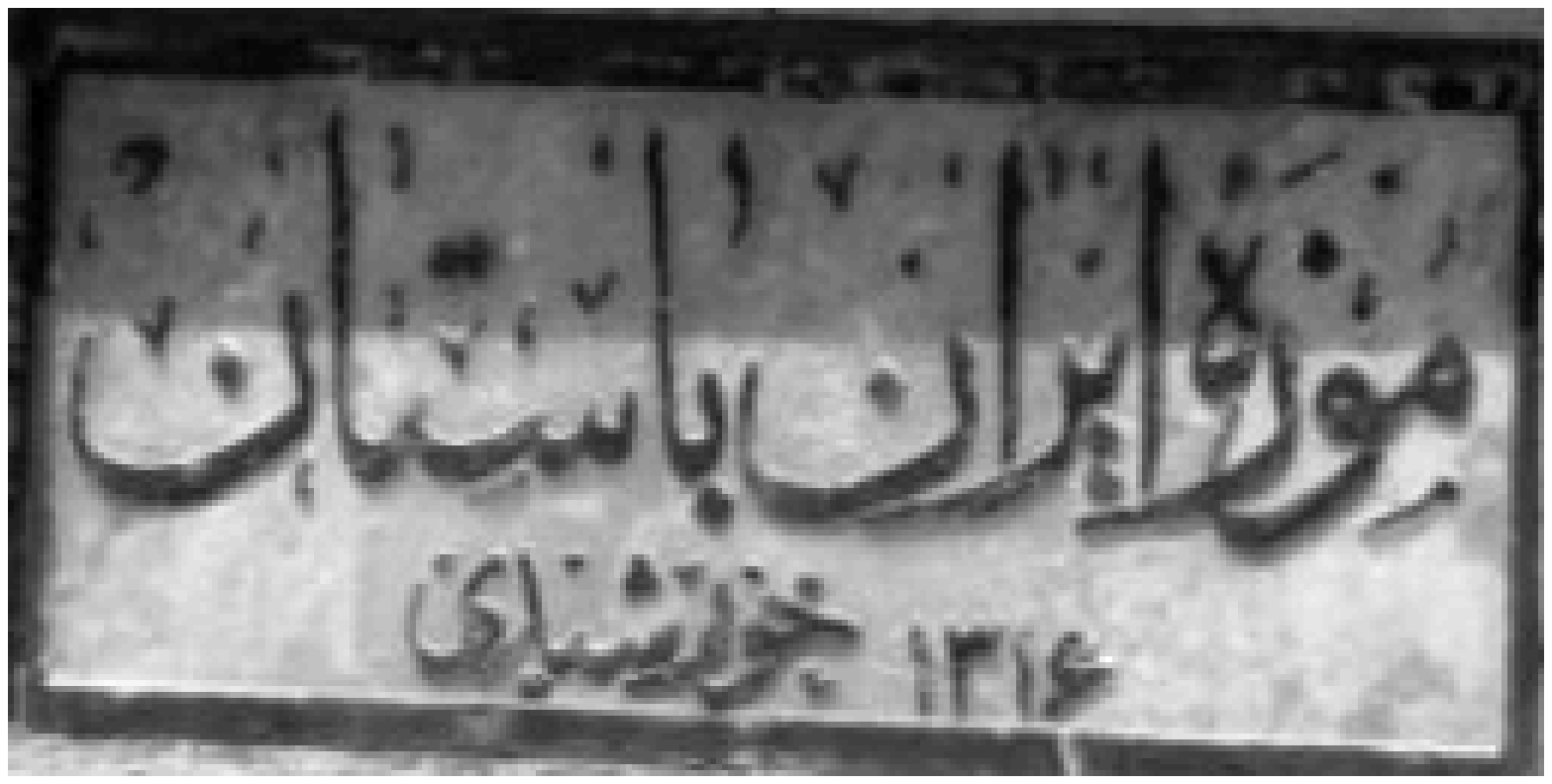} \\
\includegraphics[width=50mm,height=25mm]{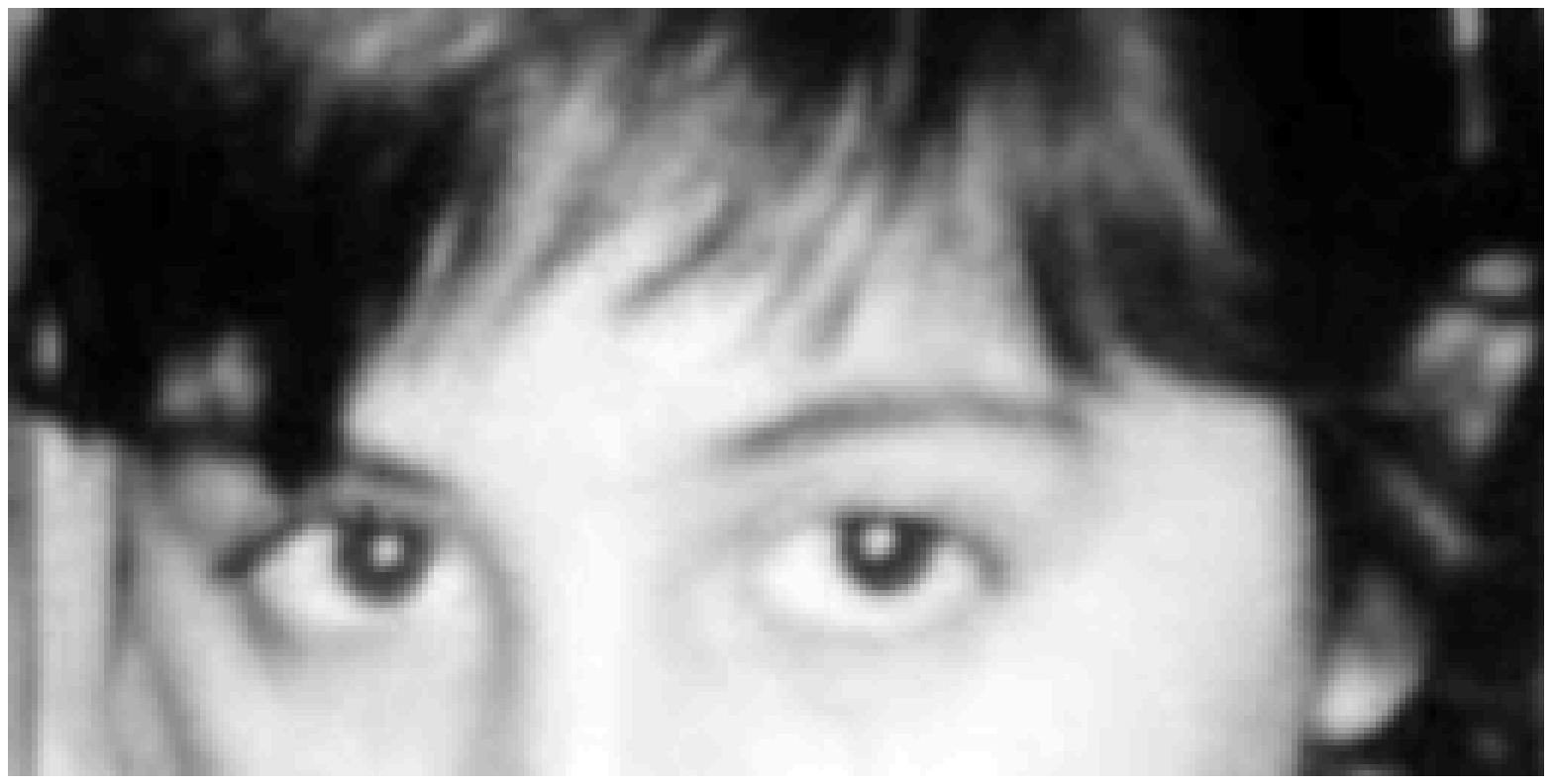} \\
\includegraphics[width=50mm,height=25mm]{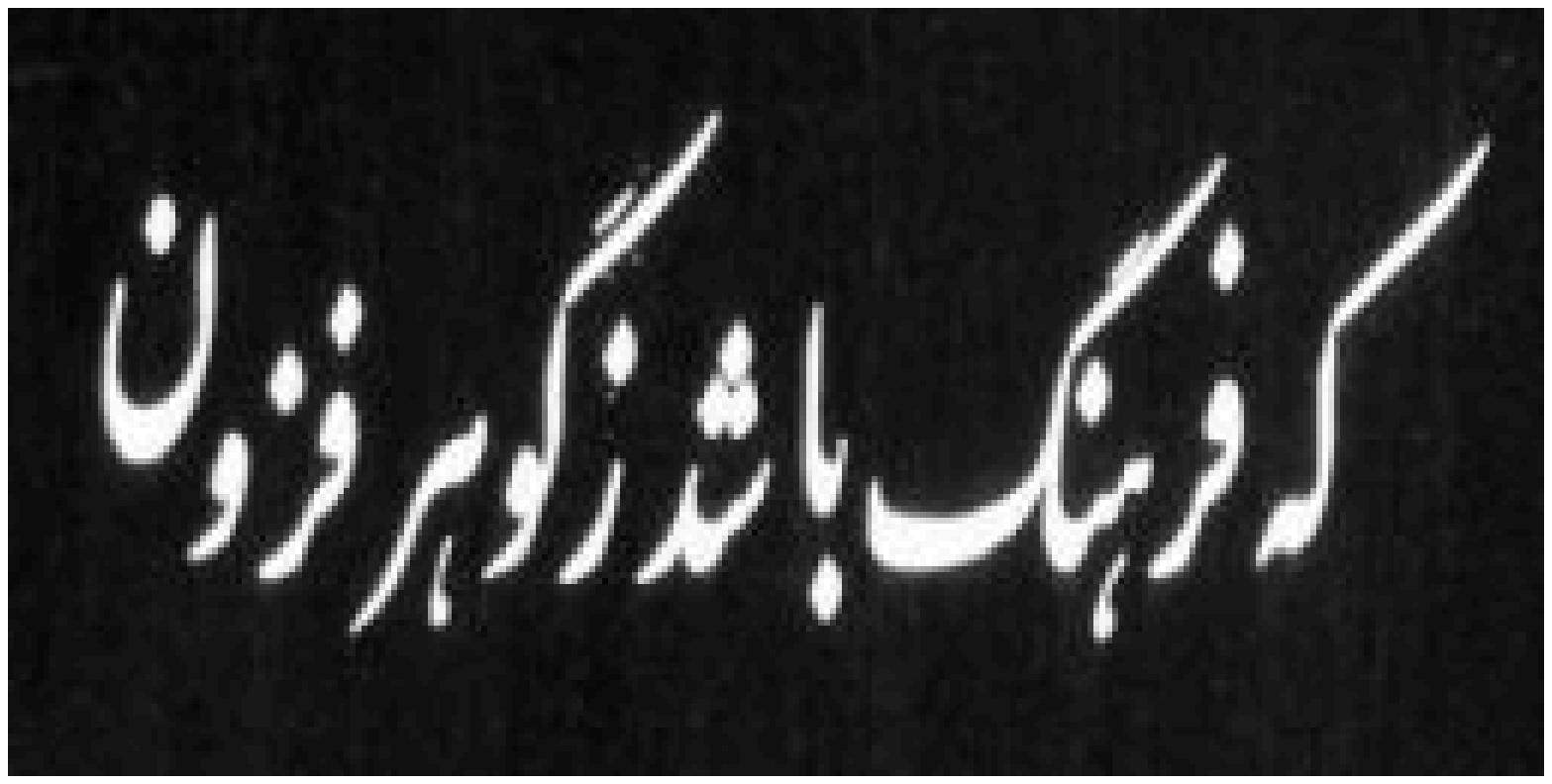}
\etabu
\etabu
\caption{Blind image separation.}
\label{Fig8}
\efig

\subsection{Multi Inputs Single Output inverse problems}
A Multi Inputs Single Output (MISO) system is a particular case of MIMO when we have only one input:
\beq\label{eq24b}
g_i(\sb) = \sum_j \int h_{i}(\sb,\rb) f(\rb)  \d{\rb} + \epsilon_i(\rb)
\eeq

\subsubsection{MIMO sources localisation and estimation}
One example of MISO inverse problem is a non destructive testing (NDT) for detection and evaluation of the defect created due to an impact on a surace of an object using microwave imaging where two images are obtained when a rectangular waveguide scans this surface two times. In the first scan the rectangular waveguide is oriented in shorter side and in the second case in longer side. By this way, two images $g_i(\rb), i=1,2$ are obtained, each has to be considered as the output of a linear system with the same input $f(\rb)$ and two different channels. This is a MISO linear and invariant systems. 
\subsubsection{Image super-resolution as a MISO inverse problem}
Another MISO system is the case of Super-Resolution (SR) imaging using a few Low Resolution (LR) images obtained by low cost cameras:
\beq\label{eq25b}
g_i(\sb) = \sum_j \int h_{i}(\sb,\rb) f(\rb)  \d{\rb} + \epsilon_i(\rb)
\eeq
where $g_i$ are the LR images and $f$ is the desired High Resolution (HR) image. The functions $h_i$ represent a combination of at least three operations: 
i) a low pass filtering effect, 
ii) a mouvement (translational or with rotation and zooming effects) of the camera and iii) a sub-sampling. 

The following figure shows one such situation. 

\bfig[htb] 
\btabu{@{}c@{}c@{}c@{}}
\btabu{@{}c@{}c@{}c@{}}
\includegraphics[width=20mm,height=20mm]{LRf01} & 
\includegraphics[width=20mm,height=20mm]{LRf02} & 
\includegraphics[width=20mm,height=20mm]{LRf03} \\ 
\includegraphics[width=20mm,height=20mm]{LRf04} & 
\includegraphics[width=20mm,height=20mm]{LRf05} & 
\includegraphics[width=20mm,height=20mm]{LRf06} \\ 
\includegraphics[width=20mm,height=20mm]{LRf04} & 
\includegraphics[width=20mm,height=20mm]{LRf05} & 
\includegraphics[width=20mm,height=20mm]{LRf06}  
\etabu
&
{\Large \btabu{@{}c@{}} {$?$} \\ $\Longrightarrow$ \etabu  }
&
\btabu{@{}c@{}} 
\includegraphics[width=60mm,height=60mm]{HR0}  
\etabu
\etabu
\caption{SR problem where a serie of LR images are used to construct a HR image.}
\label{Fig9}
\efig

The discretized version of this inverse problem can be written as  
\beq
\gb_i=\Hb_{i,j}\fb+\epsilonb_i
\eeq
where 
$\gb_i$ and $\fb$ contains samples of the ouput $g_i(t)$ and the intput $f_t(t)$ and the matrices $\Hb_{i,j}$ are Toeplitz matrices described by the impulse responses $h_{i,j}(t)$.

\subsection{Multi modality in CT imaging systems}
Using different modalities has become a main tool in imaging systems where to explore the internal property of a body one can use X rays, ultrasounds, microwaves, infra-red, magnetic resonance, etc. 
As an example, in X ray imaging, the observed radiographies give some information on the volumique distribution of the material density inside the object while the ultrasound echography gives information on the changing positions (contours) of ultrasound properties inside the object. One can then want to use both techniques and use a kind of data fusion to obtain a higher quality of images of the body. An example of such situation is given in (\ref{Fig10}).  

\bfig[htb] 
\bcc
\btabu{@{}ccc@{}}
a \includegraphics[width=35mm,height=35mm]{fxy2_256} &
c \includegraphics[width=35mm,height=35mm]{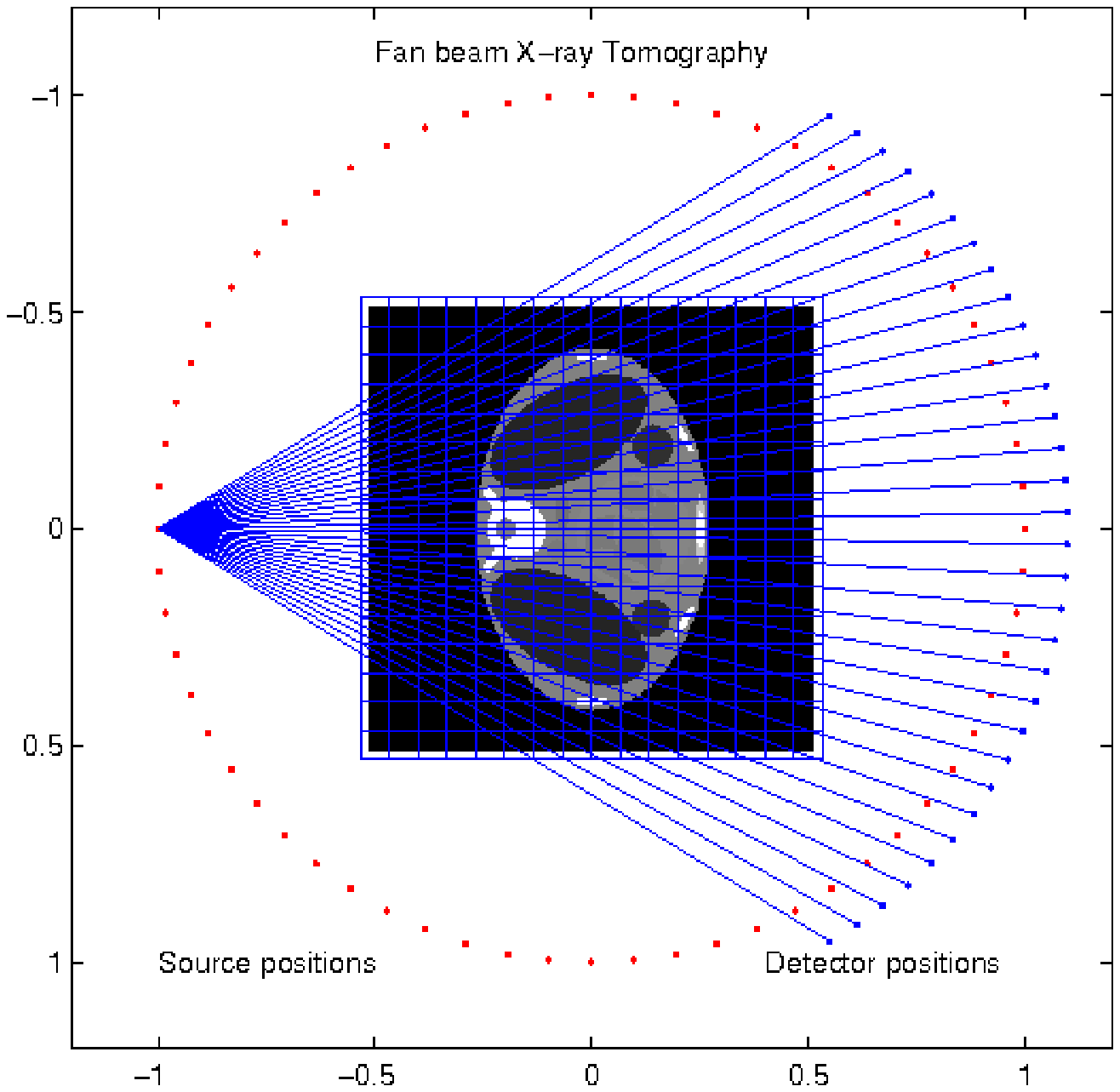} &
e \includegraphics[width=35mm,height=35mm]{p_fxy2_256} \\
b \includegraphics[width=35mm,height=35mm]{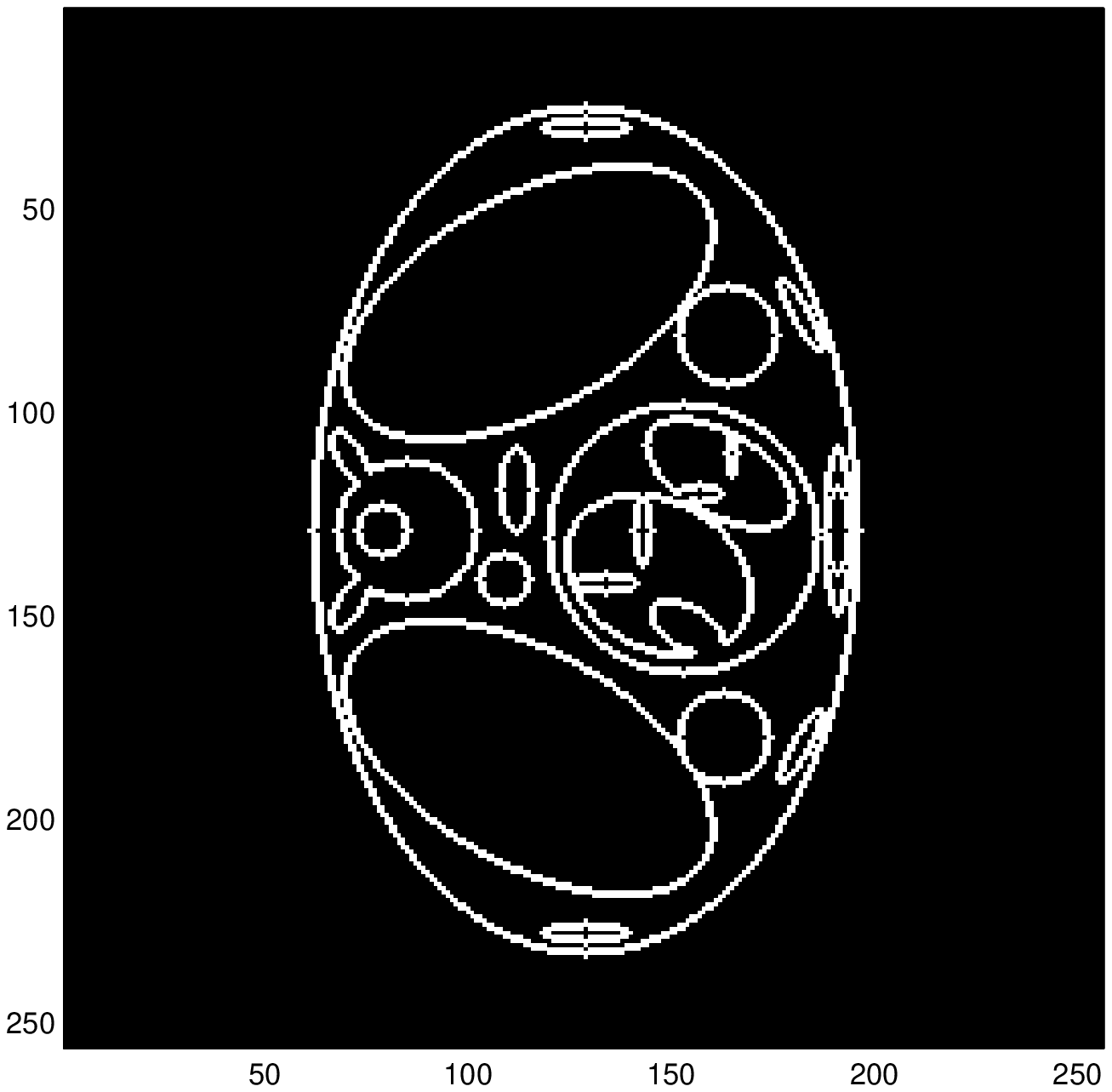} & 
d \includegraphics[width=35mm,height=35mm]{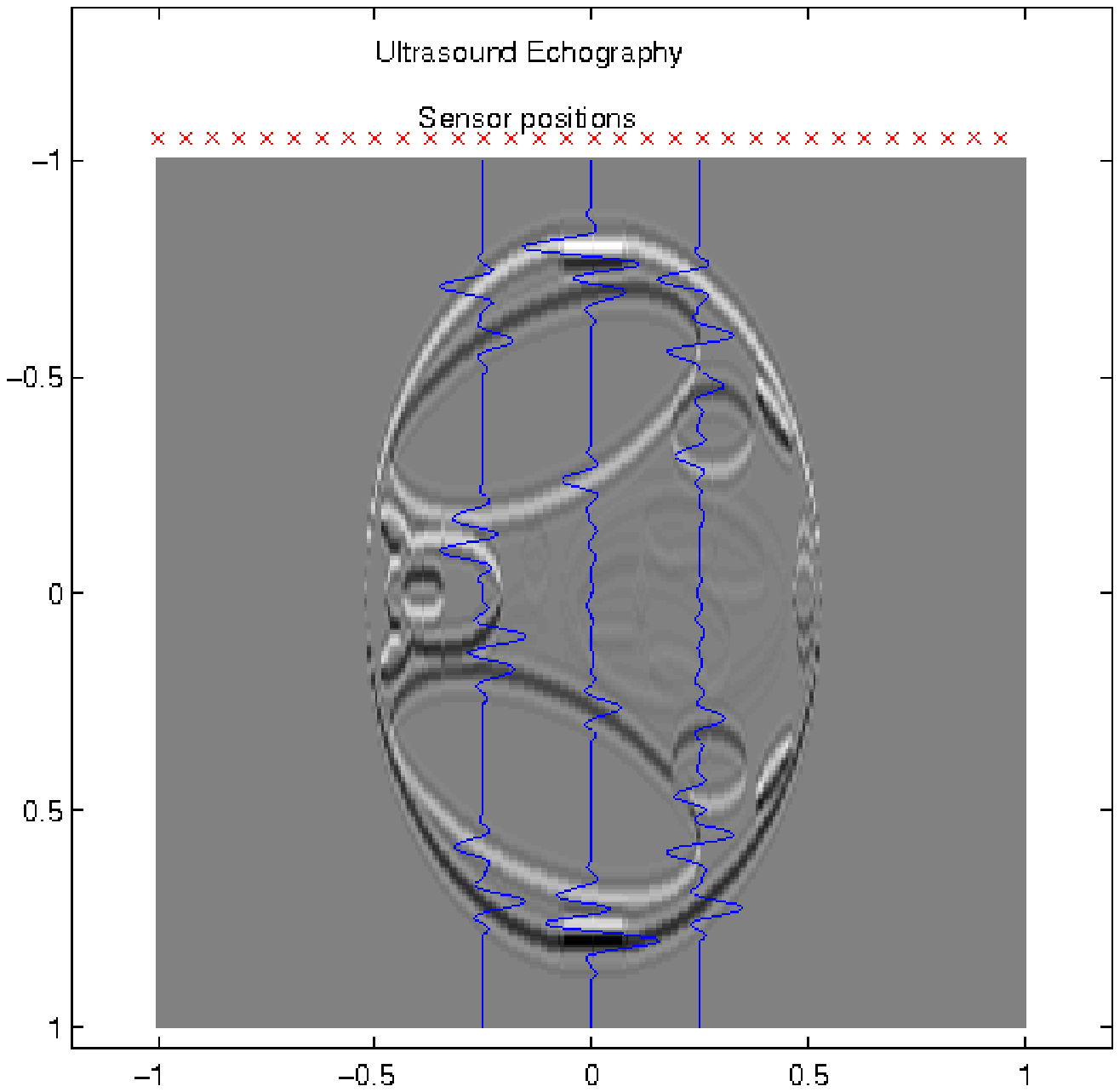} & 
f \includegraphics[width=35mm,height=35mm]{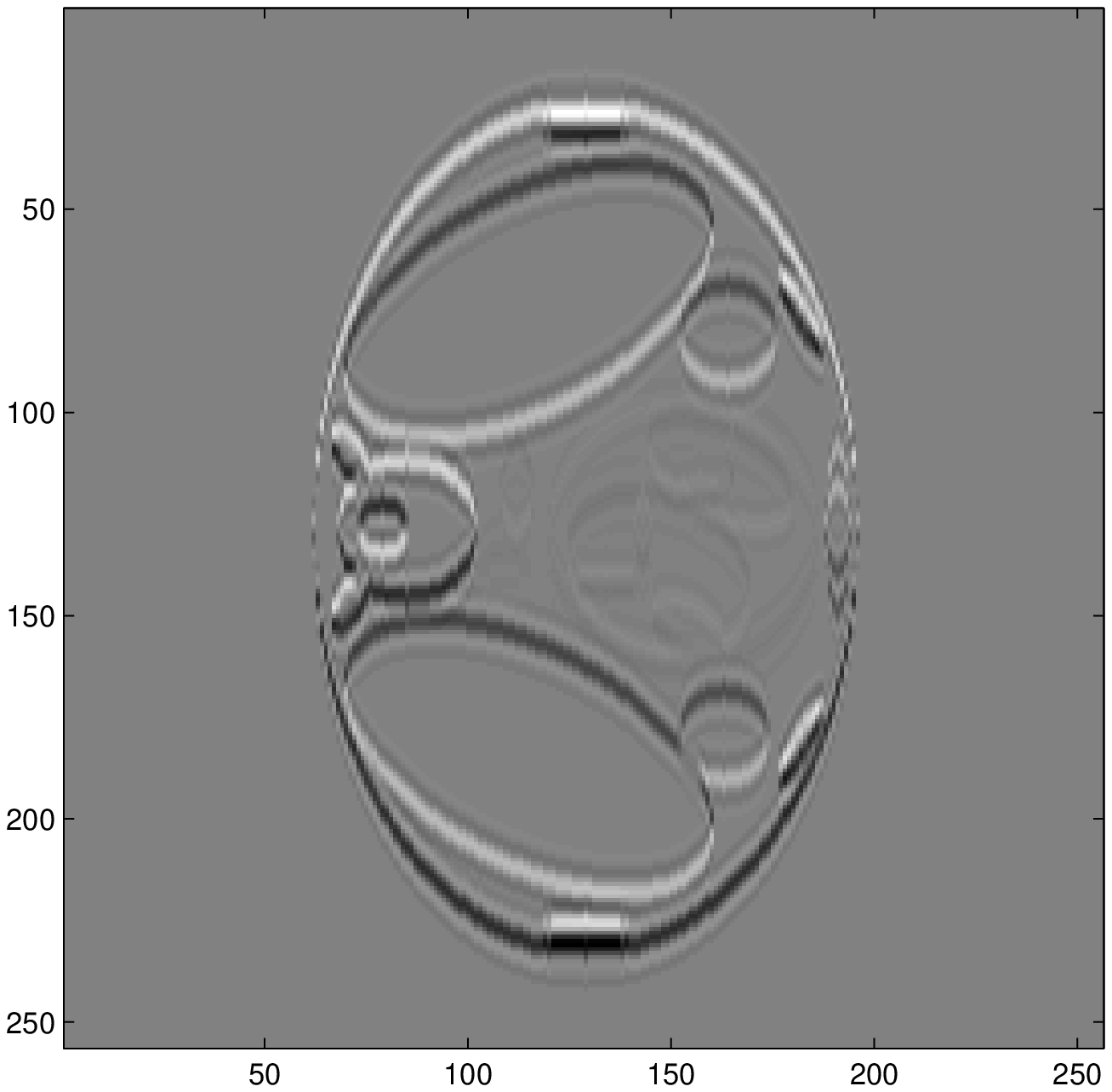} 
\etabu
\ecc
\caption[Multi modality in CT imaging systems]{Multi modality in CT imaging systems (a) Original objet, (b) Contours of the different homogeneous regions, (c) Data acquisition geometry in X ray tomographie, (d) Data acquisition geometry in ultrasound echography, (e) Observed data (sinogram) in X ray tomographie, (f) Observed data in ultrasound echographie.}
\label{Fig10}
\efig

\subsection{Fusion of X ray and ultrasound echography.}
An example of multimodality and data fusion in CT is the use of X ray radiographic data and the ultrasound echographic data is shown in Figure (\ref{Fig11}) and for more details on this application see \cite{djafari02d,djafari02c,djafari02b,djafari03b}. 

\bfig[hbt] 
\btabu{@{}c@{}c@{}c@{}c@{}c@{}}
\btabu{c}
\includegraphics[width=45mm,height=45mm]{p_fxy2_256} \\
\includegraphics[width=45mm,height=45mm]{us1_256}
\etabu
&
{\Large \btabu[b]{@{}c@{}} ~\\ ~\\ ~\\ ~\\ {$?$} \\ $\Longrightarrow$ \etabu  }
&
\btabu{c}
\includegraphics[width=45mm,height=45mm]{fxy2_256} \\
\includegraphics[width=45mm,height=45mm]{sxy0_256} 
\etabu
\etabu
\caption{Inverse problem of X ray and ultrasound data fusion.}
\label{Fig11}
\efig

\newpage
\section{Basics of deterministic inversion methods}
To illustrate the basics of the inversion methods, we start by considering the case of a Single Input Single Output (SISO) linear system:
\beq
\gb=\Hb\fb+\epsilonb
\eeq
The idea can be easily extended to the case of MISO or MIMO. For an extend details to these methods refer to \cite{djafari97m,djafari99e}.

\subsection{Match filtering}
First assume that the errors and measurement noise are negligeable and that we could choose the basis functions $\phi_i$ and $\psi_j$ could be choosed in such a way that the matrix $\Hb$ is square $(m=n)$ and self-adjoint ($\Hb'\Hb=\Ib$) (un unrealistic hypothesis). Then, the solution to the problem would be:
\beq
\fbh=\Hb' \gb
\eeq
This solution has been used in many cases. For example in deconvolution, this solution is called \emph{Matching filtering}. The main reason is that, in a deconvolution problem, the matrix $\Hb$ is a Toeplitz matrix, so is its transpose $\Hb'$. The forward matrix operation $\Hb\fb$ corresponds to a convolution {\tt conv($h,f$)}. The adjoint matrix operation $\Hb' \gb$ then also corresponds to a convolution {\tt conv($\tilde{h},g$)} where $\tilde{h}(t)=h(-t)$. 

Another example is in computed tomography (CT) where the projection data in each angle direction $\gb_i$ is related to the image $\fb$ through a projecting matrix in that direction $\Hb_i$ such that we can write:
\beq
\left[\barr{c} \gb_1 \\ \vdots \\ \gb_K\earr\right]=\left[\barr{c} \Hb_1 \\ \vdots \\ \Hb_K\earr\right]\fb+\left[\barr{c} \epsilonb_1 \\ \vdots \\ \epsilonb_K\earr\right]
\eeq
and the adjoint operation:
\beq
\fbh=\Hb' \gb=\sum_{k=1}^K \Hb_k' \gb
\eeq
corresponds to what is called \emph{backprojection}.

However, as it is mentionned, the hypothesis made here are unrealistic.  

\subsection{Direct inversion}
The next step is just to assume that the forward matrix is invertible. Then, one can try to define the solution as:
\beq
\fbh=\Hb^{-1} \gb
\eeq
But, in practice, this also is an illusion, because, even if the matrix $\Hb$ is mathematicaly invertible, it is, very often, very ill-conditionned. This means that small errors on the data $\delta \gb$ will generate great errors $\delta \fbh$ on the solution. 
This method, in deconvolution, corresponds to the analytical method of inverse filtering, which is, in general, unstable. 

In other applications, the main difficulty is that, very often the matrix $\Hb$ is even not square, \ie $m\not= n$, because the number of the measured data $m$ may not be equal to the number of parameters $n$ describing the unknown function $f$ in (\ref{eq2c}).

\subsection{Least square and generalized inversion}
For the case where $m>n$, a solution will be the least square (LS) defind as:
\beq
\fbh=\argmin{\fb}{\|\gb-\Hb\fb\|^2}
\eeq
which results to the normal equation:
\beq
[\Hb'\Hb] \fbh=\Hb' \gb
\eeq
and if the matrix $\Hb'\Hb$ is inversible ($\rang{\Hb'\Hb}=n$), then the solution is given by 
\beq
\fbh=[\Hb'\Hb]^{-1}\Hb' \gb
\eeq
When $m<n$, the problem may have an infinite number of solutions. So, we may choose one of them by requesting some particular \aprio property, for example to have \emph{minimum norme}. The mathematical problem is then:
\beq
\fbh=\argmin{\{\Hb\fb=\gb\}}{\|\fb\|^2} 
\eeq
or written differently 
\beq
\mbox{minimize~~~}\|\fb\|^2 \quad \mbox{subject to}\quad \Hb\fb=\gb
\eeq
The solution is obtained via the Lagrange multiplier method which, in this case, results to 
\beq
\left[\barr{cc} \Ib & -\Hb^t \\ \Hb & \bm{0}\earr\right] 
\pmat{\fb \\ \lambdab}
=\pmat{\bm{0} \\ \gb}
\eeq
which gives
\beq
\fbh_2=\Hb^t(\Hb\Hb^t)^{-1}\gb
\eeq
if $\Hb\Hb^t$ is invertible. 

The main difficulty in these methods is that the solution, in general, is too sensitive to the error in the data due to the ill conditionning of the matrices to be inverted. 

\subsection{Regularization methods}
The main idea in regularization theory is that a stable solution to an ill-posed inverse problem ca nnot be obtained only by minimizing a distance between the observed data and the output of the model, as it is for example, in LS methods. 
A general framework is then to define the solution of the problem as the minimizer of a compound criterion such as:
\beq
\fbh=\argmin{\fb}{J(\fb)}
\eeq
with 
\beq
J(\fb)=\Delta_1(\gb,\Hb\fb)+\lambda \Delta_2(\fb,\fb_0)
\eeq
where $\Delta_1$ and $\Delta_2$ are two distances, the first defined in the observed quantity space and the second in the unknown quantity space. $\lambda$ is the regularization parameter which regulates the compromize with the two terms and $\fb_0$ is an \aprio solution. 
Un example of such criterion is 
\beq
J(\fb)=\|\gb-\Hb\fb\|^2+\lambda \|\fb-\fb_0\|^2
\eeq
which results to
\beq
\fbh=\fb_0+[\Hb'\Hb+\lambda\Ib]^{-1}\Hb' (\gb-\Hb\fb_0)
\eeq
We may note that the condition number of the matrix to be inverted here can be controlled by appropriately choosing the value of the regularization parameter $\lambda$. 

Even if the methods based on regularization approach have been used with success in many applications, three main open problems still remains: i) Determination of the regularization parameter, ii) The arguments for choosing the two distances $\Delta_1$ and $\Delta_2$ and iii) Quantification of the uncertainties associated to the obtained solutions. Even if there have been a lot of works trying to answer to these problems and there are effective solutions such as the L-curve or the Croos Validation for the first, the two others are still open problems. The Bayesian estimation framework, as we will see, can give answers to them \cite{djafari02a}. 

\section{Bayesian estimation framework}
To illustrate the basics of the Bayesian estimation framework, let first consider the simple case of SISO system $\gb=\Hb\fb+\epsilonb$ where we assume that $\Hb$ is known. 
In a general Bayesian estimation framework, the forward model is used to define the likelihood function $p(\gb|\fb,\thetab_1)$ and we have to translate our prior knowledge about the unknowns $\fb$ through a prior probability law $p(\fb|\thetab_2)$ and then use the Bayes rule to find an expression for $p(\fb|\gb,\thetab)$ 
\beq \label{Bayes1}
p(\fb|\gb,\thetab) = \frac{p(\gb|\fb,\thetab_1) \; p(\fb|\thetab_2)}{p(\gb|\thetab}
\eeq
where $p(\gb|\fb,\thetab_1)$ is the likelihood whose expression is obtained from the forward model and assumption on the errors $\epsilonb$, 
 $\thetab=(\thetab_1,\thetab_2)$ represents all the hyperparameters (parameters of the likelihood and priors) of the problem and 
\beq 
p(\gb|\thetab) = \intg {p(\gb|\fb,\thetab_1) \; p(\fb|\thetab_2)} \d{\fb} 
\eeq
is called the evidence of the model. 

When the expression of $p(\fb|\gb,\thetab)$ is obtained, we can use it to define any estimates for $\fb$. Two usual estimators are the maximum \emph{a posteriori} (MAP) 
\beq 
\fbh=\arg\max_{\fb}\{p(\fb|\gb,\thetab)\}
\eeq
and the Mean Square Error (MSE) estimator which corresponds to the posterior mean 
\beq 
\fbh=\int \fb\; p(\fb|\gb,\thetab) \d{\fb}.
\eeq
Unfortunately only for the linear problems and the Gaussian laws where 
$p(\fb|\gb,\thetab)$ is also Gaussian we have analytical solutions for 
these two estimators. 
For almost all other cases, the first one needs an optimization algorithm 
and the second an integration one. 
For example, the relaxation methods can be used for the optimization and 
the MCMC algorithms can be used for expectation computations. 
Another difficult point is that the expressions of  $p(\gb|\fb,\thetab_1)$ and $p(\fb|\thetab_2)$ and thus the expression of $p(\fb|\gb,\thetab)$ depend on the hyperparameters $\thetab$ which, in practical applications, have also to be estimated 
either in a supervised way using the training data or in an unsupervised way. 
In both cases, we need also to translate our prior knowledge on them through a prior 
probability $p(\thetab)$. 
Thus, one of the main steps in any inversion method for any inverse problem is 
modeling the unknowns. In probabilistic methods and in particular in the Bayesian approach, 
this step  becomes the assignment of the probability law $p(\fb|\thetab_1)$. 
This point, as well as the assignment of $p(\thetab)$, are discussed the next two subsections. 

\subsection{Simple case of Gaussian models}

Let consider as a first example the simple case where $\epsilonb$ and $\fb$ are assumed to be Gauusian:
\beq
\barr{lcl}
p(\epsilonb|\sigma_{\epsilon}^2)
&=&\Nc(\zerob,\Rb_{\epsilon}=\sigma_{\epsilon}^2\Ib_0) \\ 
&\propto& \expf{-\frac{1}{2\sigma_{\epsilon}^2} \epsilonb^t\epsilonb}
\\ 
p(\fb|\sigma_f^2,\Pb_0)
&=&\Nc(\fb_0,\Rb_{f}=\sigma_f^2\Pb_0) \\ 
&\propto& \expf{-\frac{1}{2\sigma_f^2} (\fb-\fb_0)^t\Pb_0^{-1}(\fb-\fb_0)}
\earr
\eeq
Then, it is esay to to show that:
\beq
\barr{lcl}
p(\gb|\fb,\sigma_{\epsilon}^2)
&=&\Nc(\Hb\fb,\sigma_{\epsilon}^2\Ib_0) \\ 
&\propto& \expf{-\frac{1}{2\sigma_{\epsilon}^2} (\gb-\Hb\fb)^t(\gb-\Hb\fb)}
\\ 
p(\gb|\sigma_{\epsilon}^2,\sigma_f^2,\Pb_0)
&=&\Nc(\Hb\fb_0,\Hb\Rb_{f}\Hb^t+\Rb_{\epsilon})
\\
p(\gb,\fb|\sigma_{\epsilon}^2,\sigma_f^2,\Pb_0)
&\propto& \exp\bigg[-\frac{1}{2\sigma_{\epsilon}^2} (\gb-\Hb\fb)^t(\gb-\Hb\fb) \\ 
&& ~~~~~~~~-\frac{1}{2\sigma_f^2} (\fb-\fb_0)^t\Pb_0^{-1}(\fb-\fb_0)\bigg]
\earr
\eeq
and
\beq 
\barr{lcl}
p(\fb|\gb,\sigma_{\epsilon}^2,\sigma_f^2,\Pb_0)
&=&\frac{p(\gb,\fb|\sigma_{\epsilon}^2,\sigma_f^2,\Pb_0)}{p(\gb|\sigma_{\epsilon}^2,\sigma_f^2,\Pb_0)}=\Nc(\fbh,\Pbh) 
\earr
\eeq
with 
\beq
\left\{\barr{lcl}
\fbh 
&=& \fb_0 + \Rb_{f} \Hb^t  (\Hb \Rb_{f} \Hb^t+\Rb_{\epsilon})^{-1} (\gb -\Hb\fb_0) \\ 
&=& \Pbh \Hb^t \Rb_{\epsilon}^{-1} (\gb - \Hb\fb_0), \\ 
\Pbh  
&=& \Rb_{f}-\Rb_{f} \Hb^t  (\Hb \Rb_{f} \Hb^t +\Rb_{\epsilon})^{-1} \Hb \Rb_{f} \\
&=& (\Rb_{f}^{-1}+\Hb^t \Rb_{\epsilon}^{-1} \Hb)^{-1}. 
\earr\right.
\eeq
When $\fb_0=\zerob$ and noting by $\lambda=\frac{\sigma_{\epsilon}^2}{\sigma_f^2}$, these relations write:
\beq
\left\{\barr{lcl}
\fbh &=& 
\left(\Hb^t \Hb+\lambda\Pb_{0}^{-1}\right)^{-1} \Hb^t \gb=\Pbh \Hb^t \gb \\ 
\Pbh  &=& 
\sigma_{\epsilon}^2 \left(\Hb^t \Hb+\lambda\Pb_{0}^{-1}\right)^{-1}.
\earr\right.
\eeq
It is noted that, in this case, all the point estimators such as the the MAP, the posterior mean or posterior median are the same and can be obtained by
\beq
\fbh=\argmax{\fb}{p(\fb|\gb)}=\argmin{\fb}{-\ln p(\fb|\gb)}=\argmin{\fb}{J(\fb)}
\eeq
with 
\beq
J(\fb)=\|\gb-\Hb\fb\|^2+\lambda(\fb^t \Pb_0^{-1} \fb)
\eeq
Three particular cases are of interest:
\bit
\item $\Pb_0=\Ib$. This is the case where $f_j$ are assumed centered, Gaussian and \iid:
\beq
p(\fb)\propto \expf{-\frac{1}{2\sigma_f^2} \sum_j f_j^2}
\propto \expf{-\frac{1}{2\sigma_f^2} \|\fb\|^2}
\eeq
\item $\Pb_0=\Cb\Cb^t$. This is the case where $f_j$ are assumed centered, Gaussian but correlated. the vector $\fb$ is then considered to be obtained by: 
\beq
\fb=\Cb \xib 
\eeq
with $\Cb$ corresponds to a moving average (MA) filtering and $p(\xib)=\Nc(\zerob,\Ib)$. In this case, we have:
\beq
p(\fb)\propto \expf{-\frac{1}{2\sigma_f^2} \sum_j [\Cb\fb]_j^2}
\propto \expf{-\frac{1}{2\sigma_f^2} \|\Cb\fb\|^2}
\eeq
\item $\Pb_0=(\Db^t\Db)^{-1}=(\Ib-\Ab)^{-1}$. This is the case where $f_j$ are assumed centered, Gaussian and autoregressive:
\beq
\fb=\Ab \fb + \xib 
\eeq
with $\Ab$ a matrix obtained from the AR coefficients and $p(\xib)=\Nc(\zerob,\Ib)$. In this case, we have 
\beq
p(\fb)\propto \expf{-\frac{1}{2\sigma_f^2} \|\Db\fb\|^2}
\eeq
A particular case of AR model is the first order Markov chain
\beq
p(f_j|\fb_{_j})=\Nc(f_{j-1},\sigma_f^2) 
\eeq
with corresponding $\Ab$ and $\Db=\Ib-\Ab$ matrices
\beq
\Ab=\left[\barr{cccccc}
0 & 0 & . & . & 0\\ 1 & 0 & . & . & 0 \\ 0 & 1 & 0 & . & . \\ . &&&& . \\ 0 & . & . & 1 & 0
\earr\right], 
\Db=\left[\barr{cccccc}
1 & 0 & . & . & 0\\ -1 & 1 & . & . & 0 \\ 0 & -1 & 1 & . & . \\ . &&&& .\\ 0 & . & . & -1 & 1
\earr\right]
\eeq
which give the possibility to write
\beq
p(\fb)\propto \expf{-\frac{1}{2\sigma_f^2} \|\Db\fb\|^2} 
\propto \expf{-\frac{1}{2\sigma_f^2} \sum_j (f_j-f_{j-1})^2}
\eeq
\eit
These particular cases give us the possibility to extend the prior model to other more sophisticated non-Gaussian models which can be classified in three groups:
\bit
\item Separable: 
\beq
p(\fb)\propto \expf{-\alpha \sum_j \phi(f_j)}
\eeq
where $\phi$ is any positive valued function. 
\item Simple Markovian:
\beq
p(\fb)\propto \expf{-\alpha \sum_j \phi(f_j-f_{j-1})}
\eeq
where $\phi$ is any positive valued function called potential function of the Markovian model.  
\item Compound Markovian:
\beq
p(\fb|\cb)\propto \expf{-\alpha \sum_j \phi(f_j-f_{j-1},c_j)}
\eeq
where $\phi$ is any positive valued function whose expression depends on the hidden variable $\cb$. 
\eit
Some examples of the $\phi$ expressions used in many applications are: 
\beq
\phi(t)=\left\{ t^2;~ |t|^{\beta},1\le \beta\le 2;~ - t \ln t +1, t>0;~  \min(t^2,1);~ \frac{-1}{1+t^2} \right\}
\eeq 
These equations can easily be extended for the case of multi-sensor case.

However, even if a Gaussian model for the noise is acceptable, this model 
is rarely realistic for most real word signals or images. 
Indeed, very often, 
a signal or an image can be modeled locally by a Gaussian, but its energy or amplitude can be modulated, i.e.; piecewise homogeneous 
and Gaussian \cite{demoment01,nikolova98,Idier96a,Idier01a}. 
To find an appropriate model for such cases, 
we introduce hidden variables and in particular hidden Markov modeling 
(HMM). In the following, we first give a summary description of these models and then we will consider the general case of MIMO systems with prior HMM modeling. 

\subsection{Modeling using hidden variables}
\subsubsection{Signal and images with energy modulation}
A simple model which can capture the variance modulated signal or images is 
\cite{Idier01b,Idier96a,Idier00a}. 

\beq
p(f_j|d_j,\lambda)=\Nc(0,2d_j) 
\quad\mbox{and}\quad 
p(d_j|\lambda)=\Gc(3/2,\lambda) 
\eeq
where $\Gc$ is a Gamma distribution. It is then easy to show the following relations:
\beq
\barr{lcl}
p(\fb,\db|\lambda)&\propto& \expf{-\lambda \sum_j \left(\frac{f_j^2}{4 d_j^2}+d_j\right)}
\\
p(\gb|\fb)&\propto&\expf{\frac{-1}{2\sigmae^2} \|\gb-\Hb\fb\|^2}
\earr
\eeq
and
\beq
\barr{lcl}
p(\fb,\db|\gb)&\propto&\expf{-J(\fb,\db)}
\\
\mbox{with~~}
J(\fb,\db)&=&\frac{1}{2\sigmae^2} \|\gb-\Hb\fb\|^2+\lambda \sum_j \left(\frac{f_j^2}{4 d_j^2}+d_j\right)
\earr
\eeq
If we try to find the joint MAP estimate of the unknowns $(\fb,\db)$ by  optimisation successively with respect to $\fb$ when $\db$ is fixed and with respect to $\db$ when $\fb$ is fixed, we obtain the following iterative algorithm:
\beq
\barr{lcl}
\fbh&=&(\sigmae^{-2}\Hb^t\Hb+2\lambda\Db)^{-1}\Hb^t\gb \\ 
&& \quad\mbox{with}\quad 
\Db=\diag{1/(4d_j^2), j=1,\cdots,n}
\\
\widehat{d}_j&=&f_j/2
\earr
\eeq

\subsubsection{Amplitude modulated signals}
To illustrate this with applications in telecommunication signal and image processing, we consider the case of a Gaussian signal modulated with a two level or binary signal.  
A simple model which can capture the variance modulated signal or images is

\beq
\barr{lcl}
p(f_j|z_j,\lambda)&=&\Nc(z_j,2/\lambda) 
\\
\mbox{with}&&
z_j\in\{m_1=0,m_2=1\}, \\ 
P(z_j=m_k)&=&(1/2), k=1,\cdots,K=2
\earr
\eeq
It is then easy to show the following:
\beq
\barr{lcl}
p(f_j|\lambda)
&=&(1/2) [\Nc(0,2/\lambda)+\Nc(1,2/\lambda)] \\ 
&=&\sum_{k=1}^K (1/2) \Nc(m_k,\sigma_k^2=2/\lambda)
\\
p(\fb|\zb,\lambda)&\propto& \expf{-\lambda \sum_j (f_j-z_j)^2}
\\
p(f_j|z_j,\lambda)&\propto& \expf{-\lambda (f_j-z_j)^2}
\\
p(\zb|\fb,\lambda)&\propto& \expf{-\lambda \sum_j (z_j-f_j)^2}
\\
P(z_j=k|f_j,\lambda)&\propto& \expf{-\lambda (z_j-f_j)^2}
\\
p(\gb|\fb,\sigmae^2)&\propto&\expf{\frac{-1}{2\sigmae^2} \|\gb-\Hb\fb\|^2}
\earr
\eeq
and
\beq
\barr{lcl}
p(\fb,\zb|\gb,\sigmae^2,\lambda)&\propto& \expf{-J(\fb,\zb)}
\\
\mbox{with}\quad 
J(\fb,\zb)&=&\frac{1}{2\sigmae^2} \|\gb-\Hb\fb\|^2+\lambda \|\fb-\zb\|^2 
\\ 
&& +\ln (1/2) \sum_k \sum_j \delta(z_j-m_k) 
\earr
\eeq
where $\zb=[z_1,\cdots,z_N]'$. 

Again, trying to obtain the JMAP estimate $(\fbh,\zbh)$ by optimizing successively $J(\fb,\zb)$ with respect to $\fb$ and $\zb$ we obtain:
\beq
\barr{lcl}
\fbh&=&(\sigmae^{-2}\Hb^t\Hb+\lambda\Ib)^{-1} [\Hb^t\gb+\lambda\zb]
\\
\widehat{z}_j&=&\left\{\barr{ll} 1 & f_j>a \\ 0 & f_j<a \earr\right.
\earr
\eeq
where the threshold $a$ is a function of $\lambda$. 

\subsubsection{Gaussians mixture model}
The previous model can be generalized to the general mixture of Gaussians. 
We then have the following relations:
\beq
\barr{lcl}
p(f_j|z_j=k,m_k,v_k)&=&\Nc(m_k,v_k=2/\lambda_k) 
\\ 
P(z_j=k)&=&\pi_k \quad z_j\in\{1,\cdots,K\}
\\
p(f_j|\pi_k,m_k,v_k)&=&\sum_{k=1}^K \pi_k \Nc(m_k,v_k) \\ 
&=& \sum_{k=1}^K P(z_j=k) \Nc(m_k,v_k)
\\
p(\fb|\zb,\mb,\lambdab)
&\propto& \expf{-\sum_k\sum_{\{j: z_j=k\}}\lambda_k (f_j-m_k)^2}
\\
&\propto& \expf{-\sum_j\sum_k \lambda_k \delta(z_j-k) (f_j-m_k)^2}
\\
p(\gb|\fb,\sigmae^2,\mb)&\propto&\expf{\frac{-1}{2\sigmae^2} \|\gb-\Hb\fb\|^2}
\earr 
\eeq
and thus:
\beq
\barr{lcl}
p(\zb|\fb,\mb,\lambdab)
&\propto& p(\fb|\zb,\mb,\lambdab) \prod_k \pi_k^{\sum_j \delta(z_j-k)} 
\\
&\propto& \expf{-\sum_j\sum_k[\lambda_k \delta(z_j-k) (f_j-m_k)^2+\ln \pi_k]}
\\
P(z_j=k|\fb,\mb,\lambdab)&\propto& \expf{-\lambda_k (f_j-m_k)^2+\ln \pi_k}
\earr 
\eeq
and
\[
p(\fb,\zb|\gb,\sigmae^2,\mb,\lambdab) \propto \expf{-J(\fb,\zb)}
\]
with
\beq
\barr{lcl}
J(\fb,\zb)
&=&\frac{1}{2\sigmae^2} \|\gb-\Hb\fb\|^2+\sum_k\sum_{\{j: z_j=k\}} \lambda_k (f_j-m_k)^2 \\ 
&& +\sum_k \ln(\pi_k) \sum_j \delta(z_j-m_k) 
\\
&=&\frac{1}{2\sigmae^2} \|\gb-\Hb\fb\|^2+\sum_k\lambda_k \|\fb_k-m_k\oneb\|^2 \\ 
&& +\sum_k n_k \ln(\pi_k) 
\earr 
\eeq
where $\mb=\{m_1,\cdots,m_K\}$, $\lambdab=\{\lambda_1,\cdots,\lambda_K\}$,  
$n_k=\sum_j \delta(z_j-k)$ is the number of samples $f_j$ which are in the class $z_j=k$ and  $\fb_k=\{f_j : z_j=k\}$. For more details and apllications of such modeling see \cite{Snoussi00a,snoussi04c,snoussi04b,ichir06b}.

\subsubsection{Mixture of Gauss-Markov model}
In the previous model, we assumed that the samples in each class are independent. 
Here, we extend this to a markovian model:
\beq
\barr{lcl}
p(f_j|z_j=k,z_{j-1}\not=k,f_{j-1},m_k,v_k)&=&\Nc(m_k,v_k)
\\ 
p(f_j|z_j=k,z_{j-1}=k,f_{j-1},m_k,v_k)&=&\Nc(f_{j-1},v_k) 
\\ 
P(z_j=k)&=&\pi_k \quad z_j\in\{1,\cdots,K\}
\earr 
\eeq
which can be written in a more compact way if we introduce 
$q_j=1-\delta(z_j-z_{j-1})$ by 
\beq
p(f_j|q_j,f_{j-1},m_k,v_k)=\Nc(q_j m_k +(1-q_j) f_{j-1},v_k)
\eeq
which results to:
\beq
\barr{l@{}c@{}l}
p(\fb|\zb,\mb,\lambdab)
&\propto& \expf{-\sum_j\sum_k\lambda_k \delta(z_j-k) [f_j-(q_j m_k+(1-q_j) f_{j-1})]^2}
\\
&\propto& \expf{-\sum_j\sum_k\lambda_k \delta(z_j-k) [(1-q_j)(f_j-f_{j-1})^2+q_j(f_j-m_k)^2]}
\earr 
\eeq
and when combined with 
\[
p(\gb|\fb,\sigmae^2)\propto\expf{\frac{-1}{2\sigmae^2} \|\gb-\Hb\fb\|^2}
\] 
gives:
\[
p(\fb,\zb|\gb,\sigmae^2,\mb,\lambdab) \propto \expf{-J(\fb,\zb)}
\]
with
\beq
\barr{l@{}cl}
J(\fb,\zb)
&=&\frac{1}{2\sigmae^2} \|\gb-\Hb\fb\|^2
\\ 
&& +\sum_j\sum_k \lambda_k \delta(z_j-k) [f_j-(q_j m_k+(1-q_j) f_{j-1})]^2 
\\
&& +\sum_k n_k \ln(\pi_k)
\\ 
&=&\frac{1}{2\sigmae^2} \|\gb-\Hb\fb\|^2
+\sum_j(1-q_j)(\tilde{f}_j-\tilde{f}_{j-1})^2 
+\sum_k n_k \ln(\pi_k)
\\ 
&=&\frac{1}{2\sigmae^2} \|\gb-\Hb\fb\|^2
+\|\Qb\Db\fbt\|^2 
+\sum_k n_k \ln(\pi_k)
\earr 
\eeq
where $\tilde{f}_j=\lambda_{z_j}(f_j-m_{z_j})$, $\Db$ is the first order finite difference matrix and $\Qb$ is a matrix with $q_j$ as its diagonal elements. 

A particular case of this model is of great interest: $m_k=0, \forall k$ and 
$\lambda_k=\lambda, \forall k$. Then, we have: 

\beq
\barr{lcl}
p(f_j|q_j,f_{j-1},m_k,v_k)&=&\Nc((1-q_j) f_{j-1},v_k)
\\
p(\fb|\qb,\mb,\lambdab)&\propto& \expf{-\sum_j\lambda [f_j-(1-q_j) f_{j-1})]^2}
\\ 
&\propto&\expf{-\lambda \sum_j [(1-q_j) (f_j-f_{j-1})^2+q_j f_j^2]}
\earr 
\eeq
and
\[ 
p(\fb,\qb|\gb,\sigmae^2,\mb,\lambdab) \propto \expf{-J(\fb,\qb)}
\]
\mbox{with}\quad 
\beq
\barr{lcl}
J(\fb,\qb)
&=&\frac{1}{2\sigmae^2} \|\gb-\Hb\fb\|^2+\lambda \sum_j[(1-q_j) (f_j-f_{j-1})^2+q_j f_j^2]
\\
&& +\sum_k n_k \ln(\alpha_k)
\\
&=&\frac{1}{2\sigmae^2} \|\gb-\Hb\fb\|^2+\lambda \|\Qb\Db\fb\|^2+\sum_k n_k \ln(\alpha_k)
\earr 
\eeq
where $n_k=\sum_j q_j$ is the number of discontinuities (length of the contours in the case of an image) $\alpha_k=P(q_j=1)$ and $1-\alpha_k=P(q_j=0)$. 


In all these mixture models, we assumed $z_j$ independent with $P(z_j=k)=\pi_k$. However, $z_j$ corresponds to the label of the sample $f_j$. It is then better to put a markovian structure on it to capture the fact that, in general, when the neighboring samples of $f_j$ have all the same label, then it must be more probable that this sample has the same label. This feature can be modeled via the Potts-Markov modeling of the classification labels $z_j$. In the next section, we use this model, and at the same time, we extend all the previous models to 2D case for applications in image processing and to MIMO applications. 

\subsection{Mixture and Hidden Markov Models for images}
In image processing applications, the notions of contours and regions are very important. In the following, we note by $\rb=(x,y)$ the position of a pixel and by $f(\rb)$ its gray level or by $\fb(\rb)=\{f_1(\rb),\cdots,f_N(\rb)\}$ its color or spectral components. In classical RGB color representation $N=3$, but in hyperspectral imaging $N$ may be more than one hundred. When the observed data are also images we note them by $\gb(\rb)=\{g_i(\rb),\cdots,g_M(\rb)\}$. For any image $f_j(\rb)$ we note by $q_j(\rb)$, a binary valued hidden variable, its contours and by $z_j(\rb)$, a discrete value hidden variable representing its region labels. We focus here on images with homogeneous regions and use the mixture models of the previous section with an additional Markov model for the hidden variable $z_j(\rb)$.  

\subsubsection{Homogeneous regions modeling}
In general, any image $f_j(\rb), \rb\in\Rc$ is composed of a finite set $K_j$ of 
homogeneous regions $\Rjk$ with given labels $z_j(\rb)=k, k=1,\cdots,K_j$ such that 
$\Rjk=\{\rb ~:~ z_j(\rb)=k\}$, $\Rc_j=\cup_k \Rjk$ and the corresponding pixel values 
$\fbjk=\{f_j(\rb) ~:~ \rb\in\Rjk\}$ and $\fb_j=\cup_k \fbjk$. 
The Hidden Markov modeling (HMM) is a very general and efficient way to model appropriately 
such images. The main idea is to assume that all the pixel values 
${\fb_j}_k=\{f_j(\rb), \rb\in\Rjk\}$ of a homogeneous region $k$ follow a 
given probability law, for example a Gaussian $\Nc(\mjk\oneb,{\Sigmab_j}_k)$ where 
$\oneb$ is a generic vector of ones of the size ${n_j}_k$ 
the number of pixels in region $k$. 

In the following, we consider two cases:
\bit
\item The pixels in a given region are assumed iid:
\beq \label{hmm1}
p(f_j(\rb)|z_j(\rb)=k) = \Nc(\mjk,\sigmajk), \quad k=1,\cdots,K_j
\eeq
and thus
\beq \label{hmm1b}
 p(\fbjk|z_j(\rb)=k) = p(f_j(\rb), \rb\in\Rjk) = \Nc(\mjk\oneb,\sigmajk\Ib)
\eeq
This corresponds to the classical separable and monovariate mixture models. 

\item The pixels in a given region are assumed to be locally dependent:
\beq \label{hmm2}
 p(\fbjk|z_j(\rb)=k) = p(f_j(\rb), \rb\in\Rjk) = \Nc(\mjk\oneb,{\Sigmab_j}_k)
\eeq
where $\Sigmajk$ is an appropriate covariance matrix. 
This corresponds to the classical separable but multivariate mixture models. 
\eit
In both cases, the pixels in different regions are assumed to be independent: 
\beq \label{hmm2b}
p(\fb_j)=\prod_{k=1}^{K_j} p(\fbjk) = \prod_{k=1}^{K_j} \Nc(\mjk\oneb,{\Sigmab_j}_k).
\eeq

\bfig[htb] 
\bcc
\btabu{cc}
\includegraphics[width=30mm,height=30mm]{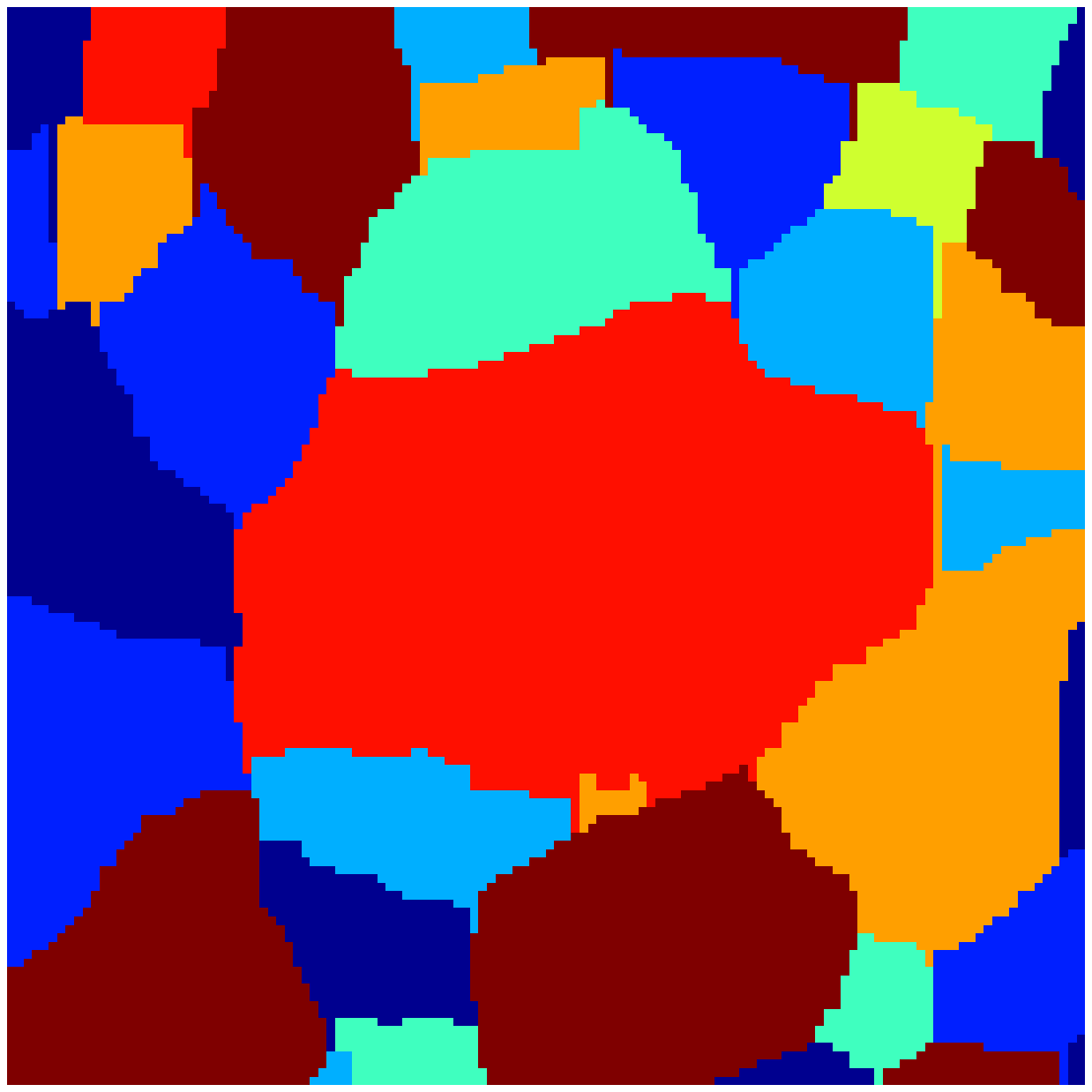}& 
\includegraphics[width=30mm,height=30mm]{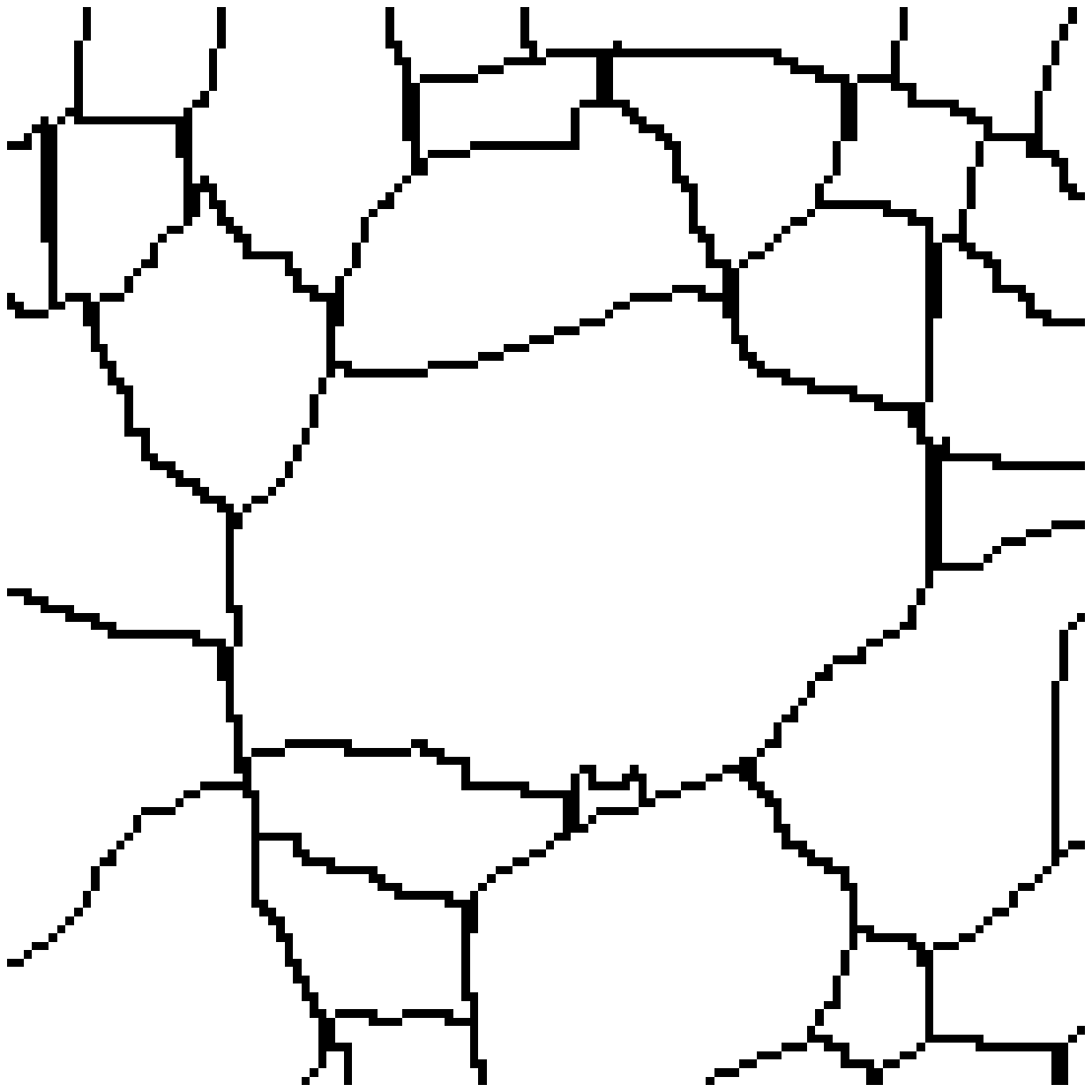}
\\ 
$z(\rb)$ & $q(\rb)$ \\
\includegraphics[width=30mm,height=30mm]{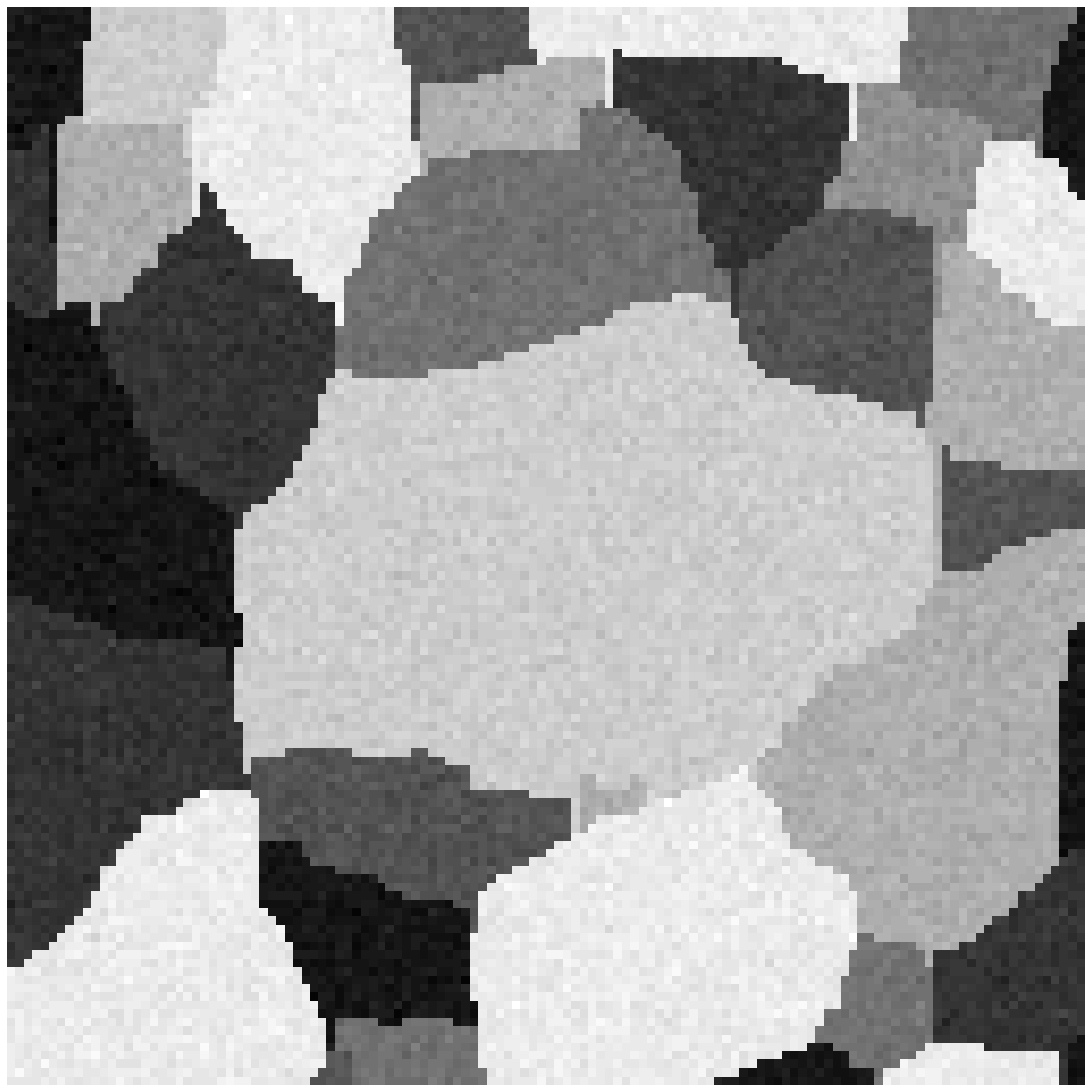}& 
\includegraphics[width=30mm,height=30mm]{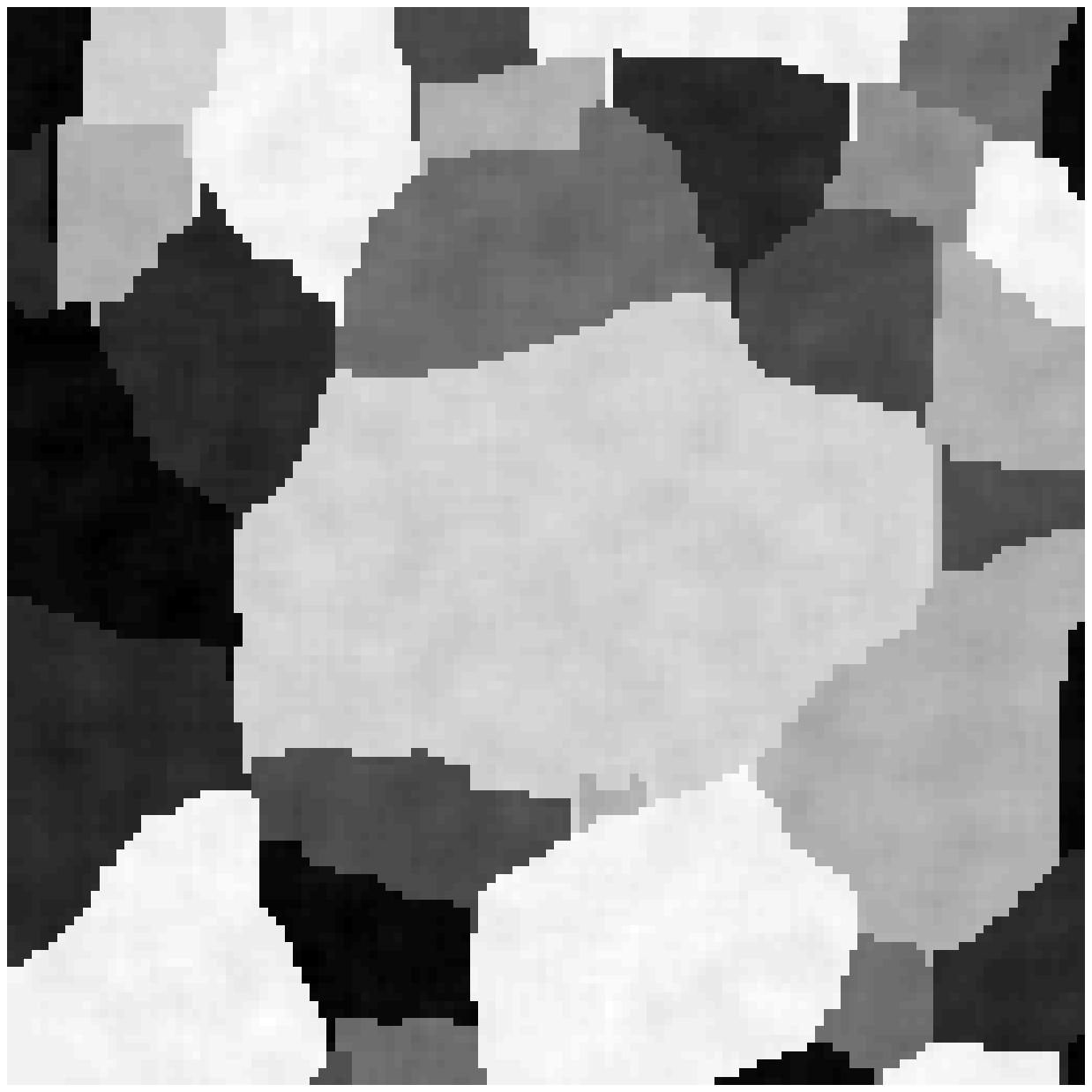}
\\ 
$f(\rb)$: Mixture of iid Gaussian & $f(\rb)$: Mixture of Gauss-Markov \\
\etabu
\ecc
\caption[Mixture and hidden Markov models for images]{Mixture and hidden Markov models for images} 
\label{Fig12}
\efig

\subsubsection{Modeling the labels}
Noting that all the models (\ref{hmm1}), (\ref{hmm1b}) and 
(\ref{hmm2}) are conditioned 
on the value of $z_j(\rb)=k$, they can be rewritten in the following general form
\beq \label{hmm3}
p(\fbjk)=\sum_k P(z_j(\rb)=k) \; \Nc(\mjk,\Sigmajk) 
\eeq
where either $\Sigmajk$ is a diagonal matrix $\Sigmajk=\sigmajk\Ib$ or not. 
Now, we need also to model the vector variables $\zb_j=\{z_j(\rb), \rb\in\Rc\}$. 
Here also, we can consider two cases:
\bit
\item Independent Gaussian Mixture model (IGM), where 
$\{z_j(\rb), \rb\in\Rc\}$ are assumed 
to be independent and 
\beq
P(z_j(\rb)=k)=p_k, \quad \mbox{with}\quad \sum_k p_k=1 \mbox{~~and~~} 
p(\zb_j)=\prod_k p_k
\eeq
\item Contextual Gaussian Mixture model (CGM), 
where $\zb_j=\{z_j(\rb), \rb\in\Rc\}$ are assumed 
to be Markovian 
\beq
p(\zb_j)\propto 
\expf{ \alpha \sum_{\rb\in\Rc} \sum_{\sb\in\Vc(\rb)} \delta (z_j(\rb)-z_j(\sb)) }
\eeq
which is the Potts Markov random field (PMRF). 
The parameter $\alpha$ controls the mean value of the regions' sizes. 
\eit

\subsubsection{Hyperparameters prior law}
The final point before obtaining an expression for the posterior probability law of all the unknowns, i.e, $p(\fbu,\thetab|\gbu)$ is to assign a prior probability law $p(\thetabu)$ to the hyperparameters $\thetabu$. Even if this point has been one of the main discussing points between Bayesian and classical statistical research community, and still there are many open problems, we choose here to use the conjugate priors for simplicity. The conjugate priors have at least two advantages:  
1) they can be considered as a particular family of a differential geometry based family of priors 
\cite{Snoussi02c,Snoussi03the,Snoussi04a}   
and  
2) they are easy to use because the prior and the posterior probability laws stay in the same family. 
In our case, we need to assign prior probability laws to the means $\mjk$, to the variances $\sigmajk$ 
or to the covariance matrices $\Sigmajk$ and also to the covariance matrices of the noises $\epsilonb_i$ 
of the likelihood functions. 
The conjugate priors for the means $\mjk$ are in general the Gaussians 
$\Nc({\mjk}_0,{\sigmajk}_0)$, those of variances $\sigmajk$ are the inverse Gammas $\Ic\Gc(\alpha_0,\beta_0)$ and those for the covariance matrices $\Sigmajk$ are the inverse Wishart's $\Ic\Wc(\alpha_0,\Lambdab_0)$. 

\subsubsection{Expressions of likelihood, prior and posterior laws}
We now have all the elements for writing the expressions of 
the posterior laws.  
We are going to summarizes them here: 
\bit
\item Likelihood: \qquad
\(
p(\gbu|\fbu,\thetabu)=\prod_{i=1}^M p(\gbu|\fbu,{\Sigmab_\epsilon}_i)
=\prod_{i=1}^M \Nc(\gbu-\fbu,{\Sigmab_\epsilon}_i)
\) \\ 
where we assumed that the noises $\epsilonb_i$ are independent, centered and Gaussian with covariance matrices ${\Sigmab_\epsilon}_i$ which, hereafter, are also assumed to be diagonal
${\Sigmab_{\epsilon}}_i={\sigma_{\epsilon}}_i^2 \Ib$. 

\item HMM for the images: \qquad
\(
p(\fbu|\zbu,\thetabu)=\prod_{j=1}^N p(\fb_j|\zb_j,{\mb}_j,{\Sigmab}_j)
\) \\ 
where we used $\zbu=\{\zb_j, j=1,\cdots,N\}$ and where we assumed that $\fb_j|\zb_j$ are 
independent. 

\item PMRF for the labels:  
\(
p(\zbu) \propto 
\prod_{j=1}^N \expf{ \alpha \sum_{\rb\in\Rc} \sum_{\sb\in\Vc(\rb)} \delta (z_j(\rb)-z_j(\sb)) }
\) \\ 
where we used the simplified notation $p(\zb_j)=P(Z_j(\rb)=z(\rb), \rb\in\Rc)$ and where we assumed $\{\zb_j, j=1,\cdots,N\}$ are independent. 

\item Conjugate priors for the hyperparameters:\\ 
\(
\barr{llllll}
p(\mjk)     =\Nc({\mjk}_0,{\sigmajk}_0),& 
p(\sigmajk) =\Ic\Gc(\alpha_{j0},\beta_{j0}),\\ 
p(\Sigmajk) =\Ic\Wc(\alpha_{j0},\Lambda_{j0}),& 
p({\sigmae}_i)=\Ic\Gc(\alpha_{i0},\beta_{i0}).
\earr
\)
\item Joint posterior law of $\fbu$, $\zbu$ and $\thetabu$
\[ 
p(\fbu,\zbu,\thetabu|\gbu) \propto p(\gbu|\fbu,\thetab_1) \; p(\fbu|\zbu,\thetab_2) \; p(\zbu|\thetab_2) \; p(\thetabu)
\]
\eit

\subsection{Bayesian estimators and computational methods}
The expression of this joint posterior law is, in general, known upto a normalisation factor. This means that, if we consider the Joint Maximum A Posteriori (JMAP) estimate
\beq 
(\fbuh,\zbuh,\thetabuh)=\argmax{(\fbu,\zbu,\thetabu)}{p(\fbu,\zbu,\thetabu|\gbu)}
\eeq
we need a global optimization algorithm, but if we consider the Minimum Mean Square Estimator (MMSE) or equivalently the Posterior Mean (PM) estimates, then we need to compute this factor which needs huge dimentional integrations. There are however three main approaches to do Bayesian computation:
\bit
\item Laplace approximation: When the posterior law is unimodale, it is reasonable to approximate it with an equivalent Gaussian which allows then to do all computations analytically. Unfortunately, very often, $p(\fbu,\zbu,\thetabu|\gbu)$ as a function of $\fbu$ only may be Gaussian, but as a function of $\zbu$ or $\thetabu$ is not. So, in general, this approximation method can not be used for all variables. 

\item Variational and mean field approximation: The main idea behind this approach is to approximate the joint posterior $p(\fbu,\zbu,\thetabu|\gbu)$ with another simpler distribution $q(\fbu,\zbu,\thetabu|\gbu)$ for which the computations can be done. A first step simpler distribution $q(\fbu,\zbu,\thetabu|\gbu)$ is a separable ones: 
\beq
q(\fbu,\zbu,\thetabu|\gbu)=q_1(\fbu) q_2(\zbu) q_3(\thetabu) 
\eeq
In this way, at least reduces the integration computations to the product of three separate ones. 
This process can again be applied to any of these three distributions, for example 
$q_1(\fbu)=\prod_j q_{1j}(\fb_j)$. With the Gaussian mixture modeling we proposed, $q_1(\fbu)$ can be choosed to be Gaussian, $q_2(\zbu)$ to be separated to two parts $q_{1B}(\zbu)$ and $q_{1W}(\zbu)$ where the pixels of the images are separated in two classes B and W as in a checker board. This is thanks the properties of the proposed Potts-Markov model with the four nearest neighborhood which gives the possibility to use 
$q_{1B}(\zbu)$ and $q_{1W}(\zbu)$ separately. For $q_3(\thetabu)$ very often we also choose a separable distribution which use the conjugate properties of the prior distributions. 

\item Markov Chain Monte Carlo (MCMC) sampling which gives the possibily to explore the joint posterior law and compute the necessary posterior mean estimates. 
In our case, we propose the general MCMC Gibbs sampling algorithm to estimate $\fbu$, $\zbu$ and $\thetabu$ by first separating the unknowns in two sets $p(\fbu,\zbu|\thetabu,\gb)$ and  $p(\thetabu|\fbu,\zbu,\gbu)$. Then, we separate again the first set in two subsets $p(\fbu|\zbu,\thetabu,\gbu)$ and $p(\zbu|\thetabu,\gbu)$.  
Finally, when possible, using the separability along the channels, separate these two 
last terms in 
$p(\fb_j|\zb_j,\thetab_j,\gb_j)$ and $p(\zb_j|\thetab_j,\gb_j)$. 
The general scheme is then, using these expressions, to generates samples $\fbu^{(n)},\zbu^{(n)},\thetabu^{(n)}$ from the joint posterior law $p(\fbu,\zbu,\thetabu|\gbu)$   
and after the convergence of the Gibbs samplers, to compute their mean and to use them as the posterior estimates. 
\eit

In this paper we are not going to detail these methods. However, in the following we propos to examine some particular cases through a few case studies in relation to image restoration, image fusion and joint segmentation, blind image separation. 

\section{Case studies}

\subsection{Single channel image denoising and restoration}
The simplest example of inversion is a single channel image denoising 
and restoration when the PSF of the imaging system is given.  
The forward model for this problem is 
\beq
\barr{r@{}cl}
g(\rb) &=& h(\rb)*f(\rb) + \epsilon(\rb), \quad \rb\in\Rc 
\mbox{~~or~~}
\gb=\Hb \fb + \epsilonb
\earr
\eeq
where the denoising case corresponds to the case where $h(\rb)=\delta(\rb)$ and $\Hb=\Ib$. 

Assuming the noise to be centered, white and Gaussian with known variance $\sigmae^2$, we have
\beq
p(\gb|\fb) = \Nc(\Hb\fb,\Sigmae) \mbox{~~with~~} \Sigmae=\sigmae^2\Ib
\eeq
The priors for this case can be summarized as follows:
\beq
p(f(\rb)|z(\rb)=k) = \Nc({m_k},{\sigma^2_k}), \quad k=1,\cdots,K
\eeq
\beq
p(\zb) = p(z(\rb), \rb\in\Rc) \propto 
\expf{ \alpha \sum_{\rb\in\Rc} \sum_{s\in\Vc(\rb)} \delta (z(\rb)-z(s)) }
\eeq
where 
\beq
\barr{r@{}cl}
\fb_k&=&\{f(\rb) ~:~ \rb\in\Rc_k\}, \quad 
\Rc_k=\{\rb ~:~ z(\rb)=k\}
\\ 
p(\fb_k|z(\rb)=k)&=&
\Nc({m_k}\oneb_k,{\Sigmab_k}) \mbox{~~with~~} 
{\Sigmab_k}={\sigma^2_k} \Ib_k
\\ 
p(\fb|\zb)&=&\prod_k \Nc({m_k}\oneb_k,{\Sigmab_k})=\Nc(\mbz,\Sigmabz) \mbox{~~with~~} \\ 
\mbz&=&[{m_1} \oneb'_1,\cdots,{m_K} \oneb'_K]' \mbox{~~and~~} 
\Sigmabz=\diag{\Sigmab_1,\cdots,\Sigmab_K}\\ 
p(m_k)     &=&\Nc({m_k}_0,{\sigma_k^2}_0),\\  
p(\sigma_k^2) &=&\Ic\Gc({\alpha_k}_0,{\beta_k}_0), \quad  
p({\sigmae}^2)=\Ic\Gc(\alphaez,\betaez)
\earr
\eeq
and the posterior probability laws we need to implement an MCMC like algorithm are: 

\beq
\barr{c@{}cl}
p(\fb|\zb,\thetab,\gb)&=&\Nc(\fbh,\Sigmabh) \\
\mbox{with} && \Sigmabh=(\Hb^t\Sigmae^{-1}\Hb+\Sigmabz^{-1})^{-1}
\\
\mbox{~~and~~} && 
\fbh=\Sigmabh \left(\Hb^t\Sigmae^{-1}\gb+\Sigmabz^{-1}\mbz\right)
\earr
\eeq
\beq
\barr{c@{}cl}
p(\zb|\gb,\thetab) &\propto& p(\gb|\zb,\thetab) \; p(\zb)
\mbox{~~where} \\
p(\gb|\zb,\thetab)&=&\Nc(\Hb\mbz,\Sigmag) \mbox{~~with~~}
\Sigmag=\Hb\Sigmabz\Hb^t+\Sigmae 
\earr
\eeq
and the posterior probabilities of the hyperparameters are:
\[
p(m_k|\zb,\fb) = \Nc(\mu_k,v_k^2) 
\mbox{~with~} 
{v_k^2}= \left(\frac{n_k}{{\sigma_k^2}}+\frac{1}{{\sigma_k^2}_0}\right)^{-1}
\mbox{and~}
\mu_k =  {v_k^2} 
\left( 
\frac{n_k \bar{f}_k}{{\sigma_k^2}}+\frac{{m_k}_0}{{\sigma_k^2}_0}   
\right) 
\]
\[
p(\sigma_k^2 |\fb,\zb) = \Ic\Gc(\alpha_k,\beta_k) 
\mbox{~with~} 
\alpha_k = {\alpha_k}_0 + \frac{n_k}{2} 
\mbox{~and~}
\beta_k  = {\beta_k}_0  + \frac{n_k \bar{s}_k}{2}
\]
\[
\mbox{where~} 
\bar{f}_k  =  \frac{1}{n_k} \sum_{\rb \in R_k} f_i(\rb) 
\mbox{~and~}  
\bar{s}_k  =  \sum_{\rb \in R_k} \left(f(\rb) - m_k\right)^2
\]
\[
p(\sigmae^2 |\fb,\gb) = \Ic\Gc(\alphae,\betae)
\mbox{~with~}
\alphae = \frac{n}{2}+\alphaez 
\mbox{~~and~~}
\betae  =  \frac{1}{2} \|\gb-\Hb\fb\|^2+\betaez
\]
\[  
n_k = \mbox{number of pixels in~} R_k 
\mbox{~~and~~}  
n=\mbox{total number of pixels}. 
\]
Here, we show two examples of simulations: the first in relation with image denoising and the second in relation with image deconvolution. In both cases, we have choosed the same input image $f(\rb)$. In the first case, we only has added a Gaussian noise and in the second case, we first blureed it with box car PSF of size $7 \times 7$ pixels and added a Gaussian noise. 
Fig.~\ref{Fig13} shows the original image, its contours and its regions. 
Fig.~\ref{Fig14} shows the observed noisy image and the results obtained by the proposed method. Remember that, in this method, we have also the estimated contours and region labels as 
byproducts. 
Fig.~\ref{Fig15} shows the observed blurred and noisy image and the results obtained by the proposed restoration method. 

\bfig[htb] 
\bcc
\btabu{ccc}
\includegraphics[width=30mm,height=30mm]{f001_2}& 
\includegraphics[width=30mm,height=30mm]{f001_q}&
\includegraphics[width=30mm,height=30mm]{f001_z}
\\ 
$f(\rb)$ & $z(\rb)$ & $q(\rb)$ 
\etabu
\ecc
\caption[Original image, its contour and its region labels]{Original image, its contour and its region labels used for image denoising and image restoration.} 
\label{Fig13}
\efig

\bfig[htb] 
\bcc
\btabu{c@{}c@{}c@{}c}
\includegraphics[width=30mm,height=30mm]{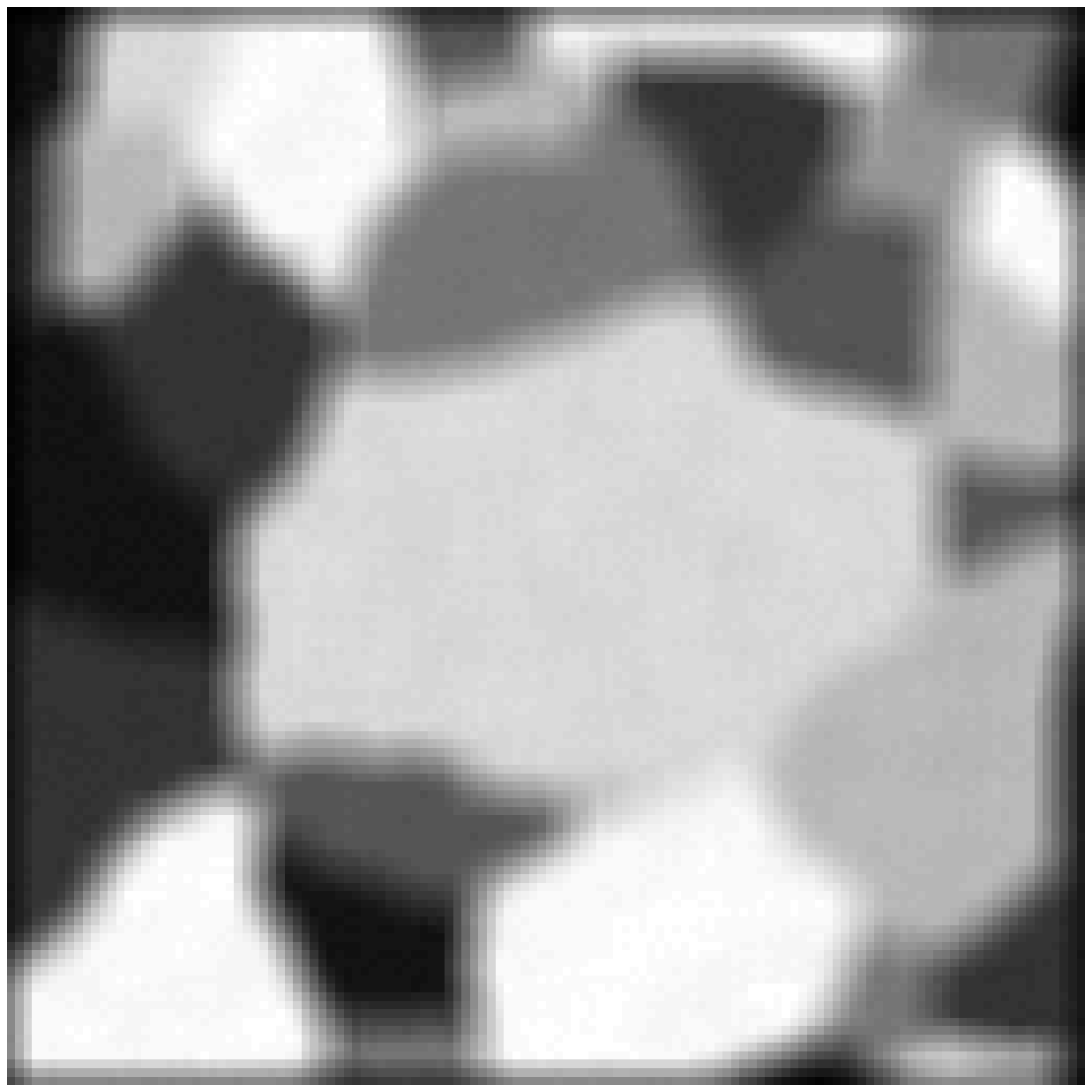}& 
\includegraphics[width=30mm,height=30mm]{f001_1}& 
\includegraphics[width=30mm,height=30mm]{f001_z}&
\includegraphics[width=30mm,height=30mm]{f001_q}
\\ 
$g(\rb)$ & $\widehat{f}(\rb)$ & $\widehat{z}(\rb)$ & $\widehat{q}(\rb)$ 
\etabu
\ecc
\caption[Observed noisy image and the results of denoising]{Observed noisy image and the results of the proposed denoising method.} 
\label{Fig14}
\efig

\bfig[htb] 
\bcc
\btabu{c@{}c@{}c@{}c}
\includegraphics[width=30mm,height=30mm]{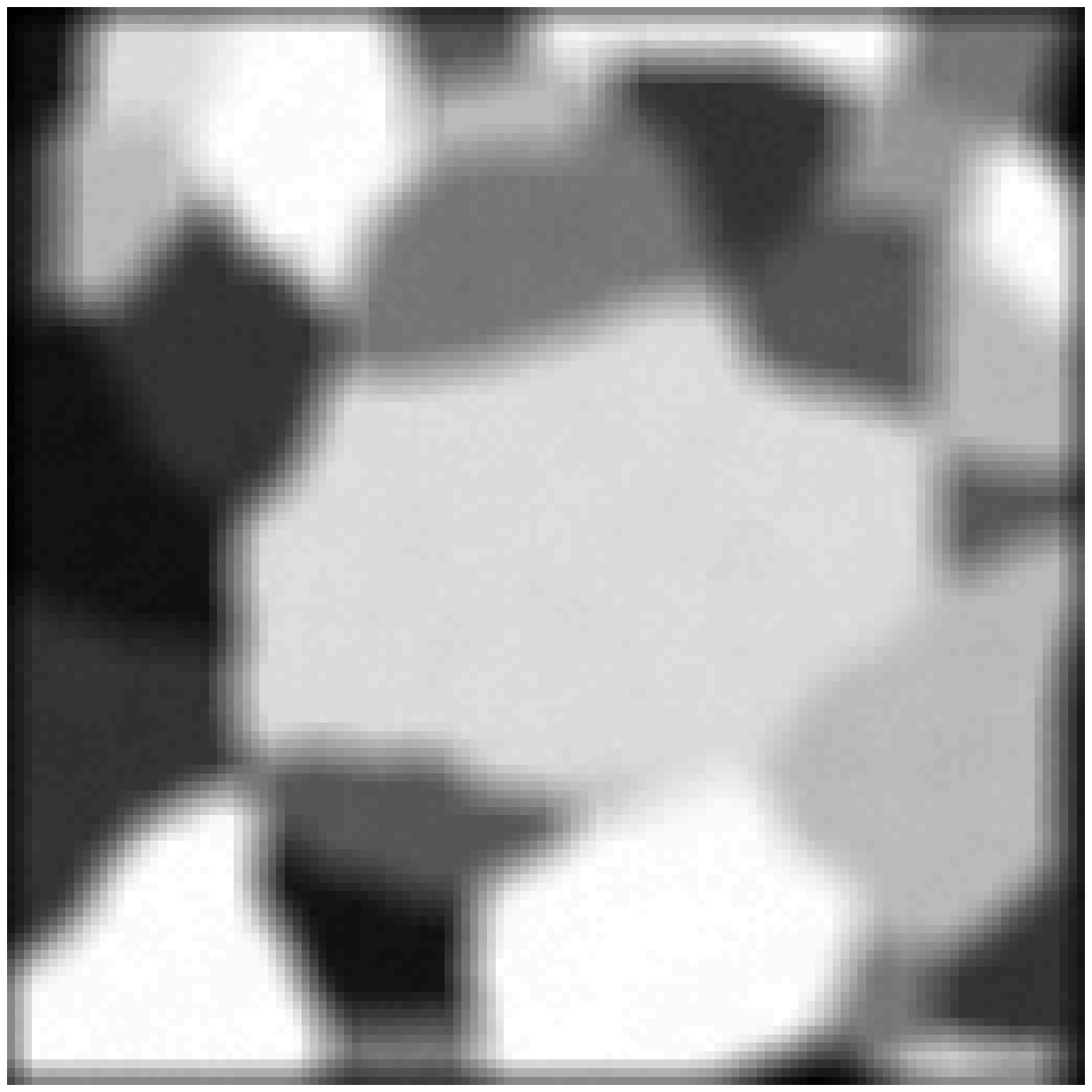}& 
\includegraphics[width=30mm,height=30mm]{f001_1}& 
\includegraphics[width=30mm,height=30mm]{f001_z}&
\includegraphics[width=30mm,height=30mm]{f001_q}
\\ 
$g(\rb)$ & $\widehat{f}(\rb)$ & $\widehat{z}(\rb)$ & $\widehat{q}(\rb)$ 
\etabu
\ecc
\caption[Observed blurred and noisy image and the results of restoration]{Observed noisy image and the results of the proposed restoration method.} 
\label{Fig15}
\efig

For other inverse problems which can be modeled as a SISO model and where such Bayesian approach has been used refer to \cite{djafari03a}

\clearpage\newpage
\subsection{Registered images fusion and joint segmentation}

Here, each observed image $g_i(\rb)$ (or equivalently $\gb_i$) is assumed to be a noisy version of the unobserved real image $f_i(\rb)$ (or equivalently $\fb_i$) 
\beq
g_i(\rb) = f_i(\rb) + \epsilon_i(\rb), \quad \rb\in\Rc, 
\mbox{~~or~~}
\gb_i=\fb_i + \epsilonb_i, 
\quad i=1,\cdots,M
\eeq
which gives
\beq
p(\gb_i|\fb_i) = \Nc(\fb_i,\Sigmae_i) \mbox{~~with~~} \Sigmae_i=\sigmae_i^2\Ib
\eeq
and
\beq
p(\gbu|\fbu) = \prod_i p(\gb_i|\fb_i) \mbox{~~with~~}
\eeq
and all the unobserved real images $f_i(\rb),~~i=1,\cdots,M$ are assumed to have a common segmentation $z(\rb)$ (or equivalently $\zb$) which is modeled by a discrete value Potts Random Markov Field (PRMF). Then, using the same notations as in previous case, we have the following relations:   

\beqnx
\barr{l@{}cl}
p(f_i(\rb)|z(\rb)=k) &=& \Nc(\mik,\sigmaik), \quad k=1,\cdots,K
\\
\fbik&=&\{f_i(\rb) ~:~ \rb\in\Rc_k\}, \quad \Rc_k=\{\rb ~:~ z(\rb)=k\}
\\ 
p(\fbik|z(\rb)=k) &=& \Nc(\mik\oneb_k,\Sigmaik) \mbox{~~with~~} 
\Sigmaik=\sigmaik\Ib_k
\\
p(\zb)&=&p(z(\rb), \rb\in\Rc) \propto
\expf{ \alpha \sum_{\rb\in\Rc} \sum_{s\in\Vc(\rb)} \delta (z(\rb)-z(s)) }
\\ 
p(\fb_i|\zb)&=&\Nc(\mbz_i,\Sigmabz_i) \mbox{~~with~~} \\ 
\mbz_i&=&[{m_i}_1 \oneb'_1,\cdots,{m_i}_K \oneb'_K]' \mbox{~~and~~} 
\Sigmabz_i=\diag{{\Sigmab_i}_1,\cdots,{\Sigmab_i}_K}
\\ 
p(\mik)     &=&\Nc({\mik}_0,{\sigmaik}_0)\\   
p(\sigmaik) &=&\Ic\Gc(\alpha_{i0},\beta_{i0}), \quad  
p({\sigmae}_i^2)=\Ic\Gc(\alphaeiz,\betaeiz)
\\ 
p(\fbu|\zb)&=&\prod_i p(\fb_i|\zb)  
\earr 
\eeqnx
and all the conditional and posterior probability laws wee need to implement the proposed Bayesian methods are summarized here:
\beqnx
\barr{l@{}cl}
p(\fb_i|\zb,\thetab_i,\gb_i)&=&\Nc(\fbh_i,\Sigmabh_i) \\ 
\mbox{with}  && \Sigmabh_i=(\Sigmae_i^{-1}+\Sigmabz^{-1})^{-1}
\mbox{~~and~~}
\fbh_i=\Sigmabh_i \left(\Sigmae_i^{-1}\gb_i+\Sigmabz_i^{-1}\mbz_i\right)
\\ 
p(\zb|\gbu,\thetab) &\propto& \left(\prod_i p(\gb_i|\zb,\thetab_i)\right) \; p(z(\rb), \rb\in\Rc)
\mbox{~~with~~} \\
p(\gb_i|\zb,\thetab_i)&=&\Nc(\mbz_i,\Sigmag_i) \mbox{~~with~~}
\Sigmag_i=\Sigmabz_i+\Sigmae_i\\ 
p(\mik|\fb_i,\zb,{\sigma_i^2}_k)&=&\Nc(\muik,\vik^2) \\ 
\mbox{with} &&  
\muik =  \vik^2 \left( 
\frac{{m_i}_0}{{\sigma^2_i}_0} + \frac{n_k \bar{f_i}_k}{{\sigma^2_i}_k}
\right) 
\mbox{~~and~~}
\vik^2 = \left( \frac{1}{{\sigma^2_i}_0} + \frac{n_k}{{\sigma_i^2}_k}  \right)^{-1}
\\
p({\sigma_i^2}_k |\fb_i,\zb)&=&\Ic\Gc(\alphaik,\betaik) 
\mbox{~with~} 
\alphaik = {\alpha_i}_0 + \frac{n_k}{2} \mbox{~~and~~}
\betaik  = {\beta_i}_0 + \frac{\bar{s}_i}{2}
\\ 
\mbox{where}&& 
\bar{f_i}_k  =  \frac{1}{n_k} \sum_{\rb \in R_k} f_i(\rb) 
\mbox{~and~}  
\bar{s}_i  =  \sum_{\rb \in R_k} \left(f_i(\rb) - \mik\right)^2
\\
p(\sigmae_i^2 |\fb_i,\gb_i)&=&\Ic\Gc(\alphaei,\betaei)
\mbox{~with~}
\alphaei = \frac{n}{2}+\alphaeiz 
\mbox{~~and~~}
\betaei  =  \frac{1}{2} \|\gb_i-\fb_i\|^2+\betaeiz
\\  
n_k&=&\mbox{number of pixels in~} R_k, \quad 
n=\mbox{total number of pixels}. 
\earr
\eeqnx

For more details on this model and its application in medical image fusion as well as in image fusion for security systems see 
\cite{Feron05a,Feron05b}.

\bfig[htb] 
\btabu{lll}
\btabu{l}
$\gb_1$ \includegraphics[width=30mm,height=15mm]{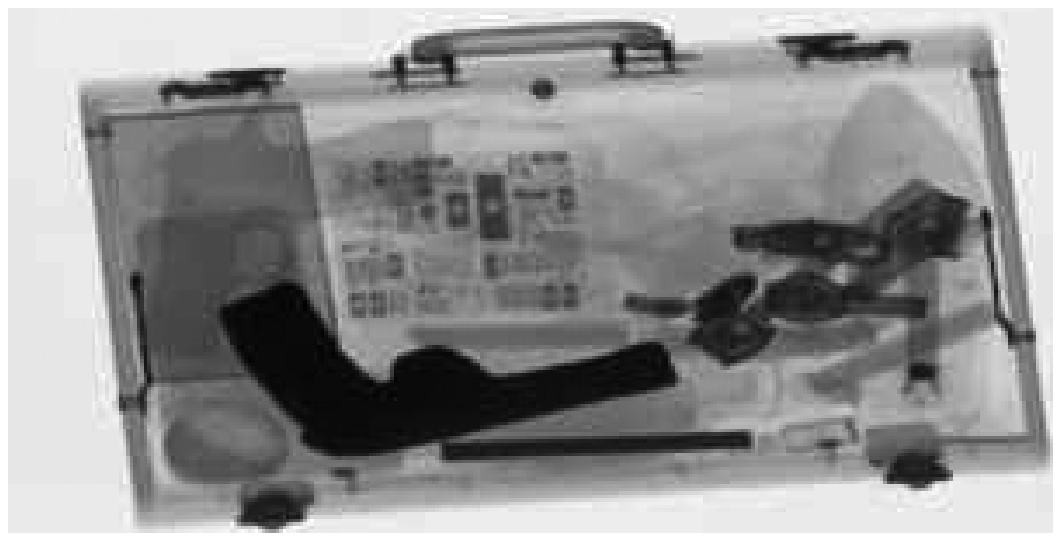}\\ 
$\gb_2$ \includegraphics[width=30mm,height=15mm]{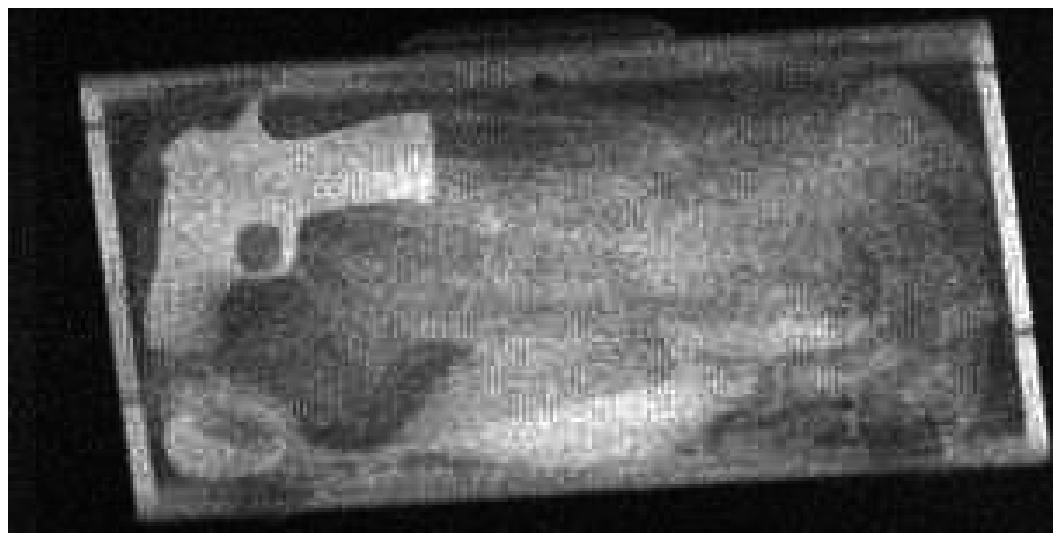}
\etabu
$\longrightarrow$
\btabu{l}
$\fbh_1$ \includegraphics[width=30mm,height=15mm]{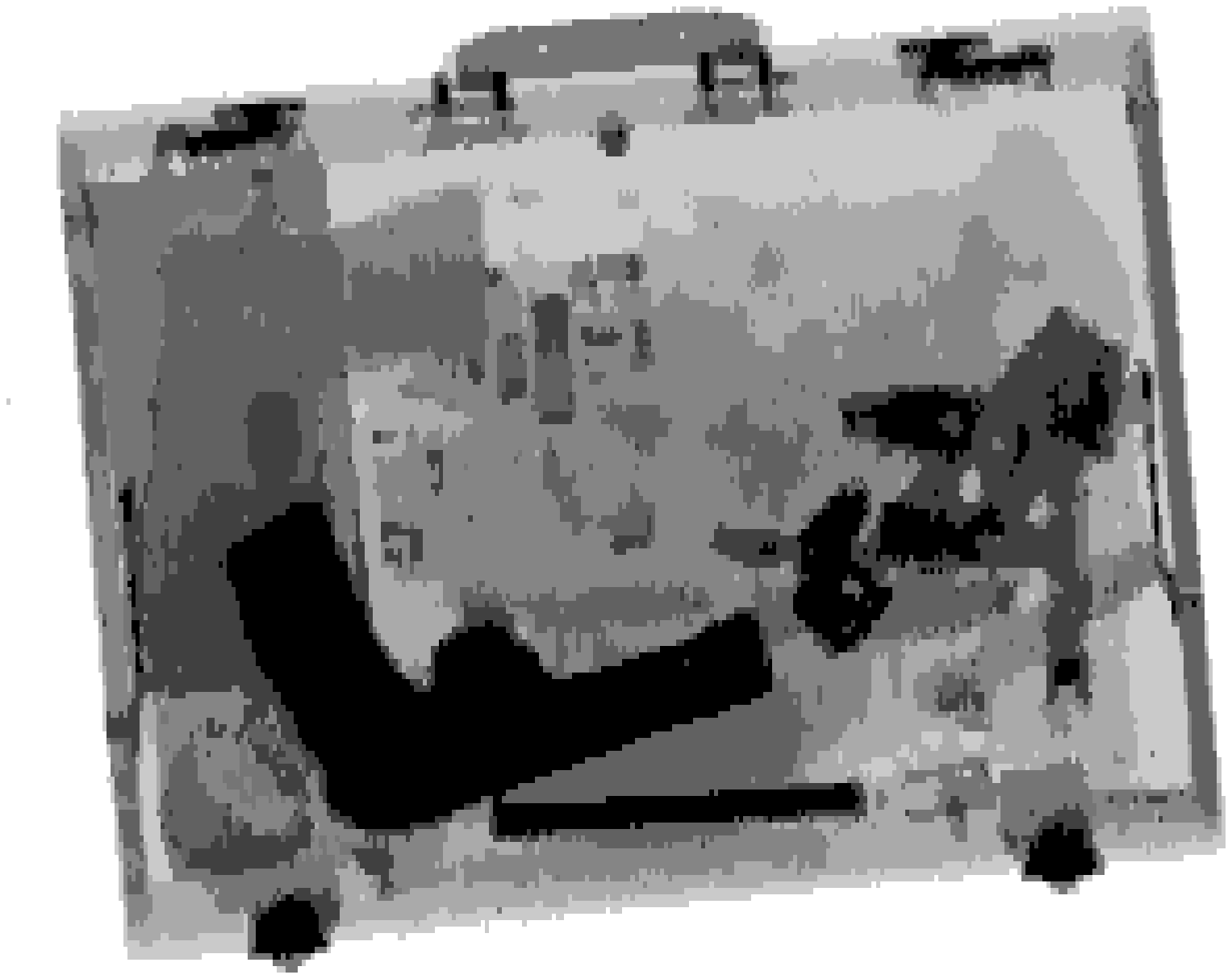}\\ 
$\fbh_2$ \includegraphics[width=30mm,height=15mm]{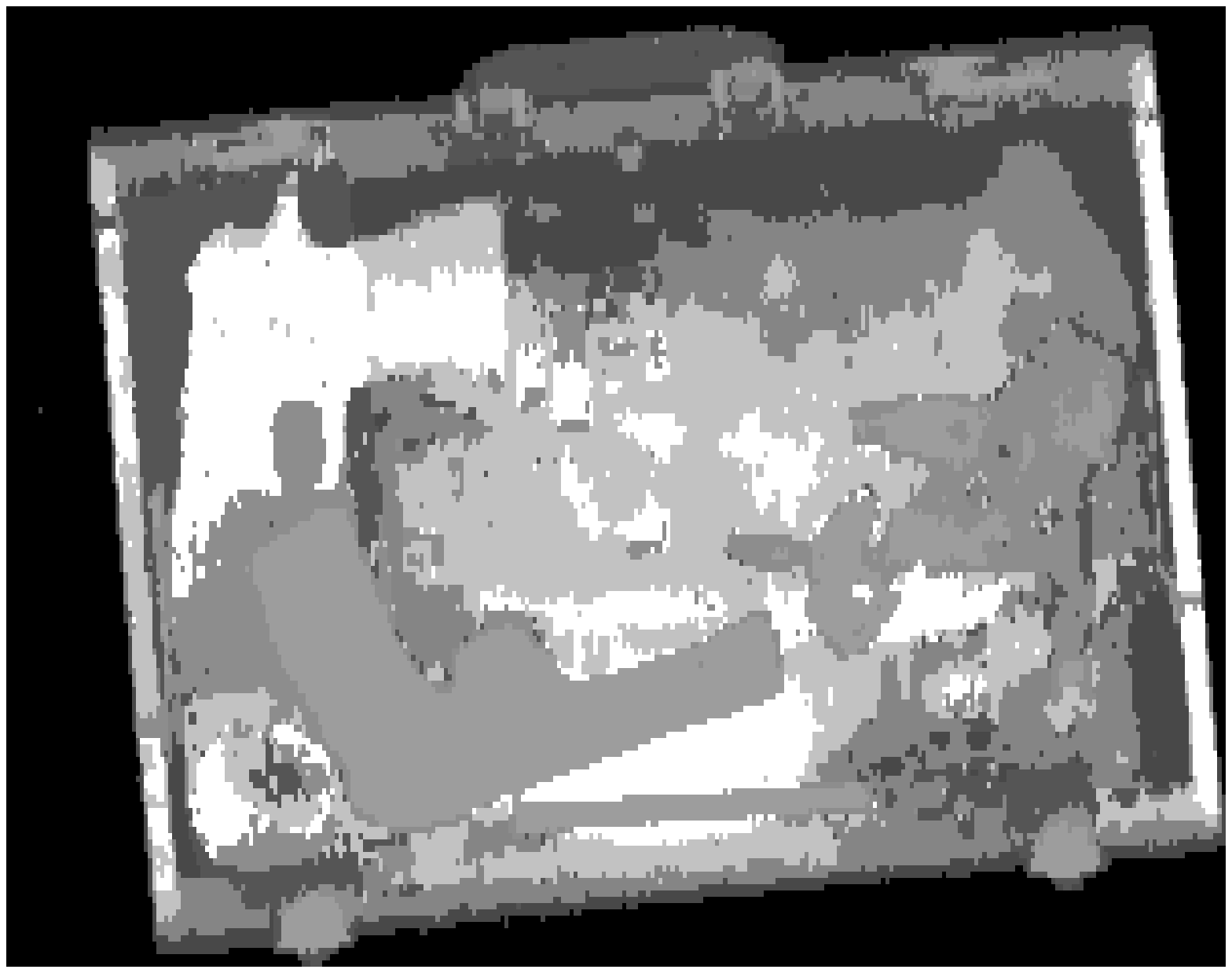}
\etabu
\btabu{l}
$\widehat{\zb}$ \includegraphics[width=30mm,height=15mm]{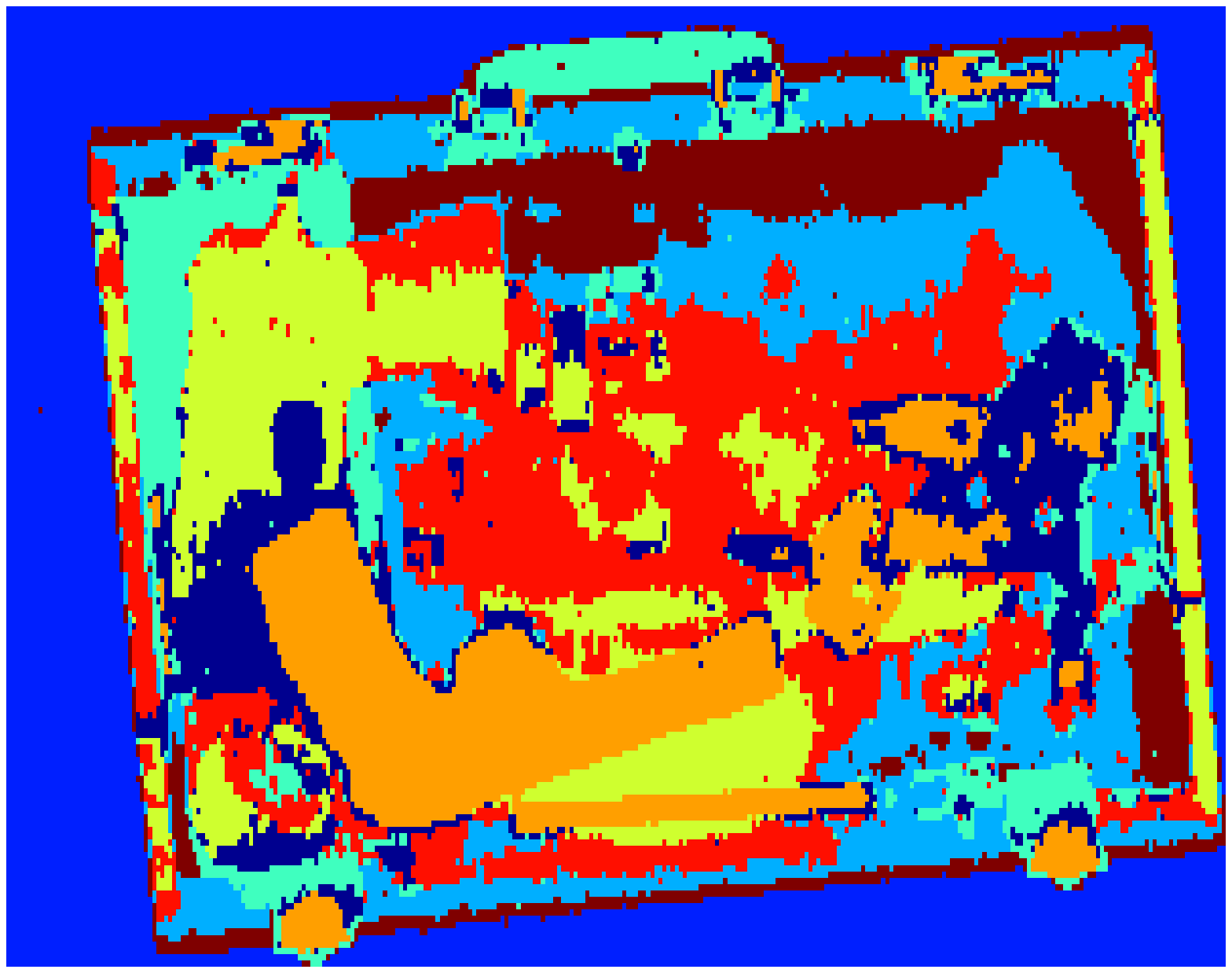}
\etabu
\etabu
\caption{Image fusion and joint segmentation of two images 
from a security system measurement. } 
\label{Fig16}
\efig

\subsection{Joint segmentation of hyper-spectral images}
The proposed model is the same as the model of the previous section except for the last equation 
of the forward model which assumes that the pixels in similar regions of different images are 
independent. For hyper-spectral images, this hypothesis is not valid and we have to 
account for their correlations. This work is under consideration.  

\subsection{Segmentation of a video sequence of images}
Here, we can not assume that all the images in the video sequence have the same segmentation 
labels. However, we may use the segmentation obtained in an image as an initialization for the 
segmentation of next image. 
For more details on this model and to see a typical result see 
\cite{Brault04}.

\subsection{Joint segmentation and separation of instantaneous mixed images}
Here, the additional difficulty is that we also have to estimate the mixing matrix $\Ab$.
For more details on this model and to see some typical result in joint segmentation and 
separation of images see
\cite{Snoussi04a,mohammadpour04ce,mohammadpour05ca,bali06cc,bali06ca,bali06cb}.

\section{Conclusion}
In this paper we first showed that many image processing problems can be presented as 
inverse problems by modeling the relation of the observed image to the unknown desired 
features explicitly. 
Then, we presented a very general forward modeling for the observations and 
a very general probabilistic modeling of images through a hidden Markov modeling (HMM) which 
can be used as the main basis for many image processing problems such as: 
1) simple or multi channel image restoration, 
2) simple or joint image segmentation, 
3) multi-sensor data and image fusion, 
4) joint segmentation of color or hyper-spectral images and 
5) joint blind source separation (BSS) and segmentation. 
Finally, we presented detailed forward models, prior and posterior probability law expressions for the implementation of MCMC algorithms for a few cases of those problems 
showing typical results which can be obtained using these methods.  

\def\sca#1{{\sc #1}}
\def\bibdir{/home/djafari/Tex/Inputs/bib/amd/}
\bibliographystyle{IEEEcs}
\bibliography{bibenabr,revuedef,revueabr,baseAJ,baseKZ,gpipubli,\bibdir amd_art,\bibdir amd_ca,\bibdir amd_ci}

\begin{thebibliography}{10}

\bibitem{Hadamard01}
J.~Hadamard,
\newblock ``Sur les problèmes aux dérivées partielles et leur signification
  physique,''
\newblock {\em Princeton Univ. Bull.}, vol. 13, 1901.

\bibitem{Demoment85}
G.~Demoment,
\newblock ``D\'econvolution des signaux,''
\newblock Cours de l'\'Ecole sup\'erieure d'\'electrit\'e 3086, 1985.

\bibitem{Andrews77}
H.~C. Andrews and B.~R. Hunt,
\newblock {\em Digital Image Restoration},
\newblock Prentice-Hall, Englewood Cliffs, \sca{nj}, 1977.

\bibitem{Hunt71}
B.~R. Hunt,
\newblock ``A matrix theory proof of the discrete convolution theorem,''
\newblock {\em \uppercase{ieee} {T}rans. {A}utomat. {C}ontr.}, vol. AC-19, pp.
  285--288, 1971.

\bibitem{Hunt72a}
B.~R. Hunt,
\newblock ``A theorem on the difficulty of numerical deconvolution,''
\newblock {\em \uppercase{ieee} {T}rans. {A}utomat. {C}ontr.}, vol. AC-20, pp.
  94--95, 1972.

\bibitem{Hunt72b}
B.~R. Hunt,
\newblock ``Deconvolution of linear systems by constrained regression and its
  relationship to the {W}iener theory,''
\newblock {\em \uppercase{ieee} {T}rans. {A}utomat. {C}ontr.}, vol. AC-17, pp.
  703--705, 1972.

\bibitem{Djafari97n}
A.~Mohammad-Djafari,
\newblock ``Binary polygonal shape image reconstruction from a small number of
  projections,''
\newblock {\em Elektrik}, vol. 5, no. 1, pp. 127--138, 1997.

\bibitem{Djafari99a}
A.~Mohammad-Djafari and C.~Soussen,
\newblock ``Compact object reconstruction,''
\newblock in {\em Discrete Tomography: Foundations, Algorithms and
  Applications}, G.~T. Herman and A.~Kuba, Eds., chapter~14, pp. 317--342.
  {B}irkhauser, Boston, \sca{ma}, 1999.

\bibitem{djafari02d}
A.~Mohammad-Djafari,
\newblock ``Bayesian approach with hierarchical markov modeling for data fusion
  in image reconstruction applications,''
\newblock in {\em Fusion 2002, 7-11 Jul., Annapolis, Maryland, USA}, {J}uly
  2002.

\bibitem{djafari02c}
A.~Mohammad-Djafari,
\newblock ``Fusion of x ray and geometrical data in computed tomography for non
  destructive testing applications,''
\newblock in {\em Fusion 2002, 7-11 Jul., Annapolis, Maryland, USA}, {J}uly
  2002.

\bibitem{djafari02b}
A.~Mohammad-Djafari,
\newblock ``Hierarchical markov modeling for fusion of x ray radiographic data
  and anatomical data in computed tomography,''
\newblock in {\em Int. Symposium on Biomedical Imaging (ISBI 2002), 7-10 Jul.,
  Washington DC, USA}, {J}uly 2002.

\bibitem{djafari03b}
A.~Mohammad-Djafari,
\newblock ``Fusion bay\'esienne de donn\'ees en imagerie x et ultrasonore,''
\newblock in {\em GRETSI 03, France}, {S}ep. 2003.

\bibitem{djafari97m}
A.~Mohammad-Djafari,
\newblock ``Solving inverses problems: From deterministic to probabilistic
  approaches,''
\newblock in {\em Seminar in Electrical Eng. Dept. of Purdue University,
  \sca{in}}, {D}ec. 1997.

\bibitem{djafari99e}
A.~Mohammad-Djafari, N.~Qaddoumi, and R.~Zoughi,
\newblock ``A blind deconvolution approach for resolution enhancement of
  near-field microwave images,''
\newblock in {\em Mathematical modeling, Bayesian estimation and Inverse
  problems, SPIE 99, Denver, Colorado, USA}, F.~Pr\^eteux, A.~Mohammad-Djafari,
  and E.~Dougherty, Eds., 1999, vol. 3816, pp. 274--281.

\bibitem{djafari02a}
A.~Mohammad-Djafari, J.-F. Giovannelli, G.~Demoment, and J.~Idier,
\newblock ``Regularization, maximum entropy and probabilistic methods in mass
  spectrometry data processing problems,''
\newblock {\em Int. Journal of Mass Spectrometry}, vol. 215, no. 1-3, pp.
  175--193, {A}pr. 2002.

\bibitem{demoment01}
G.~Demoment, J.~Idier, J.-F. Giovannelli, and A.~Mohammad-Djafari,
\newblock ``Probl\`emes inverses en traitement du signal et de l'image,''
\newblock vol. TE 5 235 of {\em Trait\'e T\'el\'ecoms}, pp. 1--25. {T}echniques
  de l'{I}ng\'enieur, Paris, France, 2001.

\bibitem{nikolova98}
M.~Nikolova, J.~Idier, and A.~Mohammad-Djafari,
\newblock ``Inversion of large-support ill-posed linear operators using a
  piecewise {G}aussian \sca{mrf},''
\newblock {\em \uppercase{ieee} {T}rans. {I}mage {P}rocessing}, vol. 7, no. 4,
  pp. 571--585, {A}pr. 1998.

\bibitem{Idier96a}
J.~Idier, A.~Mohammad-Djafari, and G.~Demoment,
\newblock ``Regularization methods and inverse problems: an information theory
  standpoint,''
\newblock in {\em 2nd International Conference on Inverse Problems in
  Engineering}, Le Croisic, France, {J}une 1996, pp. 321--328.

\bibitem{Idier01a}
J.~Idier, Ed.,
\newblock {\em Approche bayésienne pour les problèmes inverses},
\newblock Traité \sca{IC}2, Série traitement du signal et de l'image, Hermès,
  Paris, 2001.

\bibitem{Idier01b}
J.~Idier,
\newblock ``Convex half-quadratic criteria and interacting auxiliary variables
  for image restoration,''
\newblock {\em \uppercase{ieee} {T}rans. {I}mage {P}rocessing}, vol. 10, no. 7,
  pp. 1001--1009, {J}uly 2001.

\bibitem{Idier00a}
J.~Idier,
\newblock {\em Probl\`emes inverses en restauration de signaux et d'images},
\newblock {H}abilitation à diriger des recherches, Universit\'e de Paris-Sud,
  Orsay, France, {J}uly 2000.

\bibitem{Snoussi00a}
H.~Snoussi and A.~Mohammad-Djafari,
\newblock ``Bayesian source separation with mixture of {G}aussians prior for
  sources and {G}aussian prior for mixture coefficients,''
\newblock in {\em Bayesian Inference and Maximum Entropy Methods},
  A.~Mohammad-Djafari, Ed., Gif-sur-Yvette, France, {J}uly 2000, Proc. of
  MaxEnt, pp. 388--406, {A}mer. {I}nst. {P}hysics.

\bibitem{snoussi04c}
{{Hichem Snoussi} AND {Ali Mohammad-Djafari}},
\newblock ``Fast joint separation and segmentation of mixed images,''
\newblock {\em {J}ournal of {E}lectronic {I}maging}, vol. 13, no. 2, pp.
  349--361, April 2004.

\bibitem{snoussi04b}
{{Hichem Snoussi} AND {Ali Mohammad-Djafari}},
\newblock ``Bayesian unsupervised learning for source separation with mixture
  of gaussians prior,''
\newblock {\em {J}ournal of {V}{L}{S}{I} {S}ignal {P}rocessing {S}ystems}, vol.
  37, no. 2/3, pp. 263--279, June/July 2004.

\bibitem{ichir06b}
{{Mahieddine Ichir} AND {Ali Mohammad-Djafari}},
\newblock ``Hidden markov models for blind source separation,''
\newblock {\em {IEEE} {T}rans. on {S}ignal {P}rocessing}, vol. 15, no. 7, pp.
  1887--1899, Jul 2006.

\bibitem{Snoussi02c}
H.~Snoussi and A.~Mohammad-Djafari,
\newblock ``{I}nformation {G}eometry and {P}rior {S}election.,''
\newblock in {\em Bayesian Inference and Maximum Entropy Methods}, C.~Williams,
  Ed. MaxEnt Workshops, {A}ug. 2002, pp. 307--327, {A}mer. {I}nst. {P}hysics.

\bibitem{Snoussi03the}
H.~Snoussi,
\newblock {\em Bayesian approach to source separation. Applications in
  imagery},
\newblock Ph.D. thesis, University of Paris--Sud, Orsay, France, september
  2003.

\bibitem{Snoussi04a}
H.~Snoussi and A.~Mohammad-Djafari,
\newblock ``Fast joint separation and segmentation of mixed images,''
\newblock {\em {J}ournal of {E}lectronic {I}maging}, vol. 13, no. 2, pp.
  349--361, {A}pr. 2004.

\bibitem{djafari03a}
A.~Mohammad-Djafari,
\newblock ``Bayesian approach for inverse problems in optics,''
\newblock in {\em SPIE03, USA}, {S}ep. 2003.

\bibitem{Feron05a}
O.~F\'eron and A.~Mohammad-Djafari,
\newblock ``Image fusion and joint segmentation using an {MCMC} algorithm,''
\newblock {\em {J}ournal of {E}lectronic {I}maging}, vol. 14, no. 2, pp. paper
  no. 023014, Apr 2005.

\bibitem{Feron05b}
O.~F\'eron, D.~B., and A.~Mohammad-Djafari,
\newblock ``Microwave imaging of inhomogeneous objects made of a finite number
  of dielectric and conductive materials from experimental data,''
\newblock {\em {I}nverse {P}roblems}, vol. 21, no. 6, pp. 95--115, Dec 2005.

\bibitem{mohammadpour04ce}
A.~Mohammadpour, O.~Feron, and A.~Mohammad-Djafari,
\newblock ``Bayesian segmentation of hyperspectral images,''
\newblock in {\em BAYESIAN INFERENCE and MAXIMUM ENTROPY METHODS IN SCIENCE and
  ENGINEERING: 24th International Workshop on Bayesian Inference and Maximum
  Entropy Methods in Science and Engineering}. 2004, vol. 735, pp. 541--548,
  AIP.

\bibitem{mohammadpour05ca}
A.~Mohammad-Djafari and A.~Mohammadpour,
\newblock ``Hyperspectral image processing using a bayesian classification
  approach,''
\newblock in {\em Proceedings of PSIP 2005, Physics in Signal and Image
  Processing}. 2005, pp. 245--250, PSIP 2005, Physics in Signal and Image
  Processing.

\bibitem{bali06cc}
{{Nadia Bali} and {Ali Mohammad-Djafari}},
\newblock ``Joint dimensionality reduction, classification and segmentation of
  hyperspectral images,''
\newblock in {\em ICIP 2006}. {O}ct. 2006, ICIP06, October 8-11, Atlanta, GA,
  USA.

\bibitem{bali06ca}
{{Nadia Bali} and {Ali Mohammad-Djafari}},
\newblock ``Hierarchical markovian models for joint classification,
  segmentation and data reduction of hyperspectral images,''
\newblock in {\em ESANN 2006}. {S}ep. 2006, ESANN 2006, September 4-8, Belgium.

\bibitem{bali06cb}
{{Nadia Bali} and {Ali Mohammad-Djafari}},
\newblock ``Hierarchical markovian models for hyperspectral image
  segmentation,''
\newblock in {\em ICPR 2006}. {A}ug. 2006, ICPR06, Aug. 20-24, Hong Gong.

\end{thebibliography}

\end{document}